\documentclass[twocolumn, twocolappendix]{aastex631}

\shorttitle{AGB SFHs}
\shortauthors{Lee et al.}
\graphicspath{{./}{}}

\defcitealias{williams17}{W17}

\begin{document}

\title{Measuring Star Formation Histories from Asymptotic Giant Branch Stars I: A Demonstration in M31}

\author[0000-0002-5865-0220]{Abigail~J.~Lee}\affil{Department of Astronomy \& Astrophysics, University of Chicago, 5640 South Ellis Avenue, Chicago, IL 60637}\affiliation{Kavli Institute for Cosmological Physics, University of Chicago,  5640 South Ellis Avenue, Chicago, IL 60637}\affil{Department of Astronomy, University of California, Berkeley, CA 94720-3411, USA}

\author[0000-0002-6442-6030]{Daniel~R.~Weisz}\affil{Department of Astronomy, University of California, Berkeley, CA 94720-3411, USA}

\author[0000-0003-1218-8699]{Yi Ren}\affil{College of Physics and Electronic Engineering, Qilu Normal University, Jinan 250200, China}

\author[0000-0002-1445-4877]{Alessandro~Savino}\affil{Department of Astronomy, University of California, Berkeley, CA 94720-3411, USA}

\author[0000-0001-8416-4093]{Andrew~E.~Dolphin}\affil{Raytheon, 1151 E. Hermans Road, Tucson, AZ 85756, USA}\affil{Steward Observatory, University of Arizona, 933 N. Cherry Avenue, Tucson, AZ 85719, USA}

\begin{abstract}

We demonstrate how near-infrared (NIR) imaging of resolved luminous asymptotic giant branch (AGB) stars can be used to measure well-constrained star formation histories (SFHs) across cosmic time.  Using UKIRT $J$ and $K$-band imaging of M31, we first show excellent agreement over the past $\sim8$~Gyr between the PHAT SFH of M31’s outer disk derived from a deep optical color-magnitude diagram (CMD; $\sim3.3\times10^{7}$ stars with $M_{\rm F814W} \lesssim +2$), and our spatially-matched SFH based only on modeling AGB stars on a NIR CMD ($\sim7.7\times10^{3}$ stars with $M_{\rm J} \lesssim -6$).  We find that only $\sim1000$ AGB stars are needed for reliable SFH recovery, owing to their excellent age sensitivity in the NIR.  We then measure the spatially-resolved SFH of M31’s inner stellar halo ($D_{\rm M31, projected} \sim20-30$~kpc) using $\sim10^4$ AGB stars. We find: (i) a dominant burst of star formation across M31's stellar halo $3-5$~Gyr ago and lower level, spatially distributed star formation $\sim1-3$~Gyr ago; (ii)  $M_{\star}\approx3_{-1}^{+5}\times10^9 M_{\odot}$ formed over the past $\sim8$~Gyr.  We discuss some caveats and the enormous potential of resolved AGB stars in the NIR for measuring SFHs back to ancient epochs ($\sim14$~Gyr ago) in galaxies to large distances ($D\gtrsim20$~Mpc)  with JWST, Roman, and Euclid. 

\end{abstract}

\correspondingauthor{Abigail J. Lee}\email{abbyl@uchicago.edu}


\section{Introduction} \label{sec:intro}

Asymptotic giant branch (AGB) stars represent the final nuclear-fusing evolutionary phase for low and intermediate-mass ($0.5-8M_{\odot}$), intermediate-age ($\sim100$~Myr to $\sim14$ Gyr ago) stars. With peak spectral energy distributions at $\sim1.5\mu$m, they are among the brightest stars in the near-infrared (NIR; \citealt{Maraston06, Melborne2012, Melbourne2013}), and are known to exist in abundance in a wide variety of galaxies from low-metallicity dwarf galaxies to high-metallicity massive ellipticals (e.g., \citealt{Gallart96, dalcanton12a, rejkuba22, Anand24a, Anand24b, Hoyt2024, lee24}). Furthermore, there exists a clear mapping between some of their observable properties (i.e., NIR colors and luminosities) and their ages \citep[e.g.,][]{gallart05}.

However, despite being ubiquitous, observational challenges and the physical complexity of AGB star physics have resulted in them playing a limited role in star formation history (SFH) measurements from optical resolved star color-magnitude diagrams (CMDs) in favor of better understood phases of evolution (e.g., main sequence turnoff, sub-giant branch, red clump; \citealt{gallart05}).   In terms of observations, in the optical it is challenging to cleanly differentiate between AGB and red giant branch (RGB) stars, which are far less sensitive to age, as RGB stars with different ages can occupy similar regions of the CMD.  From a physics standpoint, accurately modeling the color and magnitudes of AGB stars on a galaxy's CMD requires a detailed understanding of complex processes in the interiors and atmospheres of AGB stars such as convection, overshooting, hot bottom burning, mass loss, dredge up, and pulsation \citep{gallart05, marigo17, pastorelli19, pastorelli20}.  The comparatively historically better-understood physics for other age-sensitive populations, like the main sequence turn-off (MSTO), subgiant branch (SGB), and red clump (RC), has made them the staple of accurate SFH modeling despite their less favorable observational properties, such as faintness and increased susceptibility to crowding (e.g., \citealt{gallart05}).

Nevertheless, a handful of pioneering studies recognized the potential of AGB stars for age determinations. \citet{Gallart96} was the first to suggest AGB stars as a means of measuring a SFH through CMD modeling, noting: ``\textit{...the potential power of AGBs as tracers of the SFH at intermediate ages ($1-10$~Gyrs). It can reasonably be expected that, in the near future when uncertainties in the stellar evolutionary models for AGBs have been overcome, AGBs will provide strong, detailed insight into the SFH for that time interval.}''
Since then, NIR observations of AGB stars have been used to map spatially resolved stellar ages in nearby galaxies \citep[e.g.,][]{Frogel83, Reid84, Wood85, Costa96, Davidge03, Davidge05, Cioni06, cioni08, melbourne10, Jung12, crnojevic13, davidge14, McQuinn2017}.  However, many of these studies relied on fairly rudimentary AGB star models and, in many cases, undertook simplified statistical approaches that did not capture the full complexity of CMD modeling or only pursued qualitative analysis. For example, \cite{javadi11} began an ambitious observational campaign to measure the SFHs of nearby galaxies by comparing the J and K band luminosity functions of their AGB populations to isochrones to estimate their initial stellar masses, and therefore ages.  For computational tractability, they adopted simplifying assumptions like a single metallicity and did not use artificial star tests (ASTs).  This methodology has been extended to a number of subsequent studies that highlight the scientific power of AGB stars in the NIR (e.g., \citealt{rezaeikh14, hamedani17}). 

A more recent approach has been to compute the ratio of AGB to RGB stars in optical imaging to estimate the quenching epoch (i.e., the age of the youngest stellar population) using theoretical models or empirical relations derived from the SFHs of Local Group (LG) galaxies with deep CMDs \citep[e.g.,][]{rejkuba22, harmsen23, Velguth24}.  Its application to several nearby galaxies both showcases the broad strengths of AGB stars for measuring important star formation timescales in galaxies outside the LG, while also highlighting limitations of AGB stars on optical CMDs.

In this paper, we combine NIR imaging, significant improvements in theoretical AGB stars models, and well-established CMD fitting techniques to measure quantitative, `non-parametric' SFHs using only AGB stars.  The increase in high-quality NIR resolved star photometry, in concert with substantial theoretical efforts, particularly by the \texttt{PARSEC-COLIBRI} group, have resulted in a new generation of AGB star models that are anchored to high quality datasets \citep{ marigo17, pastorelli19, pastorelli20} and contemporary stellar physics. These models employ far more sophisticated physics than their predecessors, such as the seminal Padova models \citep{Marigo07, marigo08}, which were the first to include a range of physical processes during the thermal pulsating (TP) AGB phase. These effects include the third dredge-up, hot-bottom burning, low-temperature opacity changes, C-to-M star transitions, mass loss, and pulsation.  More recently, the Padova group introduced the \texttt{COLIBRI} code in \citet{marigo13}, which built upon and eventually succeeded the Padova isochrones for AGB stars.   The \texttt{COLIBRI} models were further improved using NIR resolved star observations of AGB stars in nearby dwarf galaxies \citep{Rosenfield2014, rosenfield16,marigo17}. TP-AGB lifetimes were even more tightly constrained using deep NIR imaging of the Magellanic Clouds from the VMC/VISTA survey, resulting in the current \texttt{COLIBRI} TP-AGB star models that impressively reproduce a number of observational constraints \citep{pastorelli19, pastorelli20}.

Here, we take the next step in SFH modeling by combining the \texttt{COLIBRI} AGB star models with the well-tested CMD modeling framework \texttt{MATCH} \citep{dolphin02} in order to measure quantitative SFHs exclusively from NIR observations of AGB stars. \texttt{MATCH} is a widely used, mature CMD fitting code that has been used to measure SFHs for $>100$ galaxies from resolved star CMDs \citep[e.g.,][]{Skillman2003,Williams09,McQuinn10,Monelli10,Weisz2011,Cole14, Weisz2014,Lewis15,McQuinn2015,Skillman17,williams17,Savino2023,Weisz23a,McQuinn2024a,McQuinn2024b, McQuinn2024c}.  Though the incorporation of the \texttt{COLIBRI} models into \texttt{MATCH} is challenging (e.g., because of the need to correctly accounting for self-extinction in AGB stars), it provides several important benefits including a well-vetted framework for CMD analysis, robust uncertainty determination, and uniformity with \texttt{MATCH}-based SFHs of nearby galaxies based on optical CMDs, which we use to validate the reliability of our AGB star SFHs.

Using this newly established framework, we take several steps to illustrate the technique in practice.  We first fit mock data to illustrate self-consistency in our modeling.  We then measure the SFH of M31's outer disk using ground-based NIR observations.  This region was selected to overlap with the Panchromatic Hubble Andromeda Treasury (PHAT) survey \citep{Dalcanton12b, williams14, williams23}, which measured the spatially resolved SFH of the same region using deep \textit{Hubble Space Telescope} (HST) optical CMDs \citep[hereafter \citetalias{williams17}]{williams17}, providing a useful comparison point for our AGB star SFHs.  Finally, we demonstrate the scientific potential of this new method by measuring the spatially resolved SFH of M31's inner stellar halo using only ground-based NIR CMDs of AGB stars.

With this decade's revolution of NIR telescopes like the \textit{James Webb Space Telescope (JWST), Euclid}, and the \textit{Nancy Grace Roman Space Telescope (Roman)}, NIR imaging of AGB stars is already becoming abundantly available \citep{Weisz23b, Anand24a, Anand24b, Boyer24, Hoyt2024, Hunt24, lee24, Li24,   Nally24, Newman2024}, and is expected to substantially grow in the next decade. Our study lays the foundational groundwork for combining these exquisite NIR observations of AGB stars and their improved stellar models to measure SFHs of galaxies significantly farther than what is currently possible using optical CMDs.

This paper is organized as follows.  We summarize the \texttt{COLIBRI} AGB star models in \S \ref{sec:isochrones}. We describe the \texttt{MATCH}-\texttt{COLIBRI} SFH fitting methodology in \S \ref{sec:methodology}.  We introduce our M31 observations, photometry, and ASTs in \S \ref{sec:Data}.  We undertake self-consistency tests with mock data in \S \ref{sec:mock_summary} and then validate our SFHs against PHAT-based SFHs of M31's disk in \S \ref{subsec:m31}.  Finally, we measure the SFH of M31's inner stellar halo in \S \ref{subsec:m31_halo}, briefly discuss our findings and current limitations of this method in \S \ref{sec:caveats}.  We conclude by outlining future prospects of SFHs based on resolved AGB star CMDs in the NIR in \S \ref{subsec:future}.

Throughout this paper, we adopt the following M31 geometric parameters used by the PHAT team: center R.A. and Declination = ($\rm{00^h42^m44^s.3}$, $\rm{+41^{\circ}16\arcmin09\arcsec}$) and position angle ($\rm{P.A.}) =38^{\circ}$ \citep{mcconnachie05,barmby06}. We assume the inclination angle of M31 to be $78^{\circ}$ \citep{nieten06}. Unless otherwise noted, we use projected distances in M31.

\section{AGB Theoretical Isochrones}\label{sec:isochrones}

\begin{figure}
\centering
\includegraphics[width=\columnwidth]{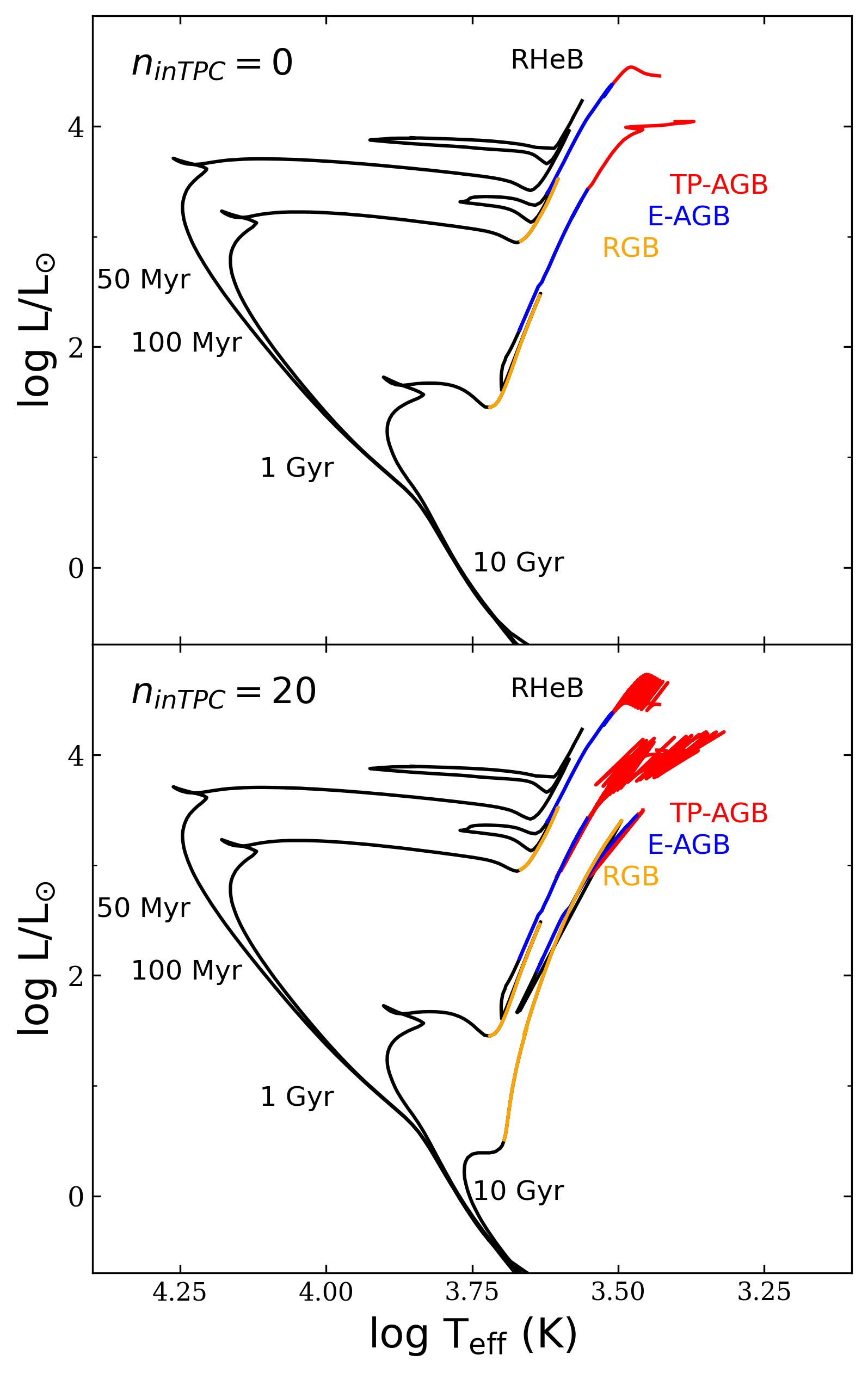}
\caption{Hertzsprung-Russel diagram for four different \texttt{PARSEC-COLIBRI} isochrones with metallicity [M/H] = 0.0~dex. The TP-AGB evolutionary stage is shown in red, the E-AGB phase is shown in blue, and the RGB phase is shown in orange. The red helium-burning (RHeB) evolutionary phase is also labeled.} The top panel shows AGB models with the number of AGB star thermal pulse cycles $\rm{n_{inTPC}}=0$ and the bottom panel shows AGB models with $\rm{n_{inTPC}}=20$.
\label{fig:hr}
\end{figure}

\begin{figure}
\includegraphics[width=\columnwidth]{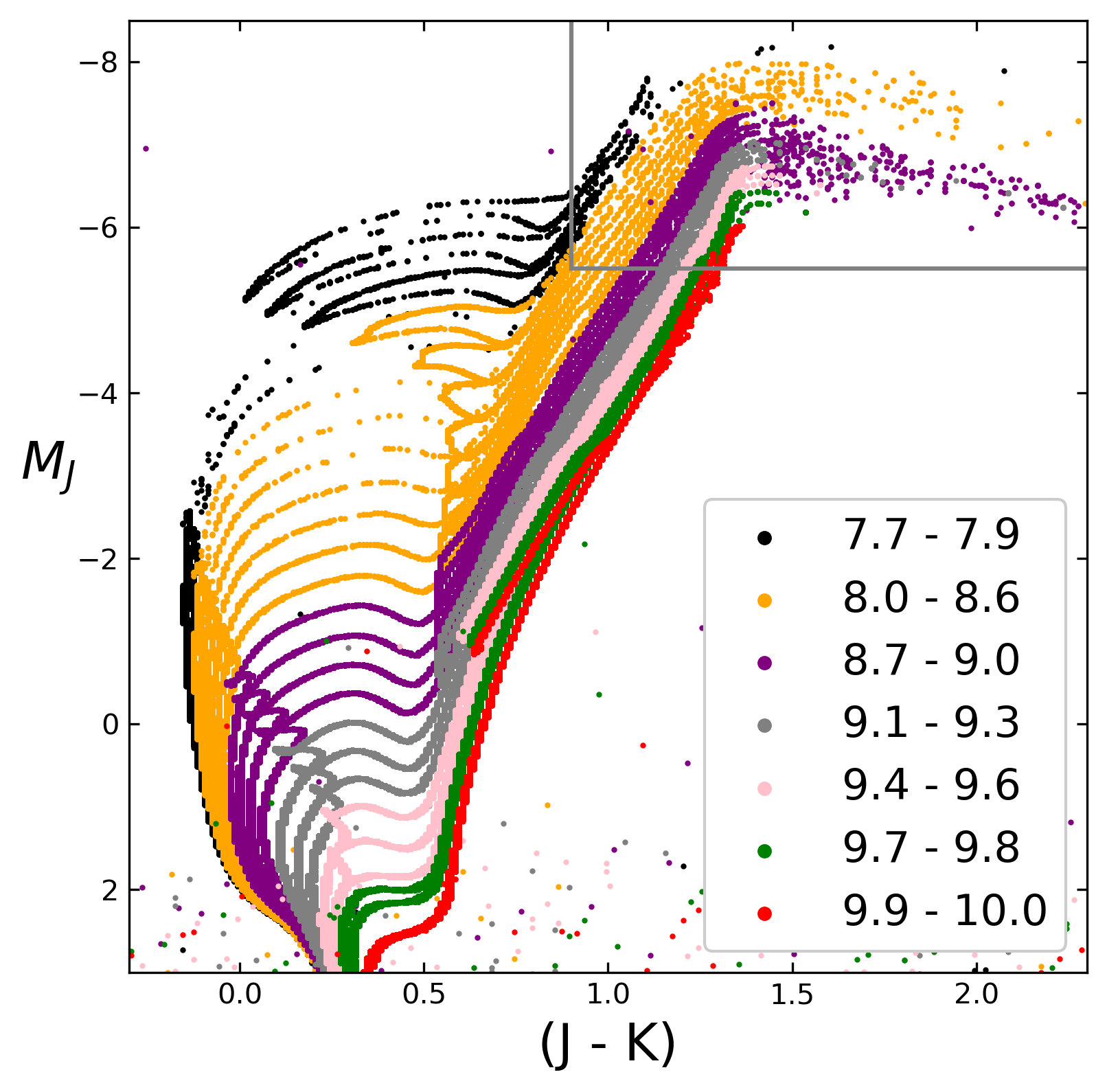}
\caption{Simulated CMDs using \texttt{PARSEC-COLIBRI} models that demonstrate the power of AGB stars (highlighted within the grey box) as age indicators from $\sim50$~Myr to $\sim10$~Gyrs. All CMDs were constructed using a constant star formation rate (SFR) of 0.1~$M_{\odot}~\rm{yr^{-1}}$ with a constant metallicity of [M/H] = 0.0~dex. The legend shows the log(age) ranges of each SSP. 
}\label{fig:isochrones}  
\end{figure}

We begin with a brief review of theoretical predictions for AGB star properties on the HR diagram and on a NIR CMD from the \texttt{COLIBRI} models.  Throughout this paper, we use the \texttt{COLIBRI} version \texttt{S\_37} isochrones (CMD version 3.7; \citealt{bressan12, chen15, chen14, tang14, marigo17, pastorelli19, pastorelli20}).\footnote{\url{http://stev.oapd.inaf.it/cgi-bin/cmd}} 

Figure \ref{fig:hr} shows select \texttt{PARSEC-COLIBRI} isochrones for select ages between 50 Myr and 10 Gyr.  The top panel shows an HR diagram with the number of AGB star thermal pulse cycles (TPC) set to $n_{\rm{inTPC}}=0$, while the bottom one has  $n_{\rm{inTPC}}=20$.  The latter value is recommended by \citet{marigo17} for realistic resolved stellar populations analysis.  We discuss this point further below.

Young AGB stars are typically the more luminous cool stars on the HR diagram, and have higher initial masses than older and fainter AGB stars.  However, an important observational property is that even faint, lower-mass AGB stars are generally more luminous than the tip of the red giant branch (TRGB) in the near-IR.  Moreover, the change in AGB age as a function of luminosity (i.e., the age gradient on the HR diagram) is much more pronounced for AGB stars than RGB stars. 

The AGB is separated into two stages: the early AGB (E-AGB) and thermally-pulsating AGB (TP-AGB). In the E-AGB, the source of energy is the He-burning shell, whereas for in the TP-AGB, the H-burning shell and He-burning shell ignite in alternating pulses \citep{habing03}. All AGB stars begin in the E-AGB phase, where their luminosities increase and temperatures decrease monotonically as a function of time \citep{Iben83}.  In the HR diagram, RGB and E-AGB stars of different ages are difficult to separate in the same area of the CMD \citep{gallart05}, as shown in Figure \ref{fig:hr}. Hence, much of our SFH information is derived from  TP-AGB stars, except for the very oldest ages for which both E-AGB and TP-AGB stars provide leverage.

The age range we plot is for illustrative purposes only.  AGB stars can exist down to $\sim100$~Myr in age; below this age, stars do not progress through the AGB star phase because they are too massive to undergo the AGB phase.  For the very oldest ages ($\gtrsim10$~Gyr ago), age information comes primarily from E-AGB stars, as lower mass stars generally do not go through the TP-AGB phase.  As shown in Figure~\ref{fig:hr}, there is decreased separation between the RGB, E-AGB, and TP-AGB (when present) phases at older ages.  In principle, it it possible to measure SFHs using only AGB stars back to $\sim14$~Gyr ago, as demonstrated by our simulations in Appendix \ref{sec:oldest_ages}.

Figure \ref{fig:isochrones} shows a NIR CMD, containing stars with a variety of ages at a fixed metallicity of [M/H] $= 0.0$~dex. 
A modest amount of simulated noise has been added to the CMD to illustrate the AGB stars of different ages are readily distinguishable in a NIR CMD. 
Specifically, we generated and applied ASTs with a 50\% completeness level at $M_J = 7$~mag.
The location of AGB stars in the $J-K$ color combination is highlighted by the grey box, 
which approximately denotes the location of the AGB stars as defined in the $\texttt{COLIBRI}$ stellar isochrones.
These same AGB stars are not easily distinguished  from RGB stars at optical wavelengths, as we illustrate in later in the context of M31.  Even in the red-optical (e.g., $\sim8000$ \AA), the spectral energy distributions (SEDs) of AGB stars are largely indistinguishable from RGB stars; shifting to even bluer wavelengths makes separation of RGB and AGB stars less tenable.  The $1-3\mu$m wavelength range appears optimal for isolating AGB stars on a CMD.
% There are clearly fewer AGB stars than RGB stars.  \citet{marigo17} predict that $\sim10^{-3}$ AGB stars form per unit stellar mass.  Thus, while AGB stars far are more age-sensitive than RGB stars and much brighter than comparably age-sensitive MSTO and SGB stars of the same age, there are fewer of them.  We estimate the minimum number of AGB stars for accurate CMD modeling in Appendix \ref{sec:how_many}.  
% AJL NOTE 2/6/25: DELETED AFTER V1 ^
Finally, we note that `extreme' AGB stars (i.e., the reddest AGB stars with $(J-K)>2.0$~mag), which  suffer from particularly complex physics (e.g., rapid and irregular mass-loss, circumstellar extinction), are so rare that they do not meaningfully contribute to the SFH determinations we describe in this paper.

%Though the physics of AGB stars are known to be complex, it is the very reddest AGB stars on the NIR CMD (and coolest on the HR diagram) that have the most complicated physics (e.g., complicated mass-loss, circumstellar extinction).  These `extreme' AGB stars are rare and do not contribute meaningully to the age determinations we describe in this paper.

%no AGB stars are also $\sim5$~mag brighter in the J band than in the V band.Furthermore, little age information can be extracted from the RGB because the colors and magnitudes of RGB stars only depend weakly on age. On the other hand, the colors and magnitudes of AGB stars depend much stronger on their age.

\begin{figure}
\centering
\includegraphics[width=\columnwidth]{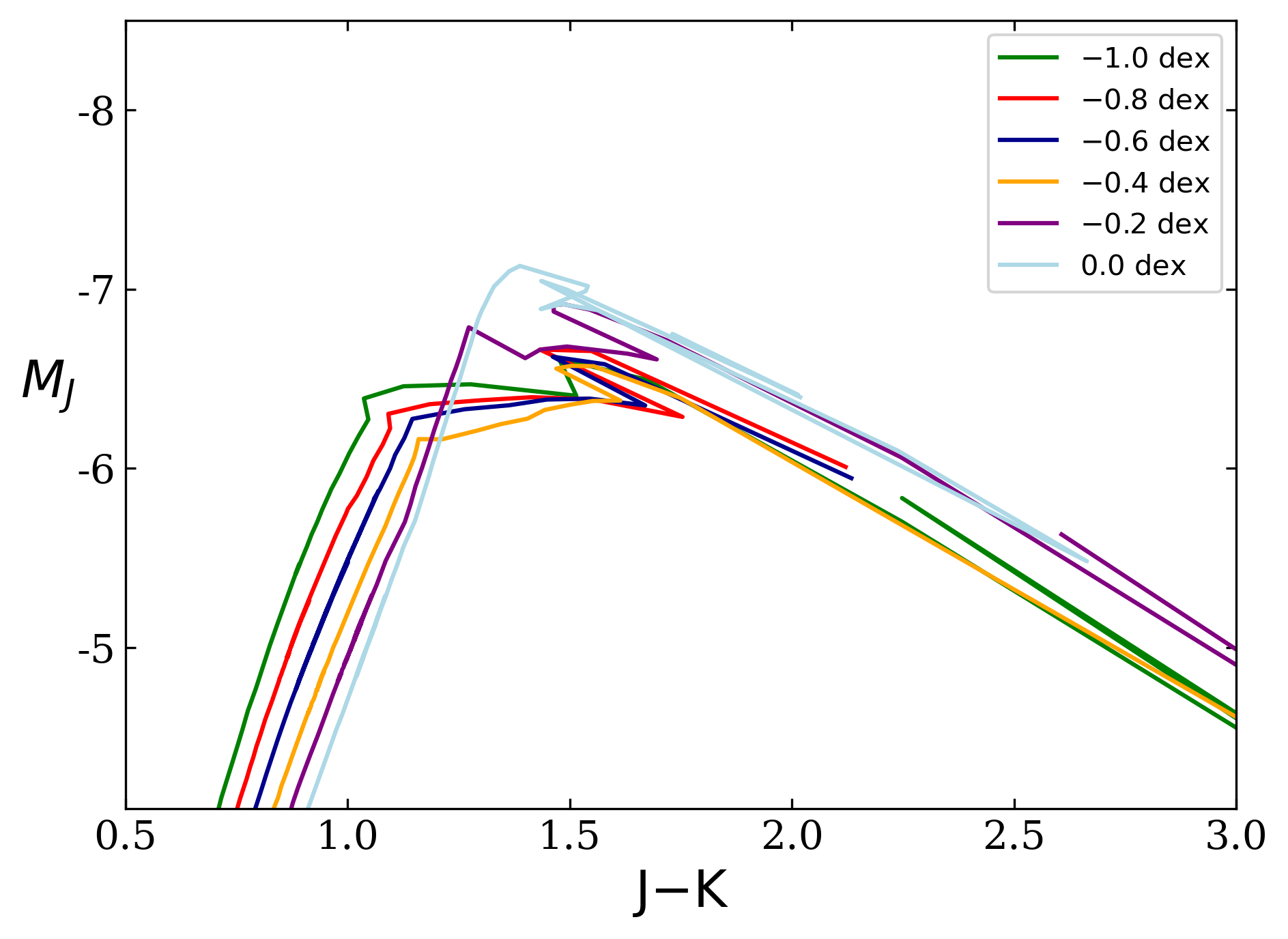}
\caption{\texttt{COLIBRI} isochrones of the asymptotic giant branch with a constant age of 1 Gyr and a range of metallicities.  Metallicity has a modest affect on AGB star positions in a NIR CMD compared to age.}
\label{fig:metal}
\end{figure}

In Figure \ref{fig:metal}  we show the behavior of AGB stars with a range of metallicities at a fixed age of 1~Gyr. Though there is clearly some dependence on metallicity, the influence of age is much more pronounced.  For a deeper demonstration of this point, Figure 11 in \cite{Parada21} shows \texttt{COLIBRI} isochrones with ages log(age) = 8.6 -- 9.4~dex (i.e, 400 Myr -- 2.5 Gyr)  and metallicities $Z_{ini}=0.001 - 0.010$ (i.e., [M/H] $\approx -1.2 -0.0$~dex). Above our cutoff selection magnitude for AGB stars of $M_J=-5.5$~mag, the average magnitude of the youngest 400 Myr AGB stars is about $M_J\approx -7.3$~mag, whereas the average magnitude of the oldest 2.5 Gyr AGB stars is about 1~mag fainter at $M_J\approx -6.0$~mag.  On the other hand, the most metal-poor AGB stars at [M/H] = -1.2~dex have an average magnitude of about $M_J\approx-6.3$~mag, whereas the average magnitude of the metal-rich AGB stars at [M/H] = 0.0~dex is much more similar at $M_J\approx-6.2$~mag. 
In summary: for AGB stars, a spread in log(age) of 0.8~dex spans  $\sim1.3$~mag, whereas a spread in metallicity of 1.2~dex spans $\sim0.1$~mag, demonstrating the age of AGB stars affects their magnitudes more significantly than their metallicities.
% Figure 10, and an associated discussion, in \citet{marigo17} provides a deeper demonstration of this point. 

\section{Methodology}\label{sec:methodology}

Having provided a qualitative foundation for the age sensitivity of AGB stars in the NIR, we now describe the process by which we formally model AGB stars on a  NIR CMD to measure a SFH.

To fit the observed CMDs, we employ the widely-used CMD fitting package \texttt{MATCH} (version 2.7, \citealt{dolphin02,dolphin12,dolphin13}).  Given a set of stellar models, a specified initial mass function (IMF), and a unresolved binary fraction, \texttt{MATCH} creates and linearly combines simple stellar population (SSP) models to create a composite CMD.  It then convolves the composite CMD with a noise model (i.e., bias, scatter, completeness) from ASTs to create a model CMD with the same observational effects as the data.  The model and observed CMD are converted into Hess diagrams (i.e., a binned CMD) and compared using a Poisson likelihood function.  \texttt{MATCH} then iterates through various weights of the SSP basis functions until the overall likelihood function is maximized.
The synthetic model CMD that best matches the observed CMD is used as the best fit SFH.  Random uncertainties on the SFH are computed using a Markov chain Monte Carlo approach, as we describe below.

Because of the complexity of AGB star models (i.e., the need for self-extinction corrections and challenge in defining equivalent evolutionary points), we cannot use \texttt{MATCH}'s native SSP model construction framework with the \texttt{COLIBRI} stellar models.  Instead, we construct custom SSP models in the desired filters directly from \texttt{COLIBRI} isochrones and then use these pre-constructed models within \texttt{MATCH}'s CMD modeling  framework.  Aside from SSP model construction, the remainder of the CMD fit is virtually identical to the commonly used modes of \texttt{MATCH} described here and elsewhere in the literature \citep[e.g.,][]{dolphin02,Skillman2003,Weisz2014}.

Starting from the \texttt{COLIBRI} models, we generate a grid of single stellar population models, each with a total stellar mass of $10^8~M_{\odot}$ (which yields $\gtrsim10^4$ AGB stars), spanning a range of ages and metallicities.  Our SSP models have a metallicity range of [M/H] $= -1.4$ to 0.0 with a spacing of 0.2~dex, and ages from $\log(\rm{age})=8.0-9.9$ dex (100 Myr to 7.9 Gyr in linear time), logarithmically spaced by 0.1~dex. 
In principle, we could generate AGB stars models back to 14~Gyr ago for SFH determination.  However, we stop at a younger lookback time in this study (1) because we were limited by the completeness of our datasets, and (2) to match the PHAT SFH we use to empirically validate our method; both reasons are described further in \S\ref{subsec:m31}.  We also note that the single stellar populations we generated are close enough to each other in log(age) and metallicity space that the artificial star based photometric errors would fill in the gaps between them. For example, a simple stellar population of log(age)=9.2~dex sufficiently represents the stellar population range from log(age)=9.15 to 9.25~dex.

Following recommendations from \citet{marigo17} for `detailed population studies aimed at accurately reproducing star counts,' we set $n_{\rm{inTPC}}=20$ (i.e., the number of thermal pulse cycles).  For comparison, and as highlighted in Figure \ref{fig:hr}, isochrones with $n_{\rm{inTPC}}=0$ contain only the quiescent points of the TP-AGB phase (i.e., any luminosity variations from the helium shell flashes are hidden). As discussed in \cite{marigo17}, isochrones with $n_{\rm{inTPC}}=5$ already contain the most important feature of the thermal pulses: the long-lived low luminosity dips. Thus, the use of $n_{\rm{inTPC}}=5$ is likely sufficient for accurate SFH construction.  Nevertheless,  we compute isochrones with $n_{\rm{inTPC}}=20$, as it only requires a modestly longer amount of time to compute and it yields slightly more accurate models. Further discussion of how the AGB models change with $n_{\rm{inTPC}}$ can be found in \cite{marigo17}. %, who also showed that increasing $n_{\rm{inTPC}}$ over 5 does not significantly change the color-magnitude range spanned by an AGB star of a given age.  

We generate the SSPs using a Kroupa IMF \citep{kroupa01} and assume only single stars. The latter is generally a poor assumption of most parts of the CMD.  However, the high luminosities of AGB stars, particularly in the NIR, mean that most binary combinations (e.g., AGB-MS star binary) do not significantly affect the colors and luminosities of AGB stars. 
However, an important caveat to this assumption is that our model ignores binaries with a surviving secondary star now on the AGB (e.g., an interacting system born with a $5M_{\odot}$ primary and a secondary now on the AGB whose current mass is larger than its initial mass). The fraction of AGB stars that originate in binaries is unclear, and determining that fraction is currently beyond the scope of this paper.

Finally, we generate models with a Reimers mass loss value of $\eta = 0.2$ for RGB stars.
% and a rotation value of YY.  While the rotation rate has little impact on the AGB lifetimes, the choice of mass loss can affect the oldest ages.  
For our fiducial models we adopt the default value recommended by \texttt{PARSEC-COLIBRI}, and discuss variations in this choice in \S \ref{sec:caveats}.

%We then ported these SSPs into \texttt{MATCH}, where version 2.7 of \texttt{MATCH} was developed specifically for this project to work with \texttt{COLIBRI} SSPs. 

%The SFH solutions were fitted with SSPs ranging in age from $\rm{log}(t)=8.0$ to 9.9 dex (10 Myr to 7.9 Gyr in linear time), logarithmically spaced by 0.1~dex. 

We import these models into \texttt{MATCH} in the desired filter set and use them to fit the observed CMD.  After determination of the best fit SFH, we calculate random uncertainties on the SFH using a hybrid Monte Carlo (HMC) approach implemented by \citet{dolphin13}, following guidance from other papers in the literature that use this approach in \texttt{MATCH} (e.g., \citealt{rosenfield16, williams17, lazzarini22, McQuinn2024a}).

Many of the SFHs measured from CMDs in the literature also feature systematic uncertainties (e.g., \citealt{Weisz2011, lazzarini22}).  These are designed to capture variations on the SFH due to choice of stellar model \citep{dolphin12} and other inputs \citep[e.g.,][]{Savino2023}.  The method for computing systematic uncertainties has been well-calibrated for optical CMDs of different depths using multiple stellar evolution models. However, because the \texttt{COLIBRI} models are the only models with the appropriate level of detail necessary for TP-AGB modeling, an inter-method comparison is currently not possible. For example, \cite{marigo17} includes a comprehensive discussion (see their Section 3.8, Figure 11) comparing the \texttt{COLIBRI} isochrones of the TP-AGB phase with the BaSTI, Padova, and MIST isochrones. In brief, the BaSTI isochrones lack the M to C-type AGB star transition, the Padova isochrones miss details related to mass loss, thermal pulse cycles, and hot-bottom burning, and the MIST isochrones apply M-type molecular opacities to C-type stars, leading the C-type stars to be significantly hotter/bluer than the C-type stars in the \texttt{COLIBRI} isochrones. Consequently, throughout this paper, we are only able to consider random uncertainties.  We discuss the issue of systematic uncertainties further in \S \ref{sec:caveats}.

\begin{deluxetable}{cccccc}
\tablecaption{UKIRT 50\% Completeness Limit for the Disk Region}\label{tab:disk_ast}
\tablehead{
\colhead{Stellar Density} & 
\colhead{J} &
\colhead{K} &
\colhead{Number of} & 
\colhead{Region}\\
\colhead{[stars $\rm{arcmin^{-2}}$]} & 
\colhead{[mag]}&
\colhead{[mag]}&
\colhead{AGB ASTs used\tablenotemark{a}
}& 
\colhead{}}
\startdata
25 & 18.7 &  17.7 & 504696 & Disk\\
\enddata
\tablenotetext{a}{Stars with injected magnitudes of $J<19$~mag, i.e., 1~mag fainter than our AGB selection.}
\end{deluxetable}

\begin{deluxetable*}{ccccc}
\tablecaption{UKIRT $50\%$ Completeness Limits over a Range of Stellar Densities for the Halo Regions}\label{tab:ast}
\tablehead{
\colhead{Stellar Density} & 
\colhead{J} &
\colhead{K} &
\colhead{Number of} & 
\colhead{Regions\tablenotemark{b}}\\
\colhead{[stars $\rm{arcmin^{-2}}$]} & 
\colhead{[mag]}&
\colhead{[mag]}&
\colhead{AGB ASTs used\tablenotemark{a}
}& 
\colhead{}}
\startdata
0 -- 6 & 19.4 &  18.4 & 89530 & 1, 5\\
6 -- 10  & 19.4 & 18.4 &  166270 &  4, 6\\
10 -- 14 & 19.2 & 18.4 &  204640 & 2, 3\\
\enddata
\tablenotetext{a}{Stars with injected magnitudes of $J<19$~mag, i.e., 1~mag fainter than our AGB selection.}
\tablenotetext{b}{Region borders shown in Figure \ref{fig:region_labels}.}
\end{deluxetable*}

%Systematic uncertainties are typically estimated by comparing the SFH measured from independent stellar libraries \citep{dolphin12}. However, because the \texttt{COLIBRI} models are the only models with the appropriate level of detail necessary for TP-AGB modeling, an inter-method comparison is currently impossible. For example, \cite{marigo17} includes a comprehensive discussion (see their Section 3.8, Figure 11) comparing the \texttt{COLIBRI} isochrones of the TP-AGB phase with the BaSTI, Padova, and MIST isochrones. To summarize, the BaSTI isochrones lack the M to C-type AGB star transition, the Padova isochrones miss details related to mass loss, thermal pulse cycles, and hot-bottom burning, and the MIST isochrones apply M-type molecular opacities to C-type stars, leading the C-type stars to be significantly hotter/bluer than the C-type stars in the \texttt{COLIBRI} isochrones.

\section{Data}\label{sec:Data}

For all AGB-based SFHs in this paper, we use  J and K images from the Wide Field Camera (WFCam), on the 3.8~m United Kingdom InfraRed Telescope (UKIRT), taken from 2005 to 2008. The data were reduced through the Cambridge Astronomy Survey Unit (CASU) 
WFCam pipeline \citep{irwin04, hodgkin09}, which provides chip-by-chip point-spread-function photometry calibrated to the Two Micron All Sky Survey \citep{skrutskie06}, along with source classification (i.e., stellar vs. non-stellar)\footnote{\url{http://casu.ast.cam.ac.uk/surveys-projects/wfcam}}.
This catalog was then further processed by \cite{ren21} for a study on red supergiants in M31. They combined all the measurements on a filter-by-filter basis by averaging the photometry for multiple measurements of the same stars.  This photometry covers over 30~kpc in projected radii across the face of M31, providing excellent coverage of AGB stars across M31's inner stellar halo.

From the raw source catalogs, we keep only sources classified as stellar in both J and K by the CASU pipeline. 
We then clean the final catalogs using photometric quality cuts that are demonstrated in the panels of Figure \ref{fig:error}. We use a constant$+$exponential function to cull the data via the the photometric uncertainty as a function of the J-band and K-band magnitudes. The specific functional form of these cuts can be found in Appendix \ref{sec:error}.

We generate ASTs to characterize the biases, uncertainties, and completeness of our data using the procedure described in 
\cite{ren24}.  We run ASTs across the entire PHAT footprint. The average stellar density and 50\% completeness limits of these ASTs is tabulated in Table \ref{tab:disk_ast}. However, because of the wide areal coverage of the halo data ($\sim7.5~\rm{degrees^2}$), running ASTs for each position of the UKIRT footprint is prohibitively  computationally expensive in the framework of this photometric routine.  Instead, we run ASTs in three representative stellar density ranges, using stellar density as an indicator of the expected crowding (and resulting photometric noise) in each range. We then measure the average stellar density of the six halo regions discussed in \S \ref{subsec:m31_halo}. We use the set of ASTs that were generated for regions of similar stellar density.  Table~\ref{tab:ast} lists the properties of the ASTs used.  
The process of generating ASTs in select regions of specific stellar density and applying them more broadly is also used by the PHAT/PHATTER teams, who faced a similar computational challenge of running ASTs across the entire extent of their M31 and M33 footprints, respectively \citep{williams17, lazzarini22}.  We summarize our ASTs in Table~\ref{tab:ast}.

\section{Star Formation Histories}
\label{sec:sfhs}

\begin{figure*}[th!]
\gridline{\fig{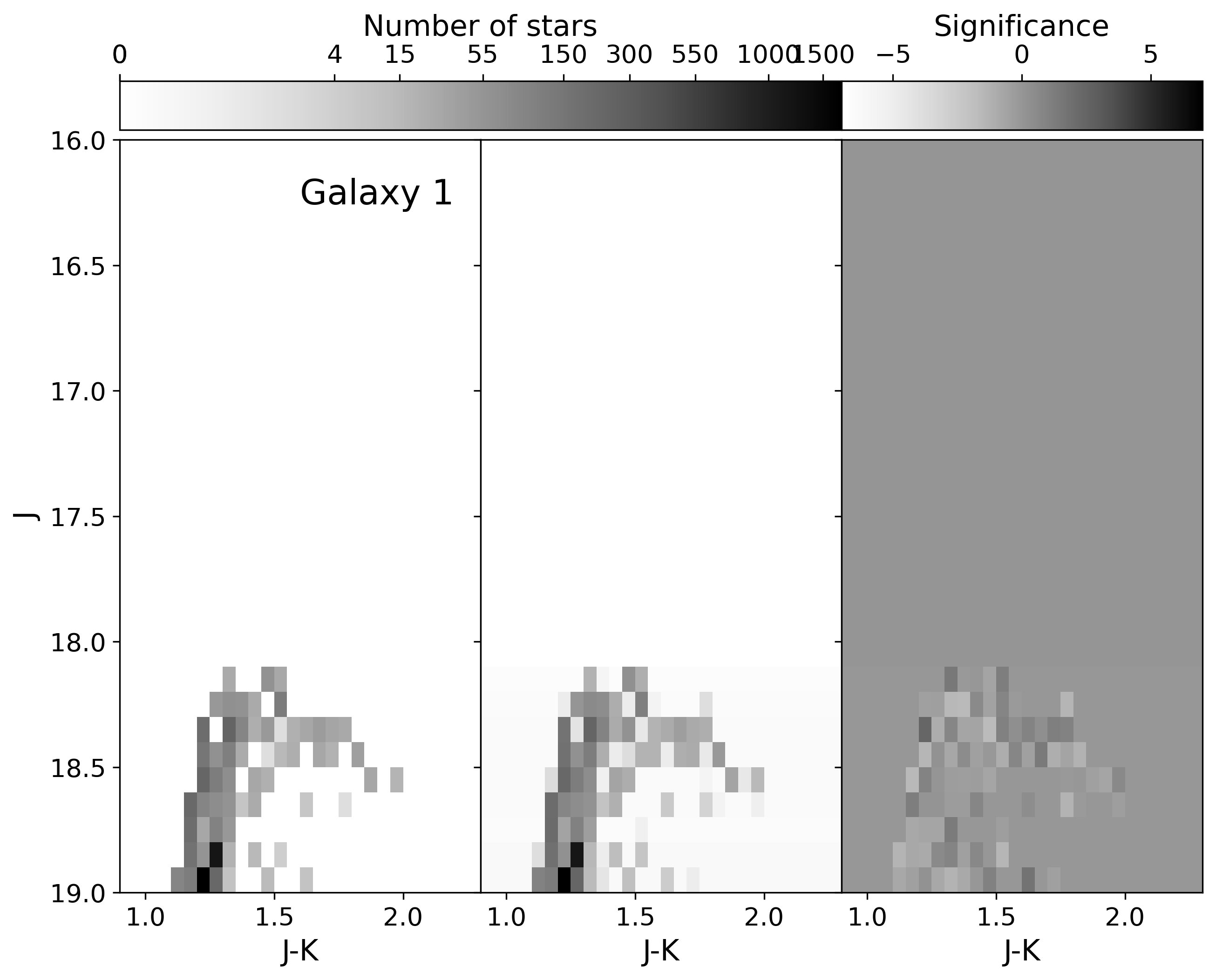}{0.333\textwidth}{} 
        \fig{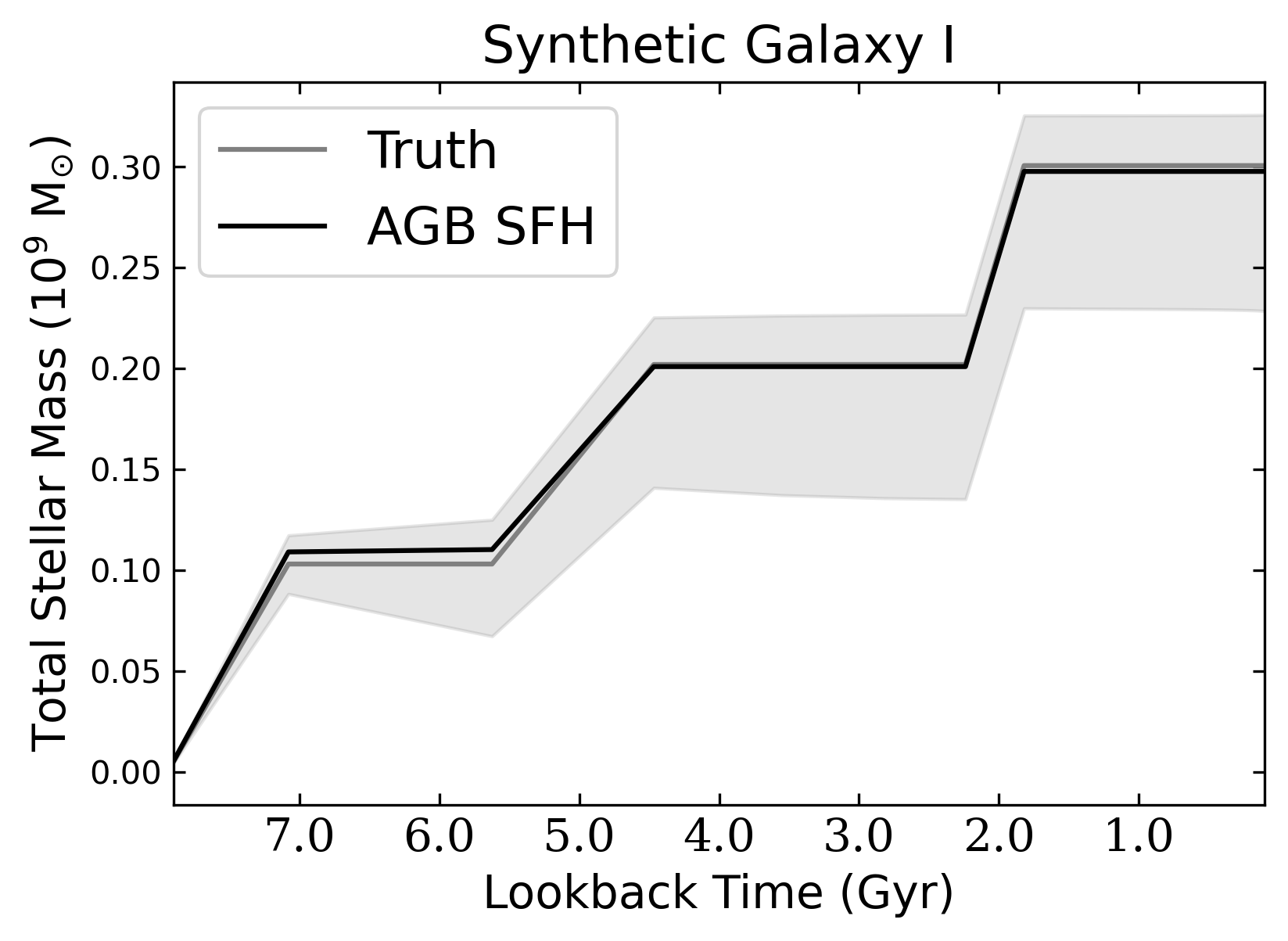}{0.333\textwidth}{}
        \fig{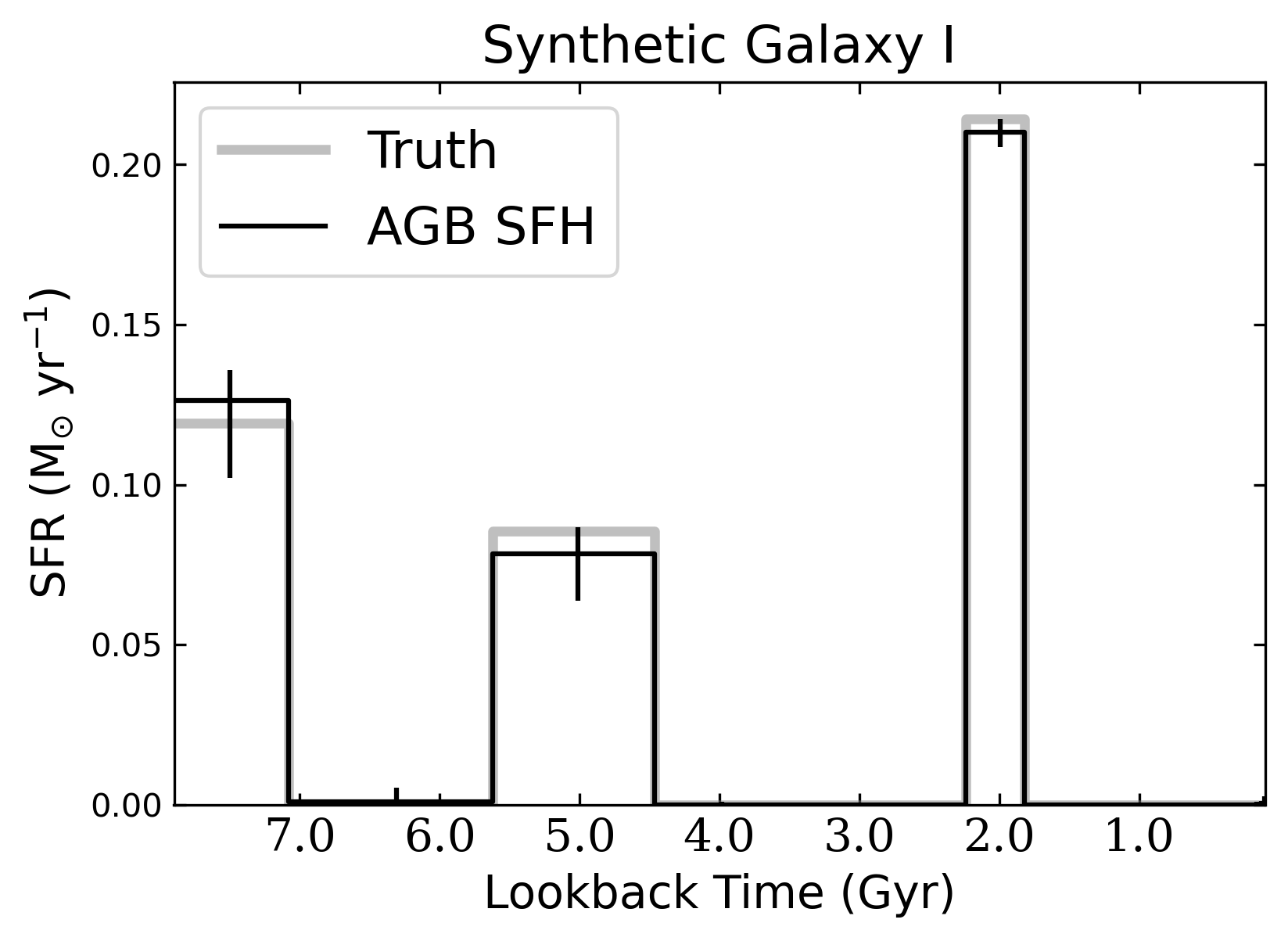}{0.333\textwidth}{}}
\gridline{\fig{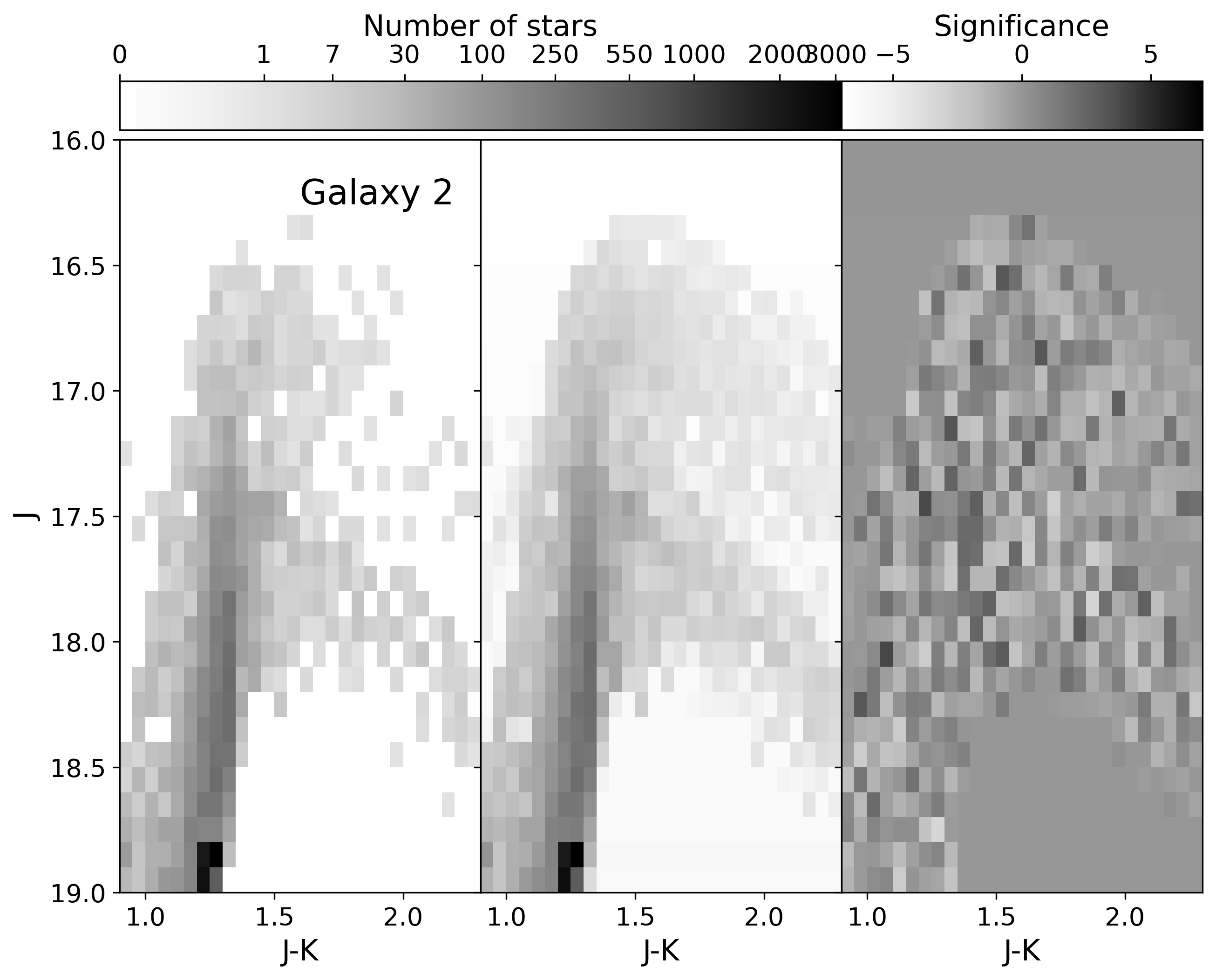}{0.333\textwidth}{} 
        \fig{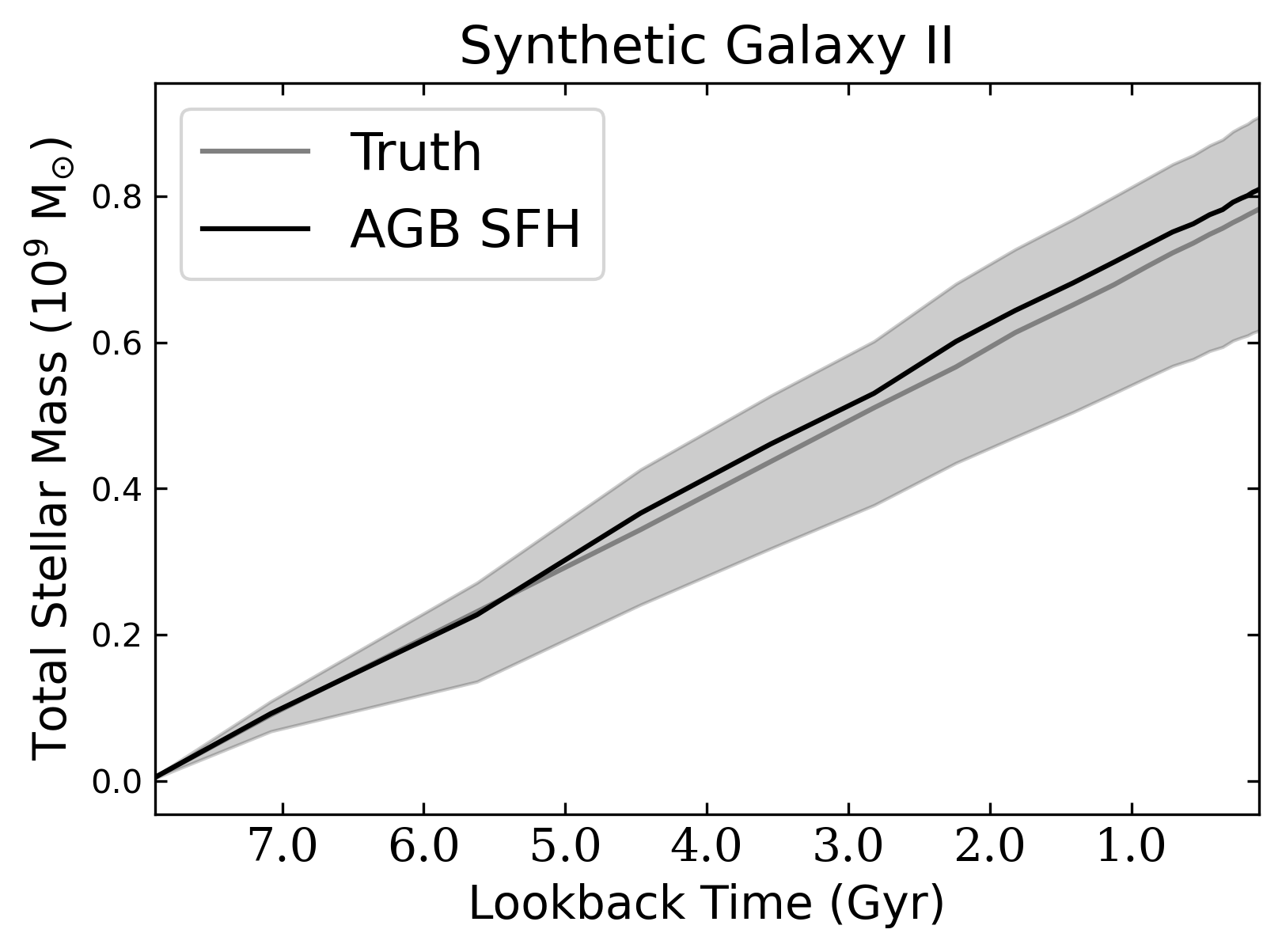}{0.333\textwidth}{}
        \fig{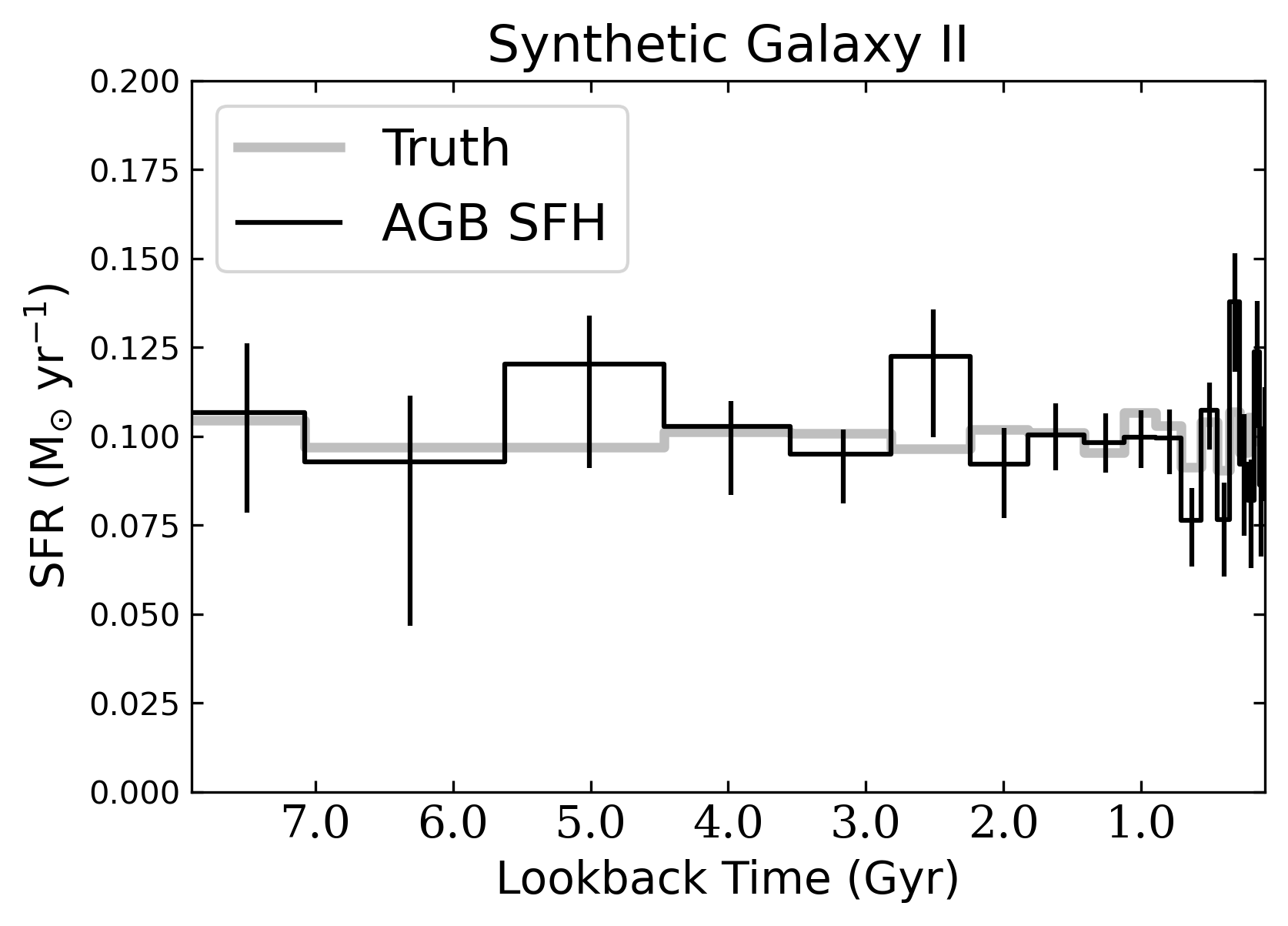}{0.333\textwidth}{}}
\caption{Comparisons between our `injected' star formation histories and `recovered' star formation histories in the two mock galaxies described in \S \ref{sec:mock_summary}. (Left) Density maps for the observed, best-fit, and residual CMDs, expressed in units of Poisson standard deviations. (Middle) Cumulative stellar mass formed as a function of time. (Right) Total star formation rate as a function of time.  These tests illustrate our ability to self-consistently recover SFHs from AGB stars.}\label{fig:mock} 
\end{figure*}

\subsection{Application to Mock Data}
\label{sec:mock_summary}

As a first test of our method for self-consistency, we generate synthetic CMDs for a variety of SFHs using our basis functions and simulated ASTs.  
For our noise model
we generate artificial stars with 50\% completeness limits in J and K at 27~mag using the \texttt{makefake} function in \texttt{MATCH}.  These fake ASTs only include the effects of photon noise and not crowding.

To measure the AGB star SFH, we run \texttt{MATCH} on the `observed' CMDs shown in Figure \ref{fig:mock}. We adopt an age range of $\log(\rm{age}) = 8.0$ to $9.9$ with a spacing of 0.1 dex and a metallicity range of [M/H] = -1.4 to 0.0 dex with a spacing of 0.2~dex.  We use well-populated basis functions ($\sim10^8 M_{\odot}$) to minimize any effects of Poisson variations, which are the not purpose of this test.

Next, we run the simulated CMDs through \texttt{MATCH} as if they were real photometric datasets. We then compute random uncertainties using HMC.  
We ran an exhaustive set of simulations with various SFHs.  For brevity, we only explore a couple of illustrative examples below for a constant and bursty SFH.  However, the conclusions are the same no matter in the input SFH: \texttt{MATCH} self-consistently recovers the input SFH using mock data, within the expected errors.

\textbf{Synthetic Galaxy I: Multi-burst SFH.} Synthetic galaxy I was created from three star formation bursts with ages of 2.0, 5.0, 7.6 Gyr and metallicities [M/H] = -0.4, -0.2, 0.0 dex, respectively. Each burst forms a stellar mass of $10^8 M_{\odot}$.  In the top panel of Figure \ref{fig:mock}, we show the ``observed'' CMD, recovered best fit model CMD, and residual CMD, along with the recovered and input star formation histories. 

\textbf{Synthetic Galaxy II: Constant SFH.} Synthetic galaxy II was created using a roughly constant star formation rate of $0.1~\rm{M_{\odot}/yr}$ from log(age)=8.0 to 9.9~dex at a fixed metallicity of [M/H]=0.0~dex.  In the bottom panel of Figure \ref{fig:mock}, we show the ``observed'' CMD, recovered best fit model CMD, and residual CMD, along with the recovered and input star formation histories. 

 The excellent agreement between all input and recovered tests indicates that \texttt{MATCH} is self-consistently able to handle our custom basis functions.  We now turn to application of this method to real data. We show two further simulated tests in Appendix \ref{sec:oldest_ages} for ages up to 14~Gyr and with JWST filters.

\begin{figure*}
\gridline{\fig{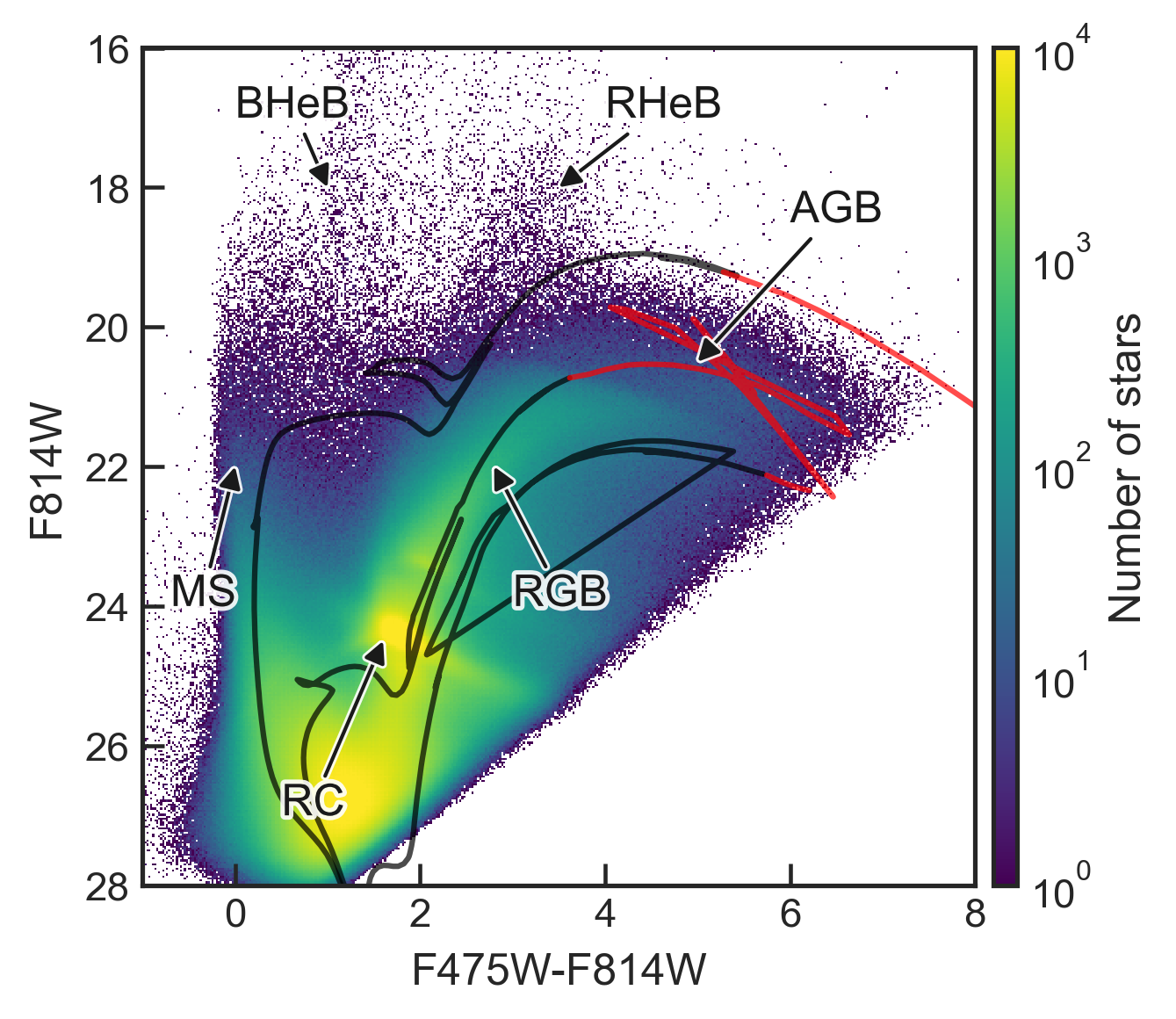}{0.47\textwidth}{}
          \fig{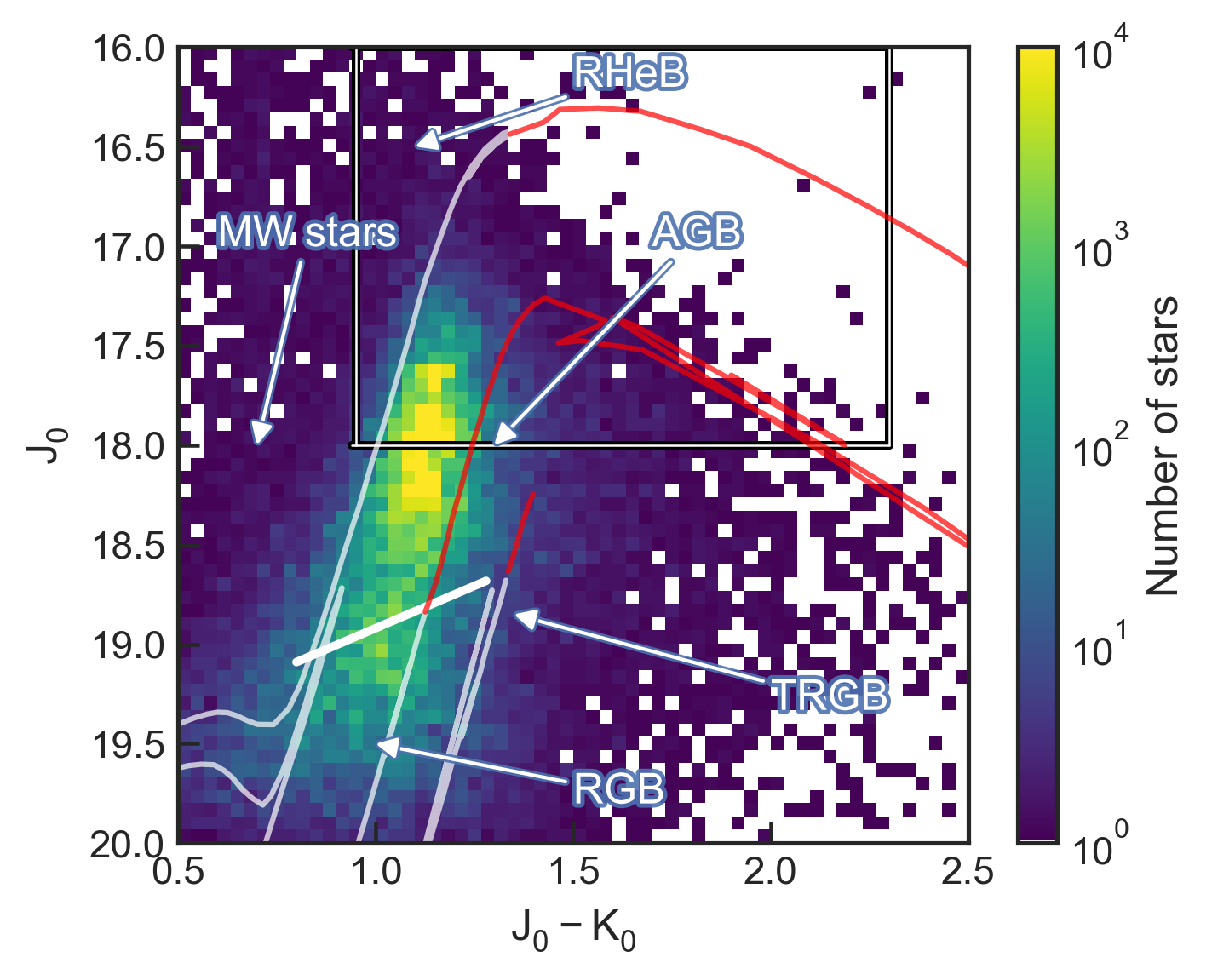}{0.5\textwidth}{}
         }
\caption{(Left) CMD of the M31 PHAT optical data analyzed in  \citetalias{williams17} for $d>11$~kpc.
(Right) CMD of our M31 UKIRT data that overlaps with the PHAT footprint analyzed in \citetalias{williams17} for $d>11$~kpc. Color indicates the number of stars in each CMD bin. The grey box shows our selection criteria for the AGB stars, which also corresponds to the grey box in Figure \ref{fig:isochrones}, shifted to M31's distance. In both CMDs, isochrones of 100 Myr, 1 Gyr, and 8 Gyr with a metallicity of [M/H]=0.0~dex are overlaid with the TP-AGB evolutionary phase highlighted in red. The PHAT optical CMD contains $\sim47$~million stars of which $\sim33$~million stars are above the 50\% completeness limit ($F814W\approx 26.5$~mag). Our NIR CMD contains $\sim 35,000$ stars and an average 50\% completeness limit of $J = 18.7$~mag.  Our selection region contains $\sim 7,700$ AGB stars from which we measure our SFH. Typical CMD features, such as BHeB (Blue Helium-Burning) stars, RHeB (Red Helium-Burning) stars, AGB stars, RGB stars, MS (Main Sequence) stars, RC stars, and MW foreground stars are labeled. }
\label{fig:cmd}  
\end{figure*}

\subsection{Comparison to PHAT}\label{subsec:m31}

A further check on consistency is to compare our AGB star SFHs to those derived from deep optical CMDs.  Ideally, we would use an optical CMD that reaches the oldest MSTO to provide an anchor for this testing.

However, the only datasets that are suitably deep, outside the Magellanic Clouds that have been used to calibrate the AGB models, are from small-area HST or JWST fields (e.g., \citealt{Skillman2003, Dolphin2003, Brown2006, Geha2015, Albers19, McQuinn2024a}) or from ground-based imaging of Milky Way (MW) satellites (e.g., \citealt{deBoer11, deBoer12a, deBoer12b, deBoer14, Munoz18}).  At issue is that these fields do not contain a suitable number of AGB stars (at least $\sim1000$, as we discuss below) to enable a reliable SFH.  %As one concrete example, \citet{McQuinn2024a} present a deep JWST-based SFH of LG dwarf galaxy WLM.  The JWST field only subtends a modest fraction of the main optical body of WLM, and only includes $\sim150$ AGB stars \citep{Boyer24}.  \citet{Lee21} analyzed NIR imaging of the entirety of WLM for calibrating distance indicators and identify only $\sim300$ AGB stars in the entire galaxy.  The situation is virtually identical for all LG targets with such deep HST or JWST that we considered.  

We instead opt to compare our AGB star SFH to those measured as part of the PHAT survey.  Specifically, we use the SFH fits from \citetalias{williams17}, who measured the spatially resolved ancient SFH of the PHAT region from optical CMDs that extend to $\sim2$~mag below the RC, as shown in Figure~\ref{fig:cmd}.  Although these CMDs do not reach the 10 Gyr MSTO, and instead rely on advanced phases of evolution (e.g., the position and shape of the RC), \citetalias{williams17} took great care to validate their results against CMDs taken throughout M31's disk that did reach the oldest MSTO \citep{bernard15a, bernard15b}.  We did consider comparing results directly to other M31 SFHs that do reach the oldest MSTO \citep{Brown2006, richardson08, bernard15a}, but there are not enough AGB stars in the spatially matched HST footprints. 

\citetalias{williams17} measured the SFH of M31's outer disk up to 14 Gyr.
Recognizing the limitations of their moderately deep CMDs, \citetalias{williams17} adopt a single ancient age bin of 8-14 Gyr ago. They adopt a finer time resolution (0.1~dex) for ages younger than $\sim8$~Gyr.   
For this study, to probe the oldest AGB-star SFH for ages of $>8$~Gyr, complete, well-measured photometry of the faintest (and oldest) AGB stars are needed. Unfortunately, because our data were taken for different primary science goals,  the photometry of these oldest AGB stars have high uncertainties; the 50\% completeness limit of our data that overlaps with the PHAT footprint is at $J = 18.7$~mag. This is approximately where the oldest AGB stars lie in magnitude; all the AGB stars with ages $>8$~Gyr are fainter than J = 18.3~mag (see Figure 5). 
Therefore, we adopt J = 18.0~mag as the lower magnitude limit because the largest difference between the injected and recovered ASTs is 0.3~mag at that magnitude, and thus no AGB stars with ages $>8$~Gyr will be present above J = 18~mag. 
In conclusion, due to the current limits of the data in this study, we only measure the AGB-based SFHs back to 8 Gyr ago.  We stress this age limit is not inherent to the AGB star SFH method itself, but rather due to our desire for empirical validation.

We also limited the fitting routine to follow the chemical evolution model in \citetalias{williams17}, which involved limiting our fits to only include the metallicities allowed in Figure 8 of \citetalias{williams17} for M31's outer disk (blue line, $d>12$~kpc). In a future paper, we plan to incorporate more sophisticated chemical enrichment modeling schemes in our code.

\subsubsection{Photometric Selection}\label{subsubsec:NIRCMD}

\begin{figure}
\centering
\includegraphics[width=\columnwidth]{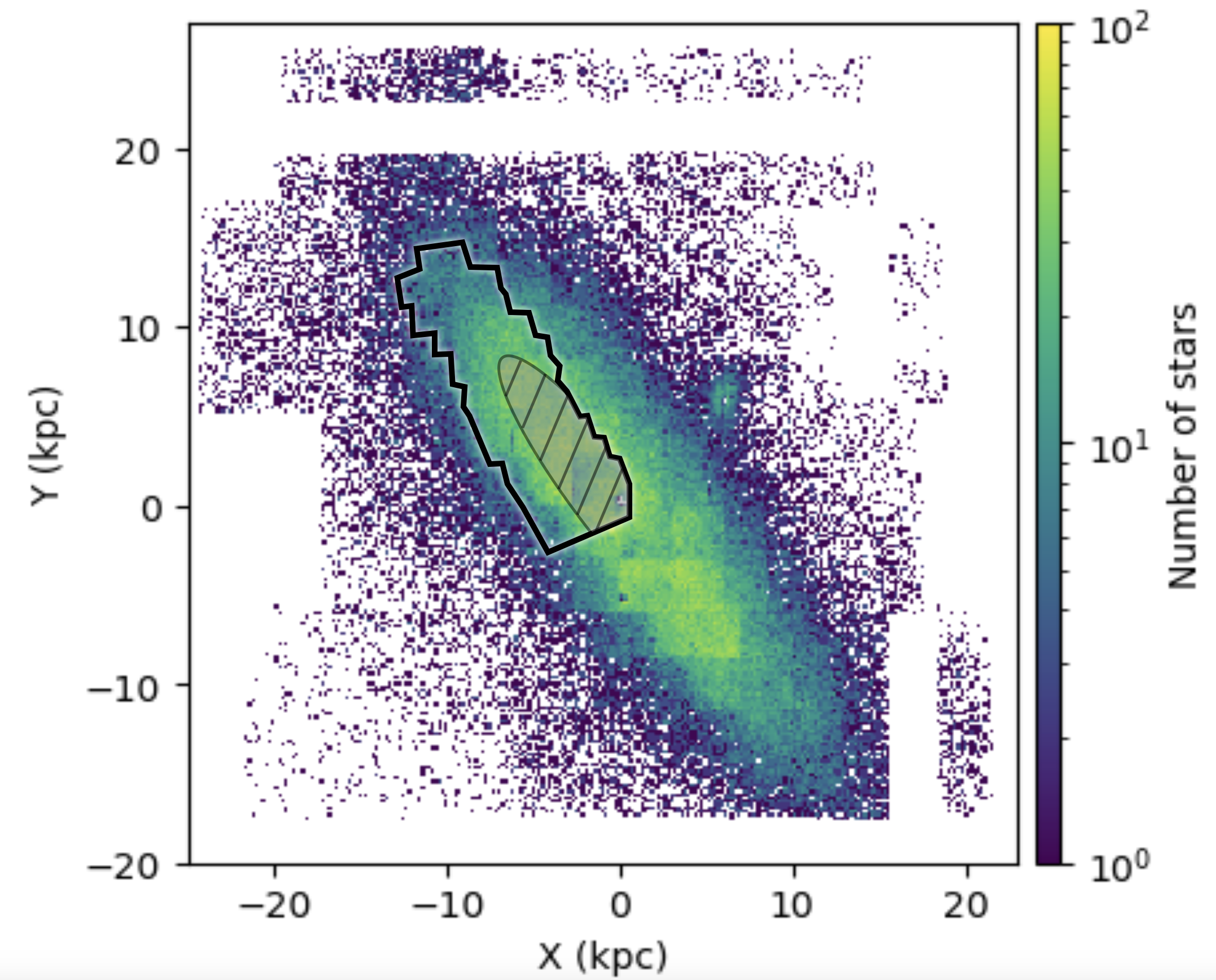}
\caption{Spatial map of AGB stars from our UKIRT data over the entire region of M31 covered by our data. The PHAT footprint is overlaid in black and includes data out to a deprojected distance of $d\lesssim20$~kpc in the Northeastern quadrant of M31}. We excluded data inside a deprojected distance of $d = 11$~kpc, denoted by the shaded grey hatched ellipse, to avoid the crowded and reddened inner regions of M31's disk that significantly affect the optical CMD.  
\label{fig:map}
\end{figure}

Beyond an age restriction, we limit this comparison to the outer region ($d>11$~kpc) of the PHAT survey, as illustrated in Figure~\ref{fig:map}.  This enables us to compare to the deepest optical data available, and avoids the need to model the complex spatially varying differential extinction that dominates most of M31's inner disk \citep[e.g.,][]{dalcanton15, Lewis15}. It also mitigates the impact of crowding on the optical CMDs and hence the SFHs.  Crowding significantly impacts the depth of the optical CMDs in M31's inner disk, reducing the fidelity of the \citetalias{williams17} SFHs. We note that while crowding affects the faintest stars in the the PHAT survey, it has minimal impact on brighter stars. For example, Figure 6 in \cite{gregersen15} demonstrates the RGB stars with $d\gtrsim
11$~kpc have $\geq90\%$ completeness in the PHAT survey. 

For $d\gtrsim
11$~kpc, the PHAT CMD has $\sim47$~million stars (Figure~\ref{fig:cmd}). Of these stars, $\sim33$~million are located above the 50\% completeness limit of $F814W\sim26.5$ and were used by \citetalias{williams17} to measure the spatially resolved SFH of this region.  We sum the SFH of each region in this area as provided in the \citetalias{williams17} paper.   

In comparison, our AGB star SFH contains $\sim7700$ AGB stars that are located within the selection box in Figure~\ref{fig:cmd}.  Unlike the the PHAT optical CMDs, our AGB star data are not generally crowded in any part of M31. The 50\% completeness limit in the inner regions ($d\lesssim
11$~kpc) is only $\sim0.5$~mag brighter than in the outskirts.  The significantly reduced sensitivity to crowding is another strength of AGB stars in the NIR for SFH determinations.

Having defined the spatial area of our sample, we next correct our NIR data for reddening. Even in the NIR, internal reddening in M31's outer disk is not negligible. The dust map generated by \citet{dalcanton15} represents one of the highest-resolution reconstructions of M31`s internal reddening. Unfortunately, it is not publicly available necessitating a different means of applying a reddening correction.

We opt to use the \cite{draine14} dust map, which is based on far-infrared and submillimeter emission observations from the \textit{Spitzer Space Telescope} and \textit{Herschel Space Observatory}. The \cite{dalcanton15} reddening map predicts $\sim2.5\times$ less extinction than the \cite{draine14} dust map, although their overall morphological agreement is excellent. Thus, we use the \cite{draine14} map with the \cite{dalcanton15} corrections applied. Specifically, the \cite{dalcanton15} corrects the \cite{draine14} maps by dividing by a factor of $R = 2.53$ and adding a bias of $b = -0.04$ to account for possible systematic biases. This transformation is then as follows: $A_{V,Draine}/A_{V,Dalcanton}=2.53 (1+0.04/A_{V,Dalcanton})$.  

We then convert to $A_J$ and $A_K$ using $A_J= 0.282A_V$ and $A_K=0.114A_V$ \citep{cardelli89}. We find the average reddening correction from the \cite{draine14} maps in the PHAT footprint outside a radial distance $d>11$~kpc to be $A_J=0.20$~mag, and $A_K=0.08$~mag. We apply this average correction to our photometry. For reference, the foreground Milky Way extinction toward M31 is $A_J=0.05$~mag, and $A_K=0.02$~mag, calculated from the \cite{schlegel98} dust maps recalibrated by \cite{schlafly11}, assuming the \cite{cardelli89} reddening law with $R_V=3.1$  

Figure \ref{fig:cmd} shows a comparison between the HST-based optical CMD of M31 at $d>11$~kpc (left) and our ground-based NIR CMD of the same region (right). As discussed in \S \ref{sec:isochrones}, it is virtually impossible to separate out AGB and RGB stars in the optical.  This point is reinforced by the AGB isochrones overlaid on the optical PHAT CMD.

In the NIR, we select AGB stars as having colors between $0.95<(J-K)<2.3$~mag and magnitudes between $16.0<J<18.0$~mag based on the locations of AGB stars isochrones, and due to the limited completeness of the fainter, older AGB stars, as discussed in Section \ref{subsec:m31}.  This photometric selection region also contains red helium-burning stars which will  provide additional age information for young ($\lesssim100$~Myr) populations. These criteria also eliminate contamination from foreground stars, which are bluer than our selection region. We also emphasize that most of the age information in the PHAT SFHs is from the position of the red clump at $F814W\approx24.2$~mag, whereas we only used AGB stars brighter than $J\lesssim 18.0$~mag in NIR wavelengths.

\begin{figure*}[t!]
\gridline{\fig{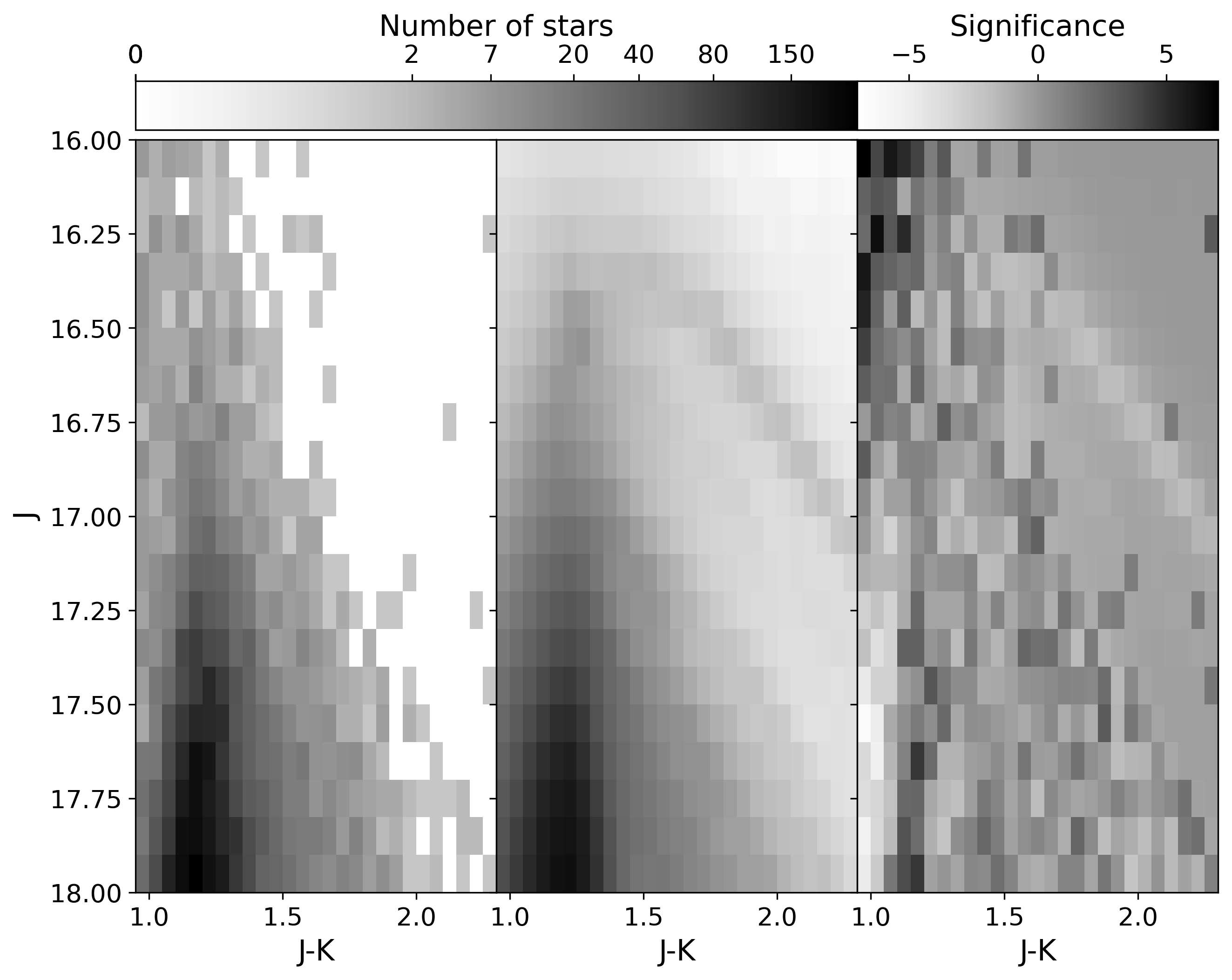}{0.333\textwidth}{} 
        \fig{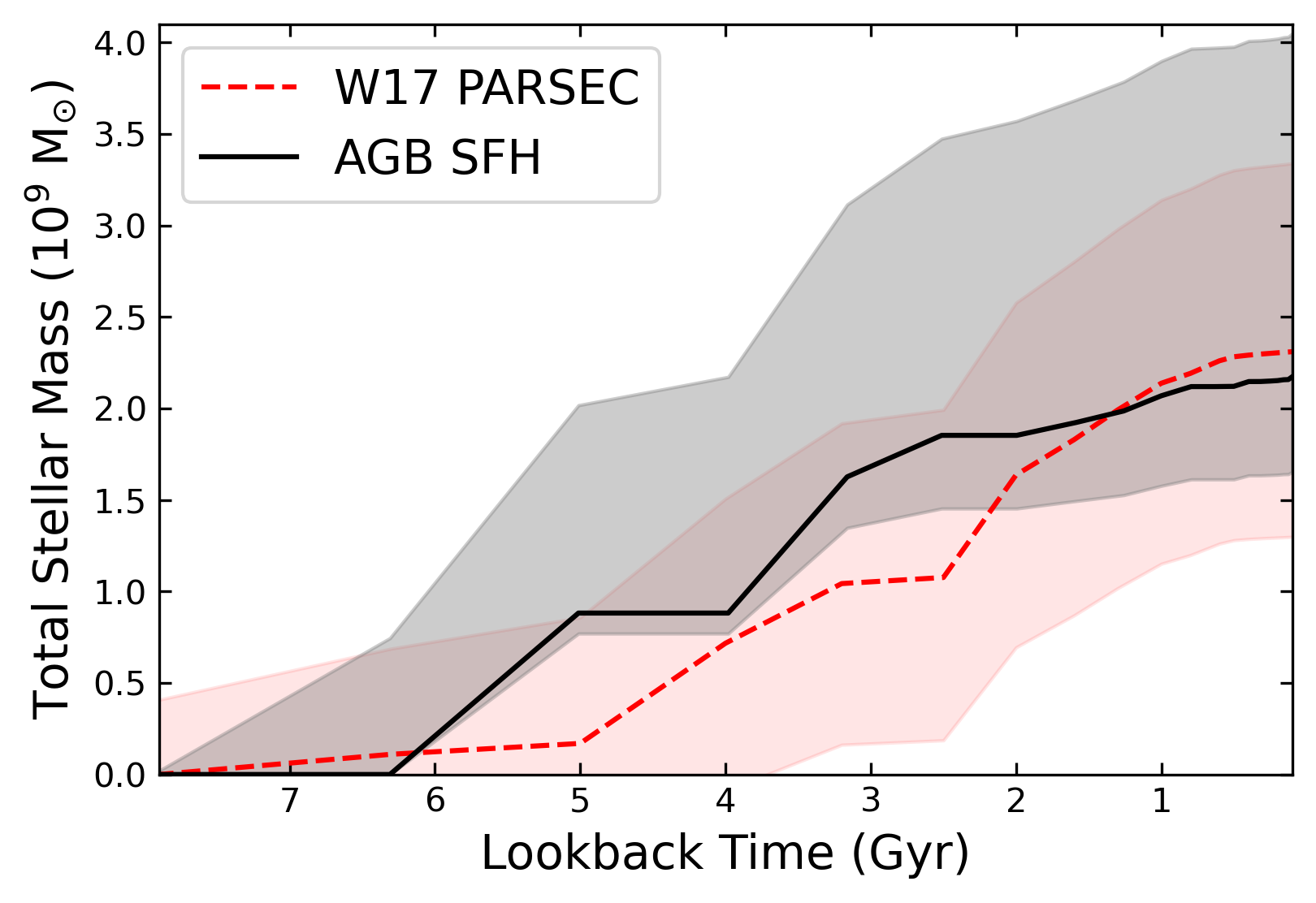}{0.32\textwidth}{}
        \fig{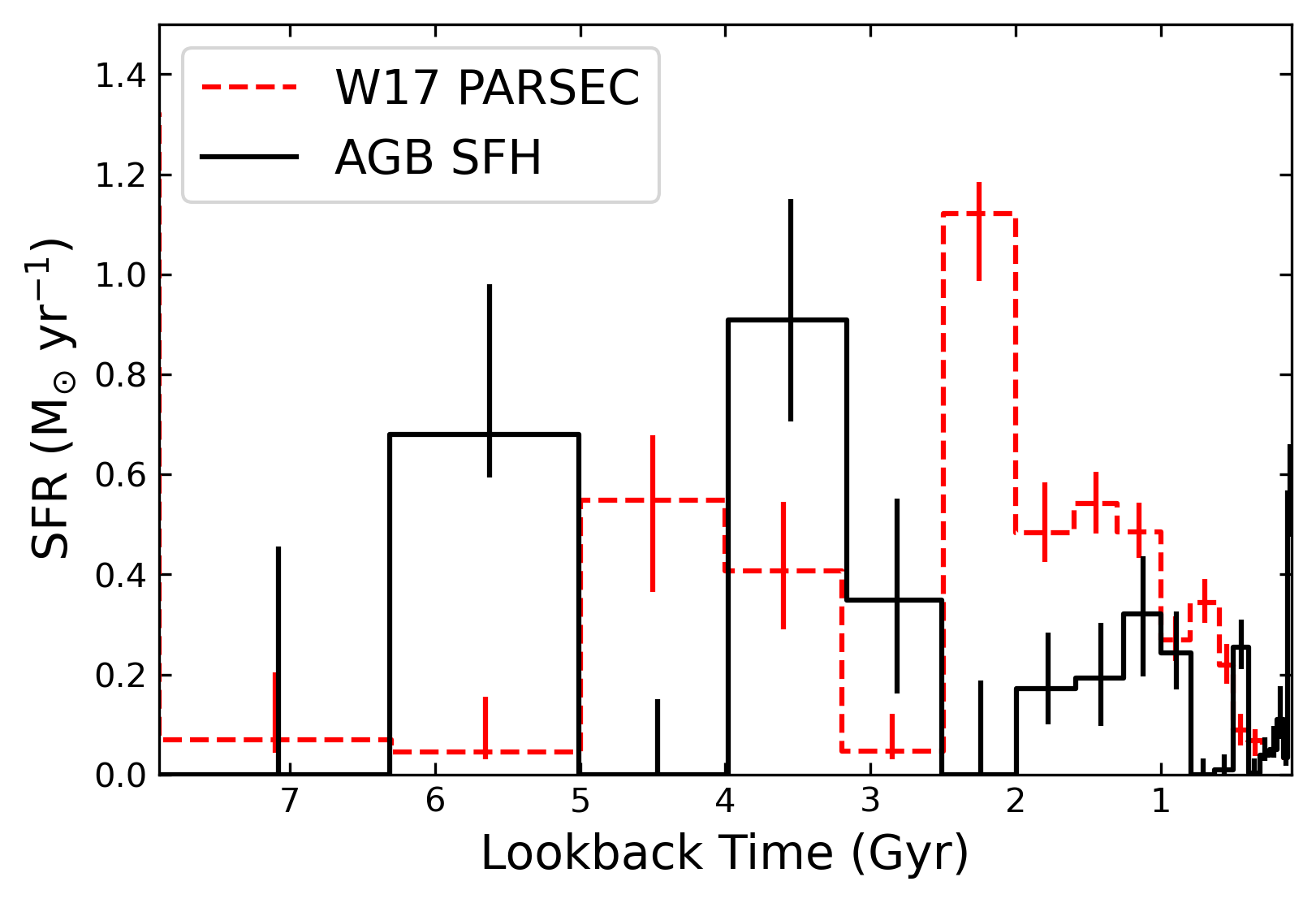}{0.333\textwidth}{}}
\caption{(Left) Hess diagrams for our AGB star SFH fit for the observed data (left panel), best-fit linear combination of SSPs (middle panel), and residual CMDs (right panel). The residuals are expressed in Poisson standard deviations. (Middle) Comparison between our measured cumulative stellar mass formed and the \citetalias{williams17} measurement using the PARSEC models. (Right) Comparison between our SFR measured from AGB stars and the SFR measured in \citetalias{williams17}.  } \label{fig:disk_compare} 
\end{figure*}

\subsubsection{Measuring the AGB star SFH in the PHAT Region}\label{subsubsec:phat_AGB}

To measure the AGB star SFH, we run \texttt{MATCH} on stars within the grey box in the UKIRT CMD shown in Figure \ref{fig:cmd}.    
For our fitting, we adopt an M31 distance modulus of $\mu_0=24.42\pm0.05$~mag ($766\pm18$~kpc; \citealt{lee23}). This distance was derived from the J-region asymptotic giant branch (JAGB) method, a standard candle based
on AGB stars, using the same UKIRT photometry studied in this paper,
and agrees to within 1$\sigma$ with other commonly adopted distances to M31 measured with the TRGB, RR Lyrae, and Cepheids \citep[e.g.,][]{mcconnachie05, Jeffery11,deGrjis14,Bhardwaj16,Savino2022}. 
We use a CMD bin size of $0.05 \times 0.10$~mag in color and magnitude, respectively. 
We then follow the CMD modeling procedure  described in \S \ref{sec:methodology}.

Figure \ref{fig:disk_compare} shows the SFH of the PHAT region based on AGB stars.  From the residual CMDs (left panel), we see that our best fit model provides a good match to the data.  
The residual significance plot, i.e., the residual weighed in each Hess diagram in units of standard deviations, shows no strong artifacts but a rather uniform `checkerboard' pattern. This indicates that there are no regions of the CMD that are systematically discrepant.
This plot shows modest discrepancies between the data and model at the bluest colors, but they are of moderate significance and are on-par with the quality of residuals in most optical CMD modeling (e.g., \citealt{Monelli10, Hidalgo2011, Weisz2011, Weisz2014, Skillman2017, Albers19, Savino2023}) and with JWST in the NIR \citep[e.g.,][]{McQuinn2024a, McQuinn2024c}.

The middle panel shows the cumulative SFH the AGB star SFH in black, and the \citetalias{williams17} best fit \texttt{PARSEC}-based SFH in red. The uncertainty envelope for both SFHs reflects the 68\% random uncertainties. We note that although \citetalias{williams17} include AGB stars in their CMD modeling they (a) are of limited utility due to overlap with the RGB and (b) are not fit with the more recent versions of the \texttt{COLIBRI} AGB models, as they were not yet available.  As discussed extensively in \citetalias{williams17}, the age information at most epochs comes from the RC and ratios of RC to RGB stars, with some contribution from AGB stars.  The two SFHs (AGB star and \citetalias{williams17}) use the same IMF, but adopt slightly different minimum mass limits for the low-mass IMF. This introduces an order of unity difference in the SFHs and stellar masses formed.  We apply this small IMF correction to the \citetalias{williams17} for the purposes of equal comparison.

The two cumulative SFHs are in excellent agreement to within $1\sigma$.  Both formed the same stellar mass.  The AGB star SFH has a burst $\sim5-6$ Gyr ago, while the optical SFH shows a burst of similar amplitude from $\sim4-5$~Gyr ago.  Both show a modest amount of stellar mass formed $2-4$~Gyr and $\sim1-2$~Gyr ago.  Though the difference in timing is not statistically significant, in future studies it will be interesting to understand if such differences could be due to information content of the data (e.g., `bleeding' of star formation into adjacent time bins due to the increased age-metallicity degeneracy in the RC).

We also compared the age-metallicity relationships (AMRs) from our fits and \citetalias{williams17} in Appendix \ref{sec:amr}.  Both are generally consistent with solar metallicity, except for a noticeable drop to [M/H]=-0.5 at $\sim1$~Gyr.  Similar decreases in metallicity are measured in other regions of M31's disk \citep{bernard15a} from CMDs that reach the oldest MSTO.  

\begin{figure}
\centering
\includegraphics[width=\columnwidth]{"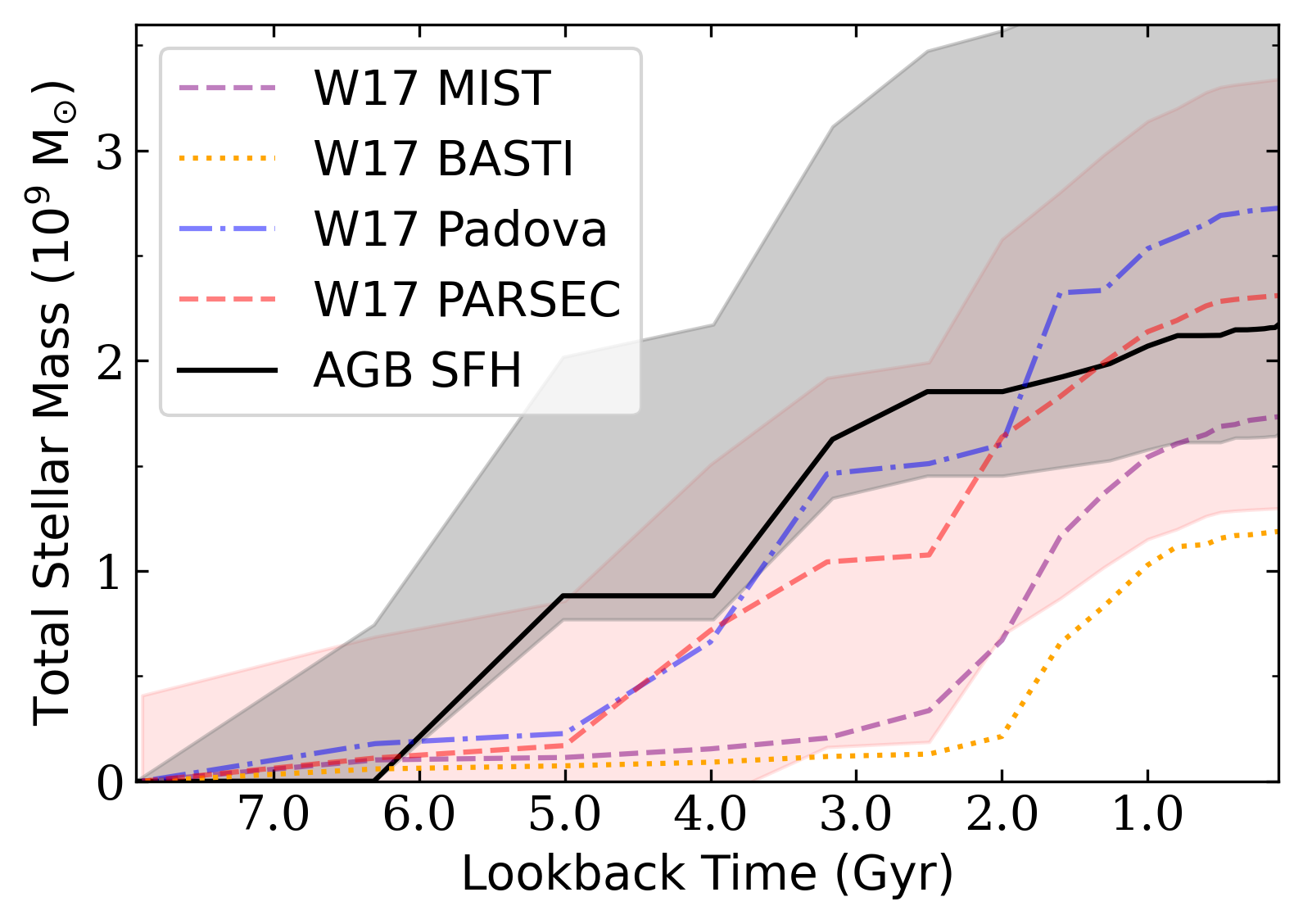"}
\includegraphics[width=\columnwidth]{"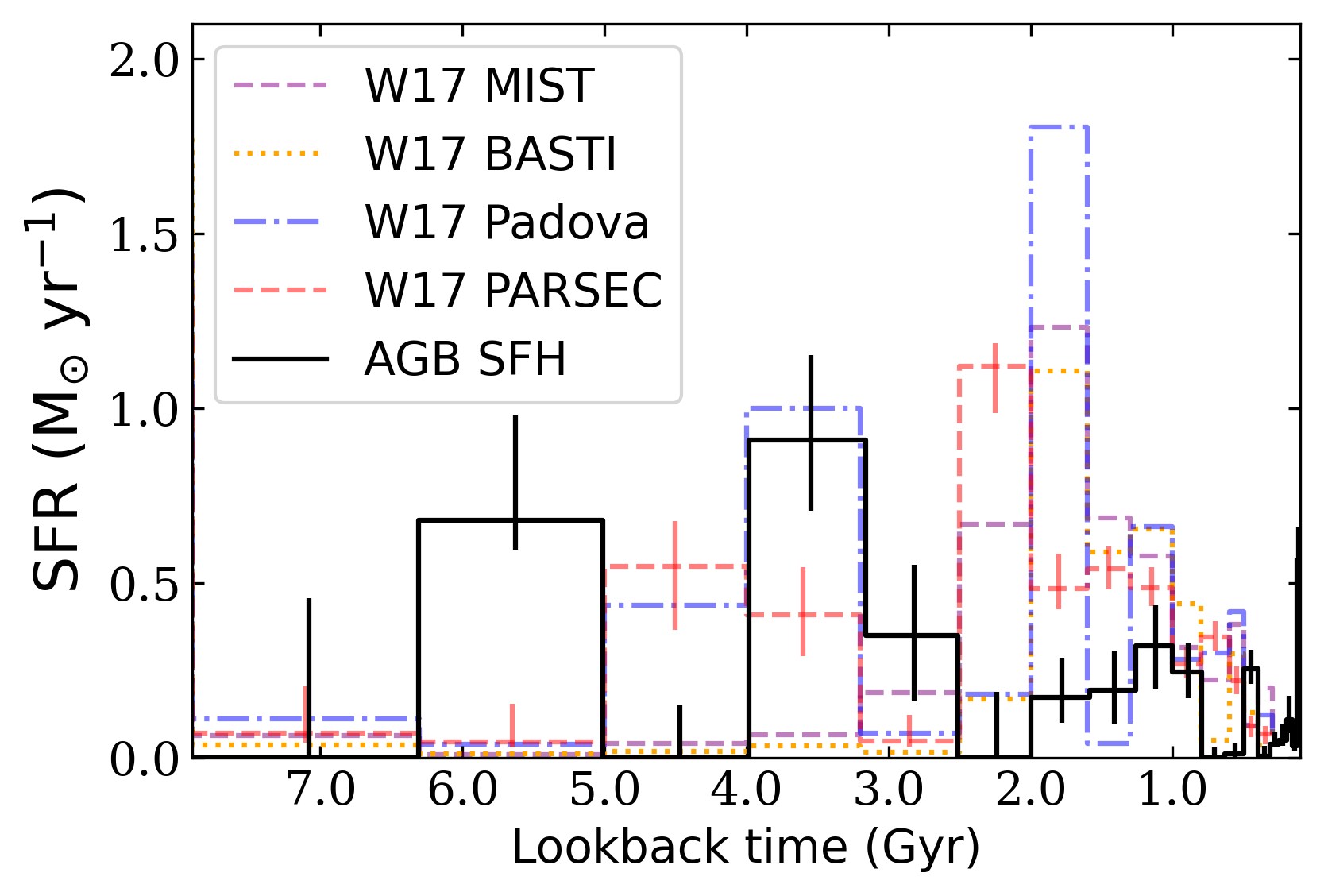"}
\caption{Same plots as Figure \ref{fig:disk_compare} now with all model sets from \citetalias{williams17}.
}
\label{fig:all_models_sfh}
\end{figure} 

The right panel of Figure~\ref{fig:disk_compare} shows the differential SFH, i.e., SFR vs time, of the AGB star and PHAT SFHs.  As with the cumulative SFHs, we find good agreement between the amplitude and timing of star formation in both SFHs, though there are a few differences.  For example, the timing and amplitude of star formation around $\sim4-6$~Gyr and $\sim2-4$ Gyr ago are slightly different.  As with the cumulative SFHs, such differences in the exact timing of star formation from CMD modeling are often observed when using different stellar models or CMD fitting codes \citep[e.g.,][]{Weisz2011, dolphin12, Cole14, Skillman2017, Savino2023}.  

To examine how significant these variations may be, in Figure \ref{fig:all_models_sfh} we compare best fit SFHs measured using different stellar models from \citetalias{williams17} with our AGB star SFH.  It is clear from the cumulative and differential SFHs that our AGB star SFH is well within the model-to-model variance from optical CMD analysis.  Each of the optical-based SFHs shows the same qualitative trend of stellar mass growth, but with slightly different timings and amplitudes of star formation episodes, as well as modest differences in the total stellar mass formed.  Within this context, our AGB-based SFHs are as robust as any of the SFHs based on deeper optical CMDs.

Our measurement is based on $\sim7700$~AGB stars.
In Appendix \ref{sec:how_many}, we quantify how the quality of the fit and statistical errors scale with the number of AGB stars, by measuring the AGB star SFH from decreasingly smaller random samples of AGB stars. In Figure \ref{fig:disk_uncertain}, we show that as few as 1000~AGB stars can deliver a reasonable (i.e., recovering bursts at roughly the correct time, albeit with much larger uncertainties on the amplitudes of these bursts) SFH compared to the \citetalias{williams17}.

% Finally, we note that our NIR CMD of the PHAT region (Figure~\ref{fig:cmd}) clearly contains AGB stars that are older than 8~Gyr.  While in principle we could extend our analysis to older ages (as discussed in Appendix~\ref{sec:oldest_ages}), for the sake of empirical verification \textbf{and because of the limitations of the UKIRT photometry used in this paper}, 

\subsection{Measuring the SFH of M31's Inner Halo}\label{subsec:m31_halo}

We now provide an illustrative use case for our AGB star SFH technique by measuring the SFH for the inner stellar halo of M31 in 6 spatially-separated regions.  M31 is long-known to have a complex stellar morphology and a halo filled with substructure (see \citealt{mcconnachie18} and references therein).  Measuring the SFH of M31's stellar halo from AGB stars may help shed new light on its formation history, while also providing a practical showcase for the power of AGB stars for measuring SFHs of stellar halos in other galaxies throughout the local Universe.

\subsubsection{Data and Extinction Corrections}

%\begin{figure}
%\centering
%\includegraphics[width=\colum%nwidth]{"CMD_halo.png"}
%\caption{CMD of our UKIRT %data outside of a %deprojected distance of 20 %kpc. Color indicates the %number of data points in %each place in the CMD. The %grey box shows our selection %criteria for the AGB stars. %Isochrones of 100 Myr, 1 %Gyr, and 8 Gyr with a %metallicity of [M/H] = 0.0 %dex are overlaid with the %AGB evolutionary phase %highlighted in red. 
%}
%\label{fig:CMD_halo}
%\end{figure} 

\begin{figure}
\centering
\includegraphics[width=\columnwidth]{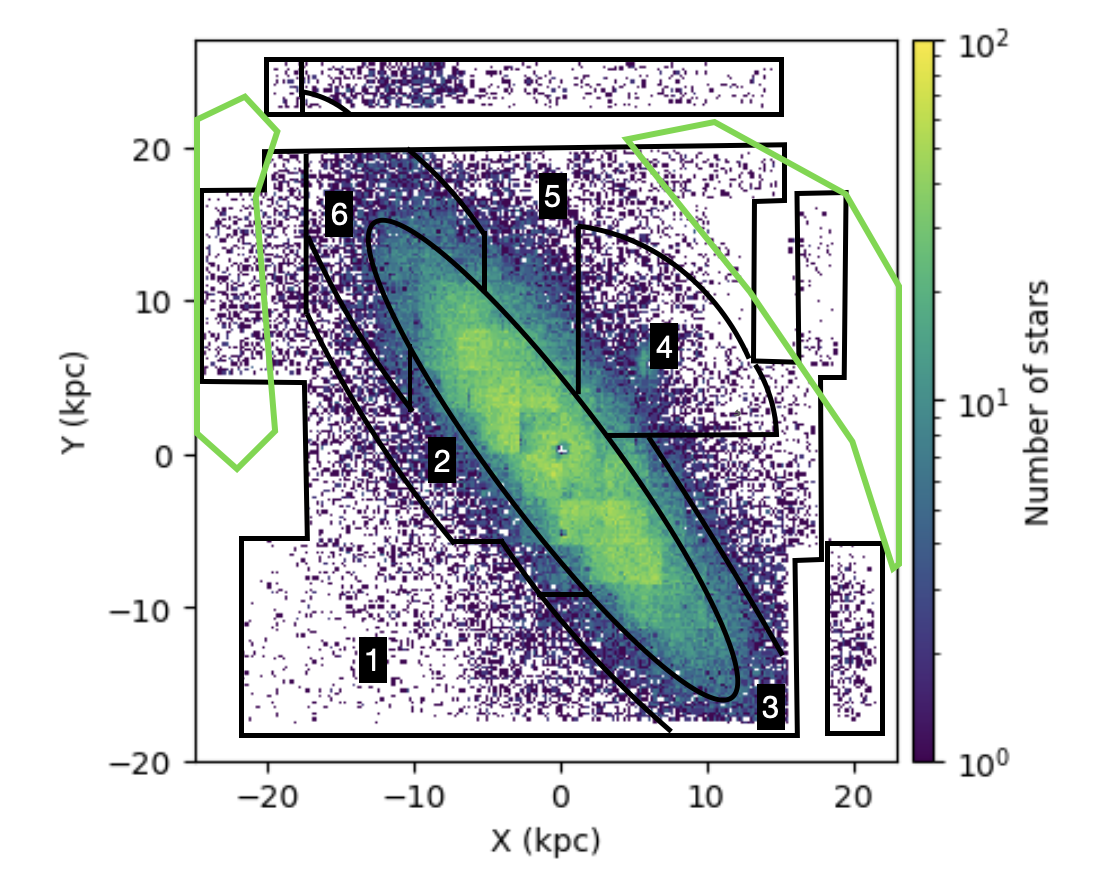}
\caption{Map of the six regions overplotted on a stellar density map (for stars with $J < 19$~mag) of M31. Regions 1, 4, and 5 were specifically designed to contain the following major substructures:} region 1 contains the Eastern Shelf, region 5 contains the Western Shelf, and region 4 contains NGC 205. Green polygons highlight the approximate positions of the the stellar structures which are a part of the GSS, and the pink ellipse corresponds to the  dwarf galaxy NGC 205 in this map. 
The positions of the AGB stars and their region labels are available as Data behind the Figure.
\label{fig:region_labels}
\end{figure}

Our photometry includes data out to a projected radius of 32~kpc, encompassing M31's inner stellar halo. We exclude data within a deprojected distance of 20~kpc, as shown in Figure \ref{fig:region_labels}, to avoid the complexity extinction in the disk \citep[e.g.,][]{dalcanton15}. Within the stellar halo footprint, we now verify that the internal reddening derived from the $A_J$ extinction map discussed in \S \ref{subsec:m31_halo} is very small and does not vary much spatially, using the the \cite{draine14} map which covers the full range of our UKIRT data in major axis, but not the full range in minor axis. However, the minor axis (northwest and southeast sections of M31's halo) are expected to contain even less dust. The average extinction values derived from the corrected map discussed in \S \ref{subsubsec:NIRCMD} outside of a deprojected distance of 20~kpc are $A_J=0.05$~mag and $A_J=0.02$~mag, which are exactly equal to the foreground Milky Way reddening corrections also discussed in \S \ref{subsubsec:NIRCMD}. For simplicity, we apply these small extinction corrections directly to our photometry before modeling the CMD, and then do not solve for extinction as part of the CMD modeling.

\begin{figure}
\centering
\includegraphics[width=\columnwidth]{"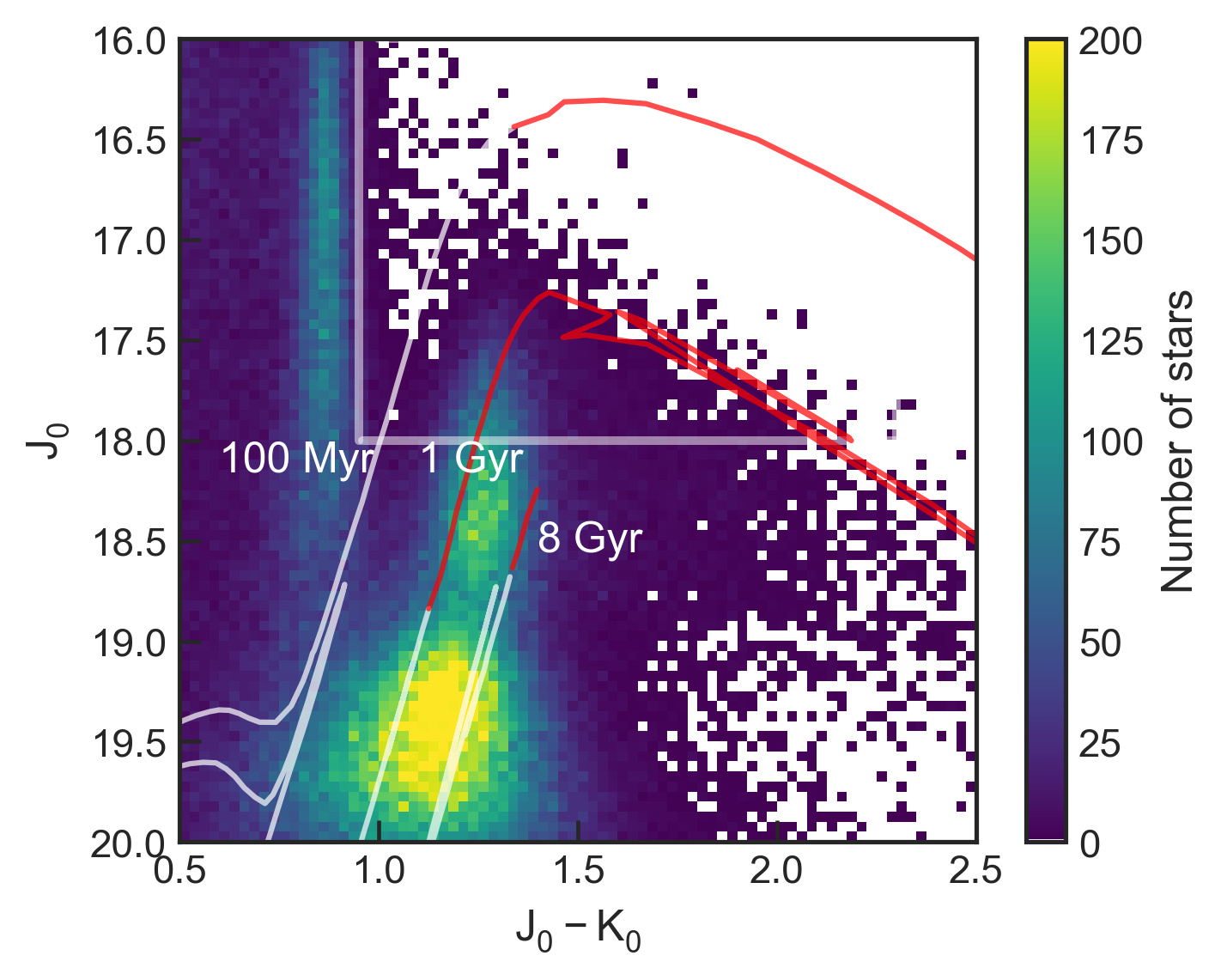"}
\caption{CMD of our UKIRT data outside of a deprojected distance of 20 kpc. Color indicates the number of data points in each place in the CMD. The grey box shows our selection criteria for the AGB stars, where the lower limit was determined by the 50\% completeness limit of the most crowded regions in the stellar halo. Isochrones of 100 Myr, 1 Gyr, and 8 Gyr with a metallicity of [M/H] = 0.0 dex are overlaid with the AGB evolutionary phase highlighted in red. 
}
\label{fig:CMD_halo}
\end{figure}

Figure \ref{fig:CMD_halo} shows the NIR CMD of M31's inner halo we used to measure its SFH ($d>20$~kpc).  The grey box indicates the selection region for AGB stars.  Select \texttt{PARSEC-COLIBRI} isochrones (0.1, 1, 8 Gyr) are included for reference.  There are $\sim7,200$ AGB stars in this region.  The average 50\% completeness limit across the inner stellar halo is $m_J\approx19.3$~mag.

\subsubsection{Spatial Selection}
\label{sec:spatial}

The large number of AGB stars in M31's inner halo allow us to measure a spatially resolved SFH.  We divide our photometry into 6 spatial regions according to the following criteria. First, as illustrated in Figure~\ref{fig:region_labels}, we prioritize placing  three major substructures from \cite{mcconnachie18} into separate regions. 
Regions 1 and 5 contain the Eastern and Western Shelves, respectively, which have been identified as being radial ``shell'' features created from the Giant Stellar Stream (GSS) progenitor's last pericentric passage \citep{ferguson02,ferguson05,fardal06,fardal08, richardson08, fardal12,bernard15b}. 
Region 4 contains NGC 205 (M110), a satellite galaxy of M31. 
We note M32, another satellite galaxy of M31, overlaps on the sky with the inner disk of M31 where crowding was very high; the 50\% completeness level in the J band across the area covered by M32 was J = 18.5~mag. Therefore, we were unable to probe the SFH of M32 using this photometry.
% \cite{dsouza18} proposed that M32 is the progenitor of the GSS, whose merger with M31 is thought to be responsible for the burst of global star formation seen in M31's disk and inner stellar halo $\sim2$~Gyr ago. }

The remaining regions were chosen to have at least $\sim1000$~AGB stars, which is near the lower limit for the number of AGB stars needed for a robust SFH recovery (see Appendix \ref{sec:how_many}). The regions were delineated via a combination of radial and azimuthal lines, resulting in the irregular region shapes seen in Figure \ref{fig:region_labels}. The goal for these remaining sections is simply to provide as much resolved spatially information as possible, particularly in the radial direction, in light of other spatial and number of star constraints.
The number of AGB stars ranged from $\sim1020$ AGB stars in region 1 to $\sim1600$ AGB stars in region 5.

\subsubsection{Star Formation History}

We run \texttt{MATCH} on each of the 6 spatial sub-regions. As discussed in Section \ref{sec:Data}, we measured the average stellar density of these 6 halo regions and used representative ASTs in three representative stellar density regions, as described in Table \ref{tab:ast}. We compute random uncertainties for each region following the HMC approach described in \S \ref{sec:methodology}.

M31's inner stellar halo is thought to be predominantly composed of substructure from past interactions \citep[e.g.,][and references therein]{mcconnachie18}.  We briefly discuss our spatially resolved SFHs in relation to some of the known substructures within M31's inner stellar halo and then comment on the global formation history of the halo.  We publish our spatially resolved SFHs as part of this paper in order to facilitate further investigation by the community.

\begin{figure*}
\centering
\includegraphics[width=\textwidth]{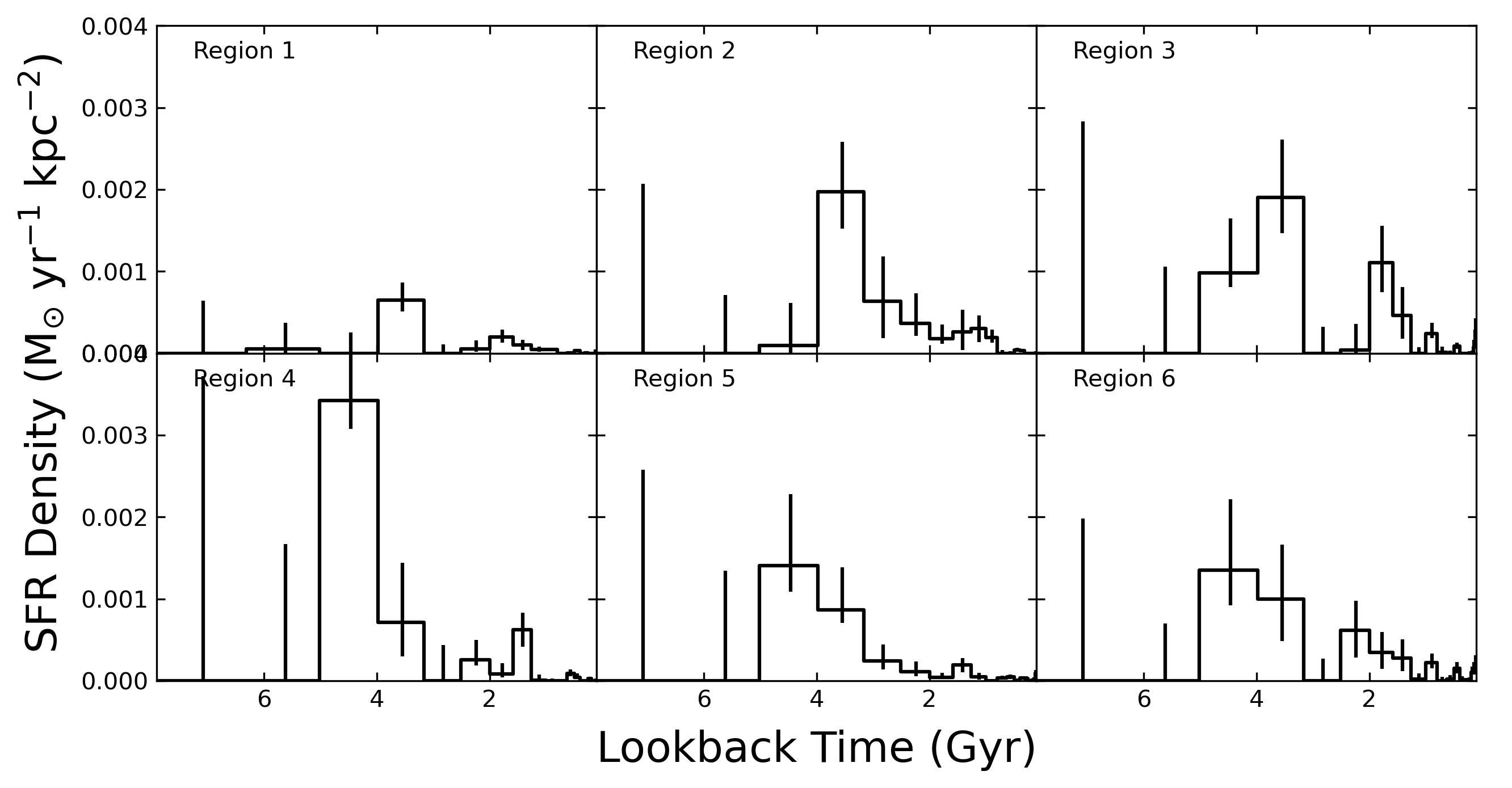}
\caption{Star formation density rate for the past 8 Gyr in the 6 regions, which are labeled in Figure \ref{fig:region_labels}. The star formation density rates  and their uncertainties of each region are available as Data behind the Figure.
} \label{fig:grid_sfr} 
\end{figure*} 

\begin{figure*}
\centering
\includegraphics[width=\textwidth]{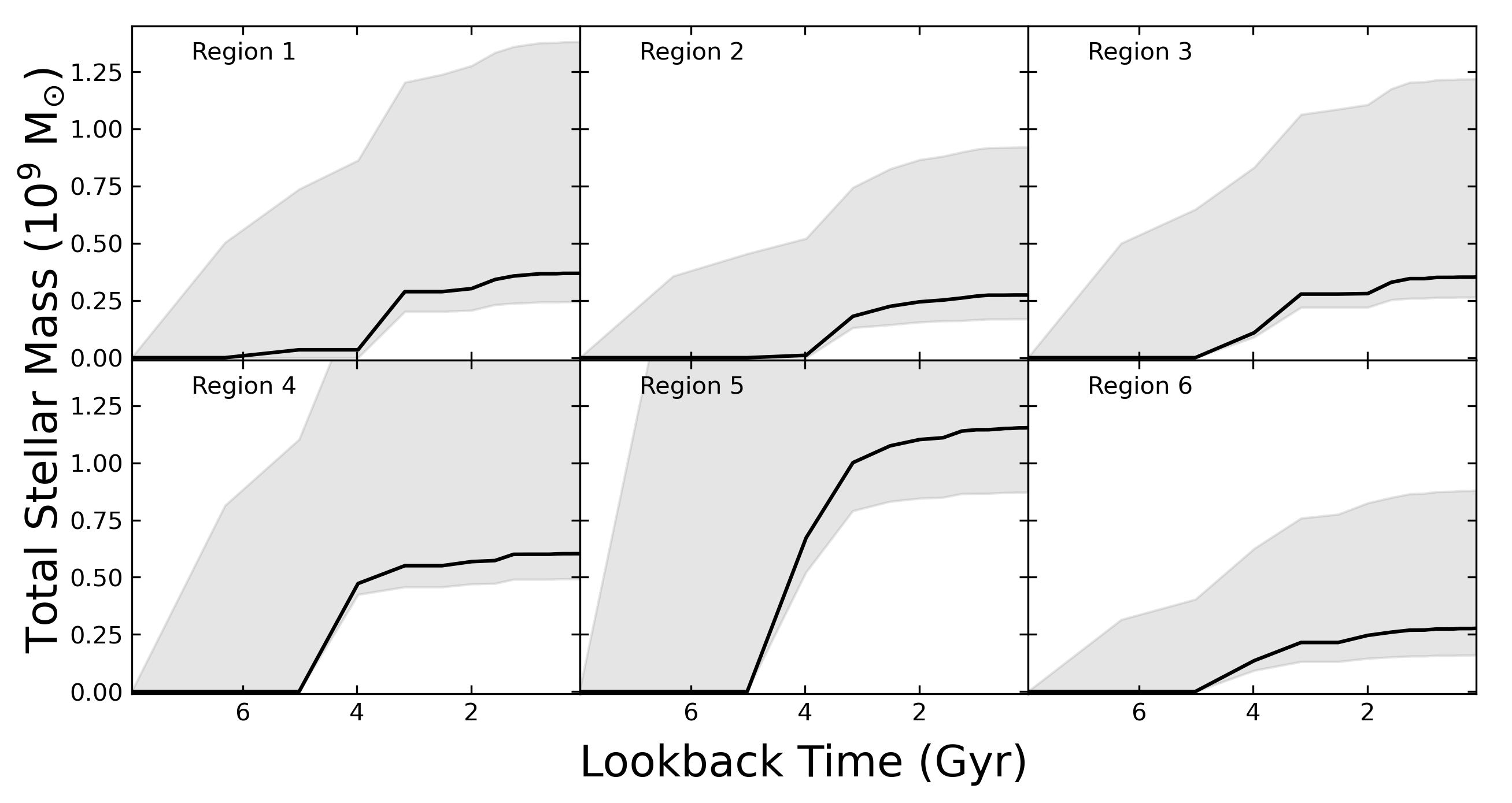}
\caption{Cumulative stellar mass formed for the past 8 Gyr in the 6 regions, which are labeled in Figure \ref{fig:region_labels}.
} \label{fig:grid_sfh} 
\end{figure*} 

Figure~\ref{fig:grid_sfr} and Figure~\ref{fig:grid_sfh} show the differential and cumulative SFHs for each region. The SFR density was calculated by normalizing the SFR by the projected area of each region. We note that there is fairly sizable variation in timing, duration, and amplitude of the SFHs as a function of position.  Because the inner stellar halo of M31 is dominated by sub-structure, and not a clear smooth component, it is perhaps not surprising to see such variations in the SFHs and stellar mass growth as the inner halo was assembled from many individual pieces, each with their own formation histories.  

In general, we find that most regions have prominent star formation $\sim3-5$~Gyr ago and lower-level star formation $\sim1-3$~Gyr ago.  Many also exhibit some level of star formation at ages of $<1$~Gyr.  
% In terms of cumulative stellar mass growth, there is a factor of $\sim3$ difference in the mass formed in the least (region 13) and most (region 3) active regions.  
% Region 3 contains NGC~205, while region 13 contains among the least amount of substructure according to the maps in \citet{mcconnachie18}.  
% It may be the closest we come to sampling the smooth underlying halo of M31. 
Expanding the NIR imaging and/or applying this method to other wide-field NIR imaging of M31 is an obvious path forward to exploring the SFH of its stellar halo at larger radii and to providing broader context with which to interpret our SFHs.

\begin{figure*}[th!]
\centering
\includegraphics[width=\textwidth]{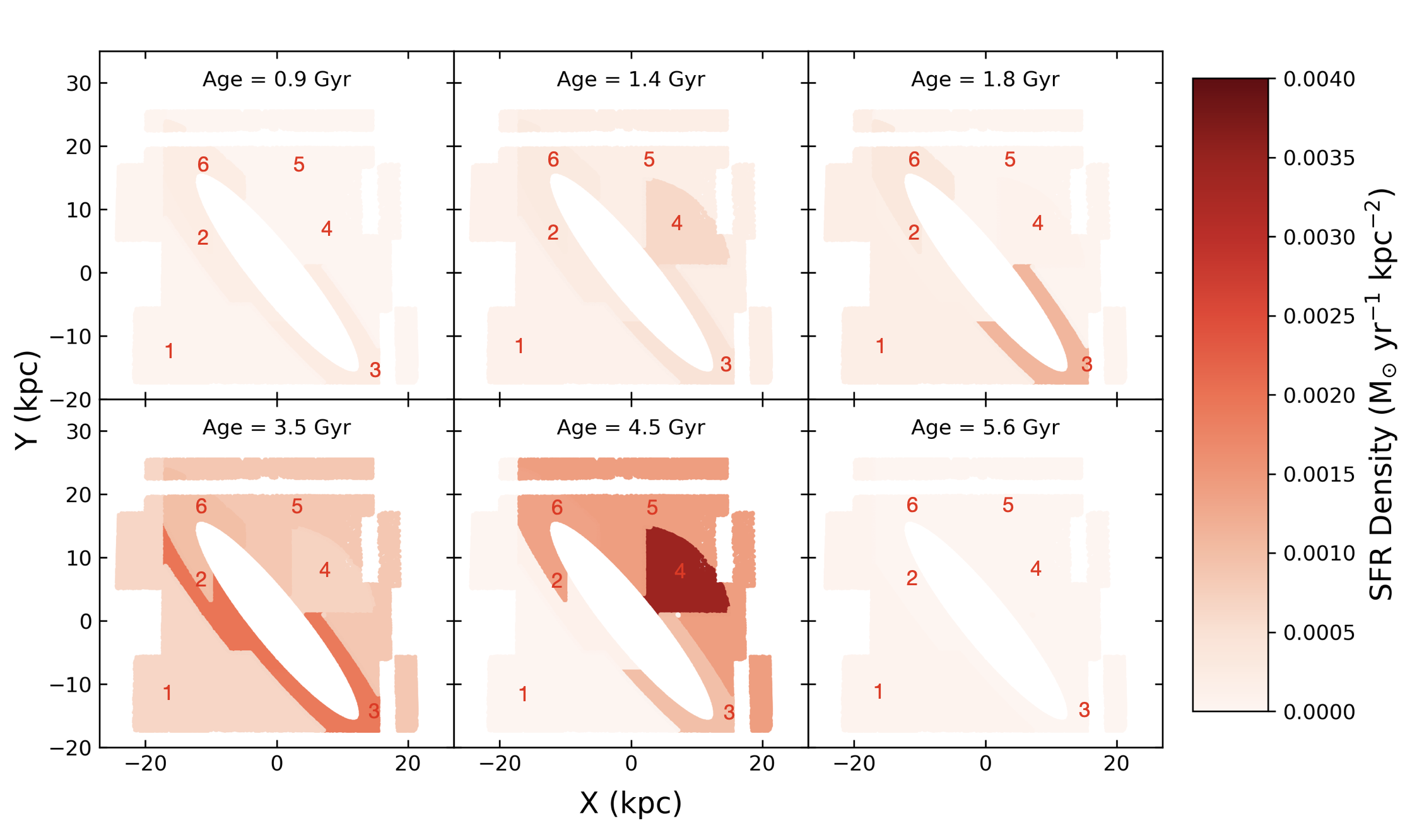}
\caption{Star formation rate density as a function of position for six different epochs.
} \label{fig:age_map} 
\end{figure*}

In Figure~\ref{fig:age_map}, we plot the spatially-resolved SFH of M31's inner stellar halo.  Each of the 6 regions is color-coded by the best fit SFR at select lookback times.  We note several interesting features. First, the clearest enhancements in SFR occurred $\sim3-5$~Gyr ago. 
Second, Region 4, which includes NGC 205, shows prominent star formation $\sim4.5$ and $\sim1.4$~Gyr ago.  The SFH of NGC~205 measured from deep HST optical imaging shows a similar enhancement in star formation activity $\sim1-4$~Gyr ago \citep{Savino25}. Older studies of NGC~205's stellar populations qualitatively support the presence of intermediate age and even young stellar populations in NGC~205's central region \citep[e.g.,][]{Richer84,Lee96,Davidge03,Davidge05,Rose05,Sharina06,Monaco09}.  
The interaction history of NGC~205 and M31 is not as well-studied as others in the M31 system, such as M32. There is a putative tidal `arc' of NGC~205 that extends to the north ( \citealt{McConnachie2004}), though its association with NGC~205 is not definitive \citep[e.g.,][]{Howley2008, Liu2024}.  
% The SFHs of NGC~205 and the arc regions appear to be anti-correlated.  
% The similarity of SFHs between NGC~205 (region 3) and region 8, may suggest an association between the two (e.g., region 8 could be material stripped from NGC~205), but the lack of obvious tidal features in that direction from ground-based data (e.g., in the PAndAS stellar maps, \citealt{Richardson11,mcconnachie18}) does not support this suggestion.
An interesting future angle would be to combine wide-area kinematics, such as measured from recent work with DESI \citep{Dey2023}, and potentially Subaru/PFS in the future, with the AGB-star based SFHs to see if there are clear relationships between ages and kinematics, analogs to the disk heating study of \citet{Dorman2015}. Finally, NGC~205 currently has no published proper motions.  An orbital history, in tandem with our SFHs, would be useful in helping to decipher the interaction history between NGC~205 and M31.

% Next, Region 16, which is in the direction of the GSS, has its highest activity levels from $\sim7$~Gyr ago.  An inner ``ring'' of regions ($D_{\rm projected} \sim 25$~kpc) all show enhanced activity $\sim2$~Gyr ago.  

\begin{figure*}
\gridline{\fig{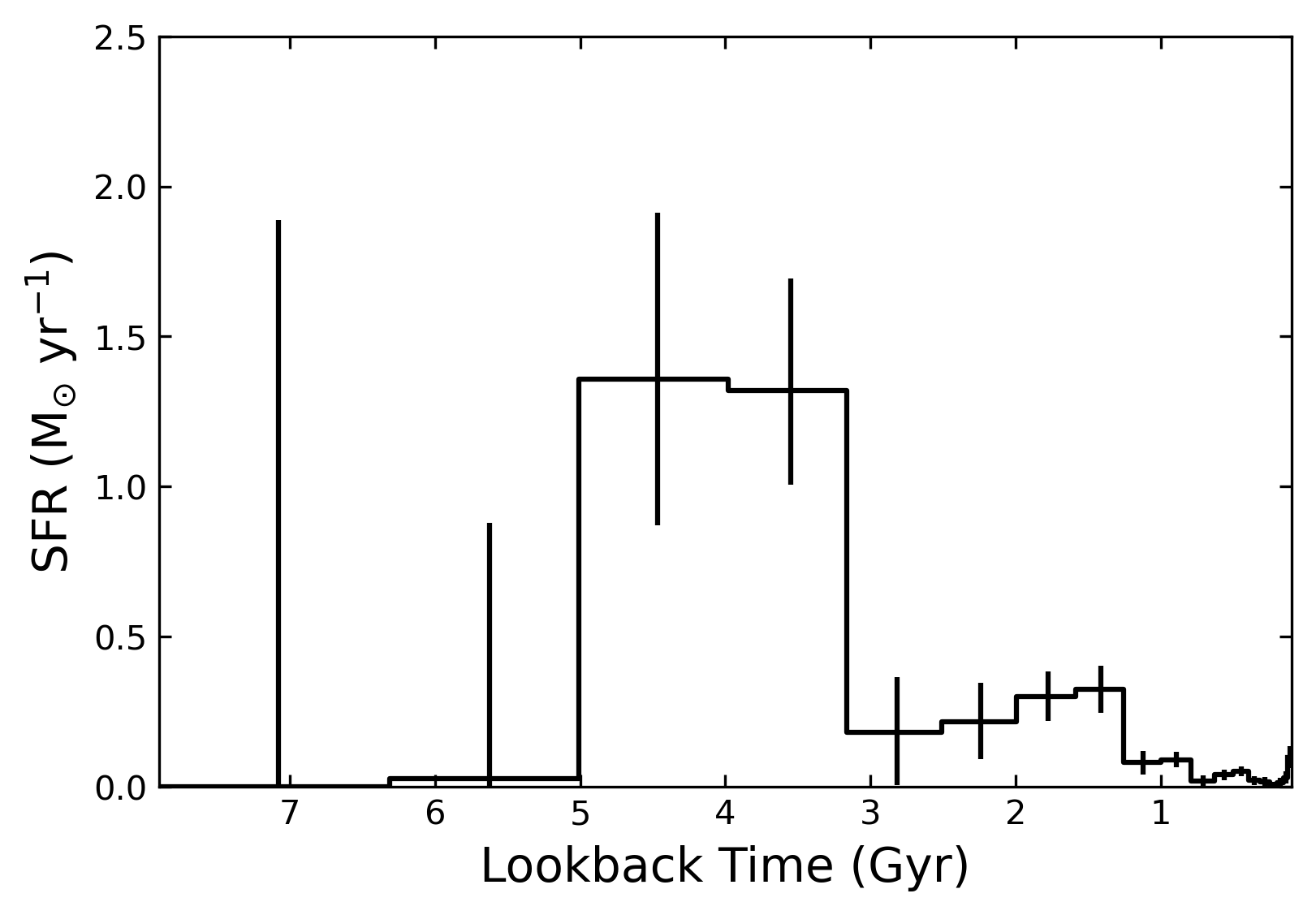}{0.5\textwidth}{}
          \fig{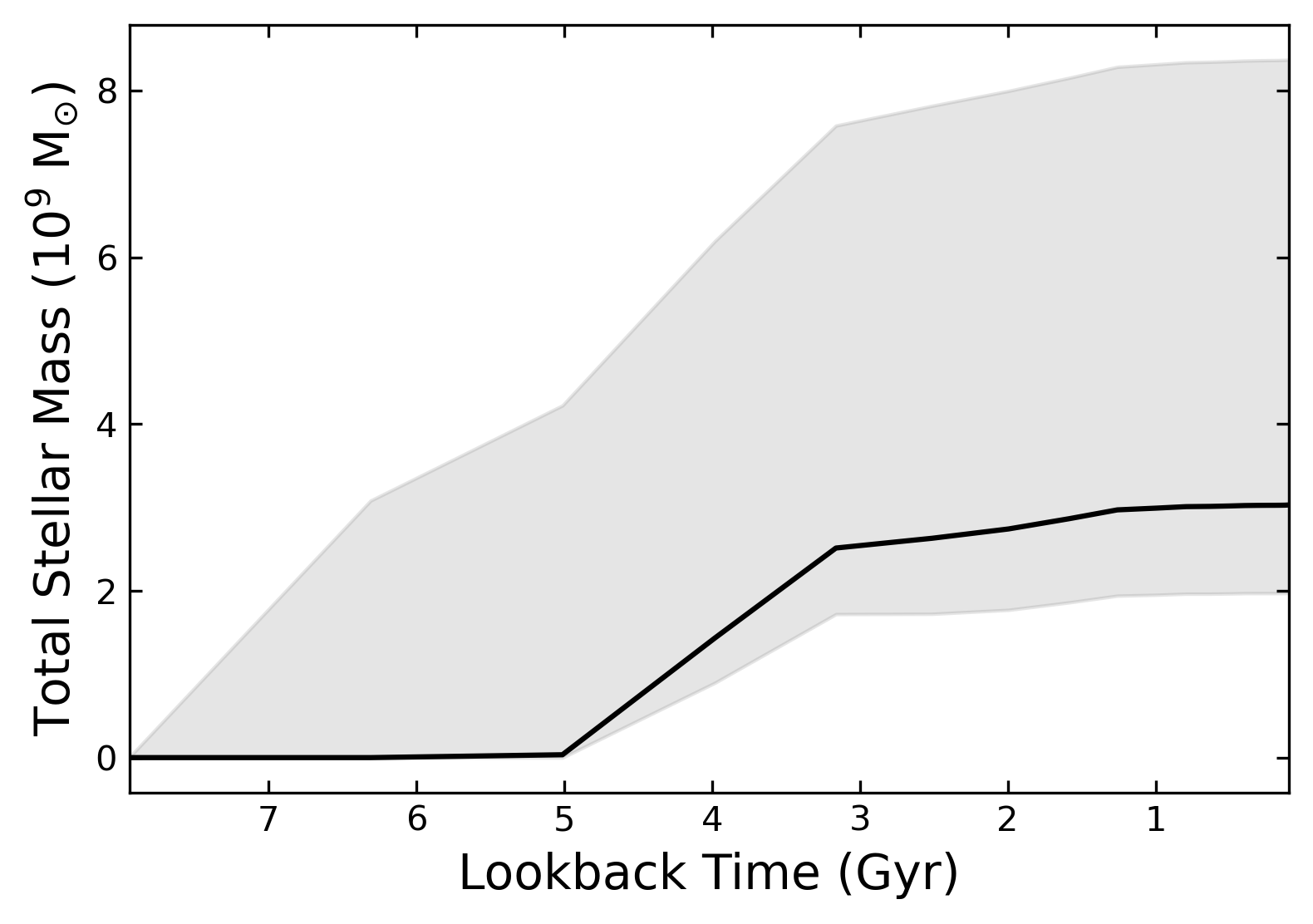}{0.5\textwidth}{}
         }
\caption{(Left) total star formation rate vs. age for the entire halo footprint. (Right) The cumulative stellar mass formed as a function of time.
}\label{fig:full_halo}  
\end{figure*}

Figure~\ref{fig:full_halo} shows the global SFH of M31's inner halo (i.e., all 6 regions summed together), along with the corresponding 68\% uncertainties.  The overall shape of the SFH is similar to many of the individual regions, with a prominent burst of star formation $\sim3-5$~Gyr ago, and steady and variable star formation from $\sim0.1-3$~Gyr ago.  Our global SFH shows that over the past $\sim8$~Gyr, the inner stellar halo of M31 formed $\sim3\times10^9 M_{\odot}$, $\sim80$\% of which appears to have occurred $\sim3$~Gyr ago.

%\section{Discussion}\label{sec:discussion}

\newcommand{\refs}{\textbf{REF}}

%\subsection{The Formation History of M31's Inner Stellar Halo}

% Another prominent sub-structure in the vicinity of M31 is M32, which is located in region 4.  Our SFH of region shows a prominent burst from $\sim5-8$~Gyr ago and lower levels from $\sim1-2$~Gyr ago. This is qualitatively similar to the SFH measured from moderately deep HST/ACS imaging \citep{Savino25}.  \citet{Monachesi2012}'s SFHs from two HST/ACS HRC fields that reach the oldest MSTO in M32 show broadly similar trends, however their coarse time resolution (i.e., 2-5~Gyr, 5-12 Gyr bins) make direct comparison challenging.  Several regions adjacent to M32 (regions 6, 7, 11, 15) also show enhanced star formation $\sim4-5$~Gyr, which is in line with some suggestions of when M32 (or its progenitor) may have started to interact with M31.  The literature is rich with plausible interaction scenarios between M32 and M31 \citep[e.g.,][]{Cepa88,Bekki01,Choi02,Ibata04,Block06,Gordon06,Dierickx14,dsouza18}. Further comparisons with predictions from detailed simulations and our SFHs may help support or rule out some of these possibilities.  A detailed exploration of these scenarios is beyond the intended scope of this paper.  As with NGC~205, a proper motion and orbital history of M32 is likely to greatly help interpretation.

%In a more global sense, our SFHs show that there was enhanced star formation close in the very inner part of the stellar halo $\sim4-5$~Gyr ago, while more regions located at larger radii have higher SFRs $\sim1-3$~Gyr ago.    

\begin{figure}
\includegraphics[width=\columnwidth]{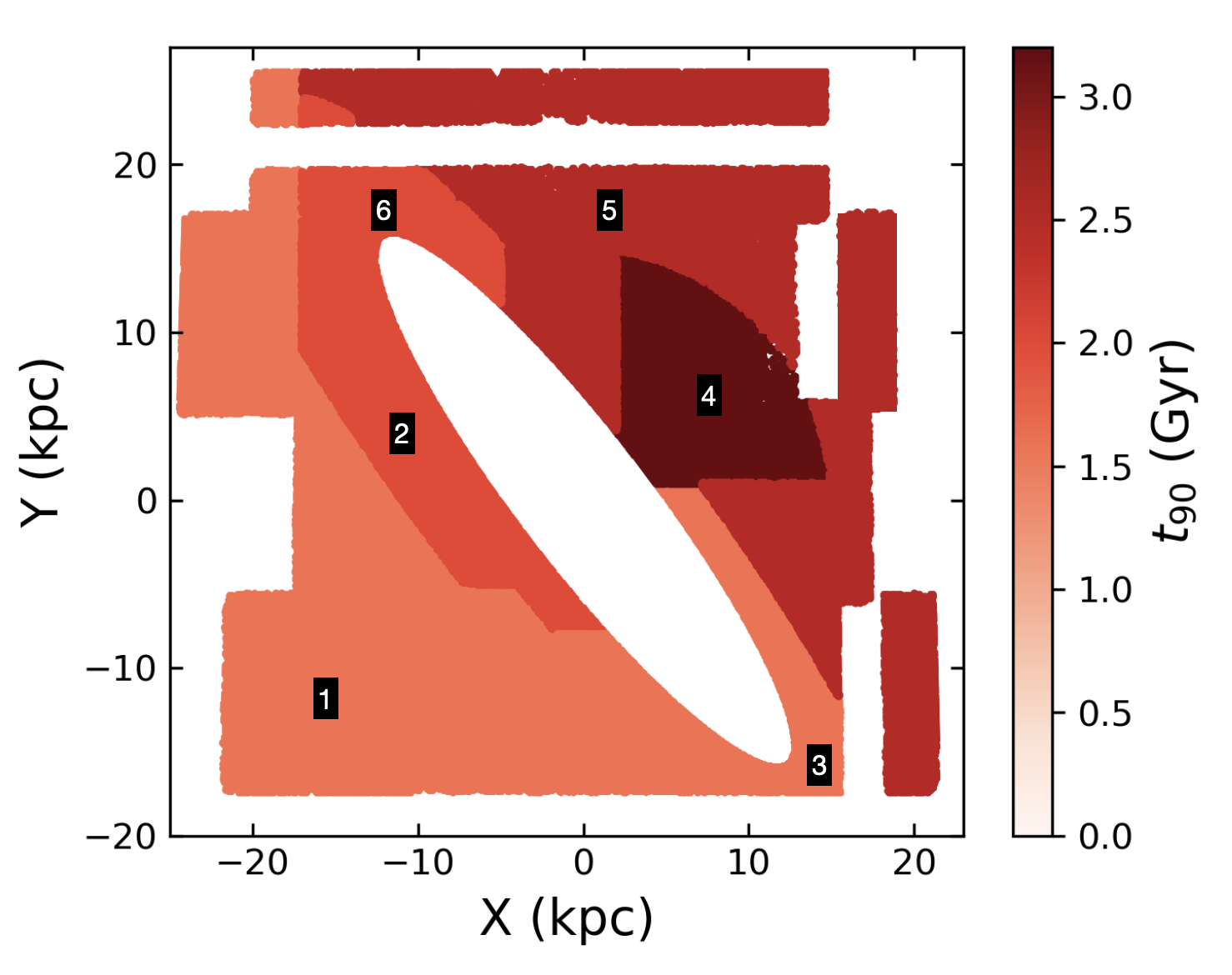}
\caption{Time at which 90\% of the stellar mass formed in each region of the inner stellar halo. The median uncertainty for $t_{90}$ across all regions was measured to be $\sim0.5$~Gyr.}
\label{fig:quench}  
\end{figure}

Figure \ref{fig:quench} shows the `quenching' time in each region.  As a proxy for `quenching' we use $t_{90}$, the lookback time at which 90\% of the stellar mass formed in each region's cumulative SFH.  
The bulk of SF ceased earlier in regions with larger values of $t_{90}$ compared with regions with smaller values of $t_{90}$.

\section{Caveats \& Limitations}\label{sec:caveats}

Despite demonstrating the power of resolved AGB stars in the NIR for SFH measurements, we emphasize a few areas of caution and items that need improvement.

First, in this paper, we only examined AGB star SFHs back to $\sim8$~Gyr ago.  This oldest age restriction was motivated by the conservative choice to empirically vet our data against another SFH and because of the limited photometry depth of this dataset.  However, in principle, as discussed in \S \ref{sec:isochrones}, it is possible to measure an AGB star SFH back to the oldest epoch of star formation (i.e., $\sim14$~Gyr ago).  While these ancient stars are unlikely to be found in the TP-AGB phase, they will appear as E-AGB stars on a NIR CMD.  Indeed, ancient MW globular clusters (GC) (e.g., NGC~6752; \citealt{Norris1981}) do have confirmed AGB stars, providing affirmation of this possibility. The limiting factor for testing this possibility is the lack of a suitable dataset (i.e., it must include the ancient MSTO and $>1000$  AGB stars).  We plan to continue this investigation in future papers.

%Further exploration of AGB

%any SFH measured using only AGB stars is limited to $\sim10$~Gyr in lookback time, due to the formation timescales of AGB stars.  It is not possible, for example, to use this approach to measure SFHs during the epoch of reionization.   Similarly, determining the quenching epoch, which is commonly done by computing the time at which 80 or 90\% of the stellar mass has formed, requires some care as the AGB stars provide no constraint on the mass formed prior to $\sim10$~Gyr ago.  One possible resolution is joint CMD modeling of the AGB stars and upper RGB, but such an exploration is beyond the scope of this paper.

Second, as illustrated in Appendix~\ref{sec:how_many}, a robust SFH requires several hundred AGB stars.  We adopt a  conservative recommendation of $\gtrsim1000$~AGB stars to recover the most accurate and precise SFH.

This minimum number is comfortably met by systems into the dwarf galaxy regime.  For example, the SMC ($M_V = -16.8$) and NGC~6822 ($M_V = -15.2$; \citealt{McConnachie2012}) each contain $>4000$~AGB stars \citep[e.g.,][]{Cioni2005, Boyer2011}, while other fainter systems have several hundred AGB stars as well \citep[e.g.,][]{dalcanton12a, Boyer2015}.  While galaxy luminosity provides a good rule of thumb for how many AGB stars to expect, the SFH is clearly an important component.  Systems that are purely ancient will have fewer AGB than those with more recent star formation (i.e., within the last several Gyr).  Even fairly luminous systems with `pauses' in the SFHs at intermediate ages \citep[e.g., WLM;][]{Albers19, McQuinn2024a} may have fewer AGB stars than their luminosities would suggest.   

%$\gtrsim700$~AGB stars above the 50\% completeness limit.  While we have not performed tests to quantify trade-off between number of AGB stars and SNR, our data indicate that an SNR of $\sim15$ and 700 AGB stars can produce a robust SFH.  

%While this minimum number is comfortably met for most nearby central galaxies, it does limit this analysis to moderately bright dwarf galaxies.  We suggest the SMC ($M_V=-16$, $M_{\star}\sim5\times10^8 \, M_{\odot}$; \citealt{McConnachie2012}) is a reasonable dividing line such that galaxies fainter or lower-mass than the SMC in general will not have enough AGB stars for a reliable SFH using our approach, while more massive galaxies will. Of course, this may depend on the particular SFH of the system, which is usually not known in advance.  One example we previously highlighted is WLM ($M_V\sim-14$, $M_{\star}\sim4\times10^7 \, M_{\odot}$; \citealt{McConnachie2012}), which only has 300 AGB stars \citep{Lee21}, but also has a relatively long period of quiescence at intermediate ages \citep[e.g.,][]{Albers19, McQuinn2024a}.  While we posit that SMC-like systems are a good rule of thumb for whether there will be enough AGB stars, this clearly depends on each galaxy's SFH and is a criteria that can be further refined.

Third, we did not calculate systematic uncertainties for our AGB star SFHs, due to the lack of multiple available AGB stellar models.  \citet{dolphin12} presents a means to estimate uncertainties on SFHs measured from resolved star CMDs that are designed to capture variations in the SFHs due to choice in stellar models.  The method for computing these systematics relies on being able to measure SFHs on test datasets using different stellar models.  Because \texttt{COLIBRI} uniquely provide suitable AGB star models, it is not possible to compute analogous systematic uncertainties for our AGB star SFHs.   Even within the \texttt{COLIBRI} models, specific choices are made that can affect the age characteristics of the AGB stars.  For example, our basis models for this analysis were generated with a Reimers mass loss value of $\eta = 0.2$ for RGB stars.  Large changes to this value can affect the evolution of AGB stars.  For example, a much higher mass-loss rate on the RGB does not produce AGB stars at the oldest ages.  It is not clear that such large mass-loss rates on the RGB are supported by data, but we nevertheless note this possibility.  This type of choice is not unique to AGB star models.  For example, the modeling of metal-diffusion in lower-mass stars can affect the lifetimes of MSTO stars, which can lead to systematically younger or older ages \citep[e.g.,][]{Hidalgo2018}.

Finally, we note that only NIR data are suitable for AGB star SFHs.  In the optical, there is too much overlap between the RGB and AGB to provide robust SFHs (e.g., see Figure \ref{fig:cmd}).  
% In limited exploration, this appears to be true even for a single optical (e.g., F814W or F090W) and NIR band.  
We have not conducted an exhaustive exploration of all possible filter combinations, but our findings thus far strongly suggest this approach is optimal with two NIR filters.  Age sensitivity is improved when both sides of the typical AGB SED peak are comfortably sampled (e.g., one band at $<1.5\mu$m and another at $>2\mu$m).
We also note that unfortunately the HST WFC3/IR F110W$-$F160W color cannot be used for determining AGB star SFHs; SSPs simulated from the COLIBRI stellar models by \cite{dalcanton12a} indicated that separating carbon-rich and oxygen-rich AGB stars with F110W$-$F160W colors is difficult because they occupy the same CMD space in that filter combination.

\section{Future Prospects}\label{subsec:future}
The method we have presented is well-suited for JWST.  JWST's exquisite sensitivity in the near- and mid-IR and high angular resolution have the potential to resolve AGB stars out to tens of Mpc. %Currently, even a single hour of intergreation time yields clearly resolved populations out to $\sim20$~Mpc.

The type of JWST dataset needed to measure AGB-star SFHs is well-matched to the JAGB (e.g., \citealt{lee24, Li24}), and to some extent, TRGB distance programs (e.g., \citealt{Anand24a, Anand24b, Hoyt2024, Newman2024}).  The JAGB method requires at least a few hundred AGB stars in the J region of the NIR CMD and is often measured in filter combinations that sample either side for the SED peak of AGB stars.  The high luminosity of AGB stars coupled with the characteristics of JWST suggest that AGB-based SFHs and JAGB distances already work out to 20~Mpc \citep[e.g.,][]{Anand24a, Anand24b, Hoyt2024, lee24} using $\sim1-3$~hours of integration time.  It is seems clear that JAGB distances and our SFH method can be extended to galaxies out to many tens of Mpc with suitably well-planned observations (e.g., to avoid too much crowding).  

Programs that aim to measure TRGB distances with JWST may also be suitable for AGB-based SFHs.  The $F090W-F150W$ filter combination provides for precise TRGBs, but is not particularly good for clearly separating all AGB and RGB stars in CMD space (see e.g., various  $F090W-F150W$ CMDs in \citealt{Anand24a, Anand24b, Newman2024}).  However, the $F115W-F277W$ filter combination appears to enable precise ($\lesssim1$.5\%) TRGB distances \citep{Newman2024} and appears to be a good filter combination of AGB star SFHs as AGB stars are both brighter in these bands, have little overlap with the RGB, and sample both sides of the SEDs peak for most AGB stars. We illustrate this further in Appendix \ref{sec:oldest_ages}.

Roman and Euclid offer similar potentials for AGB star SFHs, though their filter ranges are a little more limited than JWST.  We have not explored the specific capabilities of these facilities, but based on the $J-K$ analysis in this paper, AGB star SFHs from their NIR CMDs appear promising as well.

%combined with JWST's greater sensitivity and resolution, will recover SFHs in galaxies out to at least out to 20~Mpc, where AGB stars have already been detected with NIRCam (e.g., \citealt{lee24}). 

Beyond observational prospects, we also anticipate improvements in AGB star models. For example, improvements to the \texttt{COLIBRI} AGB stellar models are underway, as the \texttt{COLIBRI} team is actively calibrating its models in M31 and low-metallicity Local Group dwarf galaxies \citep{pastorelli20}. 

However, robust determinations of SFHs derived from AGB stars will require investigations into the full systematic uncertainties of the fits, similar to what has been done for SFHs based on optical CMDs \citep[e.g.,][]{Weisz2011, dolphin12}. As discussed in \S \ref{sec:caveats}, we highly advocate for the development of an independent set of very detailed TP-AGB models. In addition to aiding AGB star SFH studies like this one and others described in \S \ref{sec:intro}, another set of AGB models would improve our knowledge of the contribution of AGB stars to the integrated SEDs of galaxies at all cosmic epochs \citep[e.g.,][]{Melborne2012} and help with the calibration of the AGB star extragalactic distance scale, which use the P-L relations of Mira variable stars (e.g., \citealt{huang20, huang24}) or the mean luminosities of carbon stars (JAGB method; e.g., \citealt{Madore20, Ripoche20, Zgirski21, lee24}).

%For the JAGB method in particular, by directly comparing the JAGB magnitude to maps of AGB star formation history, we can constrain any potential age affects of this distance indicator. A better understanding of the systematics of the JAGB method and Mira P-L relation will also improve the precision of their measurements of the Hubble constant. 

\section{Conclusions}\label{sec:conc}

In this work we demonstrated a new approach to measuring quantitative SFHs of galaxies using NIR observations of AGB stars, \texttt{COLIBRI} stellar libraries, and the \texttt{MATCH} CMD modeling package. Our main takeaways from this paper are:

\begin{itemize}

    \item  The potential of resolved AGB stars for measuring SFHs has long been known. Theoretical models, specifically \texttt{COLIBRI}, now include many detailed physical processes lacking from other models, and have been anchored by NIR observations of resolved stellar populations in the Local Group.
    
    \item The \texttt{COLIBRI}  models show that AGB stars are highly sensitive to age and only moderately sensitive to metallicity.  Ages of AGB stars range from $\sim0.1$ to $\sim13$~Gyr.  We show that there is only a modest degeneracy between age and metallicity for AGB stars in the NIR.

    \item There are several advantages to measuring SFHs from resolved AGB stars in the NIR. First, AGB stars are much brighter ($M_J \lesssim -5$) than optical age sensitive features (e.g., red clump, MSTO).  AGB stars in the NIR are also less impacted by dust.   In the NIR AGB stars occupy a distinct region of the CMD with little-to-no overlap with other populations (e.g., RGB stars, MW foreground).  In the optical, AGB stars are challenging to separate from other phases of evolution that are not as age sensitive (e.g., the RGB) compromising their utility for SFH work.  

    \item Using synthetic data, we show that arbitrary SFHs can be recovered with the widely-used CMD fitting package \texttt{MATCH} using only AGB stars on a NIR CMD.  

    \item We use only AGB stars from ground-based UKIRT NIR photometry  to measure the SFH over the past 8~Gyr of M31's NE outer disk ($11<d<20$~kpc) covered by PHAT.  We find our SFH to be in excellent agreement with the area-matched PHAT SFH from \citetalias{williams17}, which was derived from an optical HST-based CMD.   The lookback time of this comparison is set by the fidelity of the PHAT SFH and the depth of our ground-based UKIRT data. Our simulations show that theoretically, AGB stars-based SFHs can be recovered to ages older than 8 Gyr ago. 
    
    \item The high luminosities of AGB stars in the NIR and their relative rarity mean that crowding effects are minimal, even across the entirety of M31's disk using UKIRT imaging (i.e., a modest-sized ground-based facility).

    \item By sub-sampling the number of AGB stars used for comparison against the PHAT SFH, we show that $\sim1000$ AGB stars can provide a robust SFH recovery.  We attribute this to the strong age-sensitivity of AGB stars in the NIR. 

    \item As a practical illustration of our AGB star SFH methodology, we measure the spatially resolved SFH of M31's inner stellar halo ($D_{\rm M31, projected} \sim20-30$~kpc. We find a nearly global burst of star formation $\sim3-5$~Gyr ago and much lower level, spatially distributed recent star formation 1 -- 3 Gyr ago.   We find a total of $M_{\star}\sim3\times10^9 M_{\odot}$ formed in M31's inner stellar halo in the last 8~Gyr.

    \item We use synthetic data to show that (a) SFHs only using NIR observations of AGB stars can robustly be recovered to lookback times of $13$~Gyr ago and (b) JWST CMDs of AGB stars (e.g., $F115W-F277W$) can be used to recover SFH across all cosmic time with similar fidelity.

    \item We discuss several caveats in our analysis such as the need for NIR imaging and the lack of alternative AGB star models needed to estimate systematic uncertainties on the SFHs.  We provide guidance for what types of galaxies are suitable for application of this method.

    \item We discuss the enormous scientific potential for measuring SFHs in galaxies out to tens of Mpc using NIR observatiosn of resolved AGB stars with JWST, Roman, and Euclid.

    %\item AGB stars can now be used to trace the SFH of the resolved stellar populations in galaxies from 100~Myr to at least 8~Gyr, and, in principle, back to the oldest epochs of star formation. This development is thanks to recent improvements in the \texttt{COLIBRI} models resulting from precise observational constraints provided by NIR imaging of nearby galaxies including the Magellanic Clouds \citep{Rosenfield2014, rosenfield16, marigo17, pastorelli19, pastorelli20}.

    %\item ABG stars in the NIR provide several 
    
    %Because AGB stars are the reddest and brightest stars in a galaxy, they are easily identified and less afflicted by dust extinction effects than typical SFH tracers like the red clump and MSTO in optical wavelengths.
   % This methodology can be applied to the stellar halos of galaxies where reddening corrections are negligible and stellar densities are moderate. 

    %\item We first apply our technique to two synthetic galaxies, and successfully recovered the input SFH.

   % \item While our AGB star method still needs some refinement, particularly in quantifying systematic errors (discussed in \S \ref{sec:caveats}), this paper represents a major first step forward in the method of reconstructing SFHs solely from AGB stars.

\end{itemize}

\begin{acknowledgments}
We thank the anonymous referee for their very constructive comments and suggestions which improved this paper. We also sincerely thank Léo Girardi, Paola Marigo, Giada Pastorelli, and the Padova group for their extraordinary work developing the \texttt{COLIBRI} models which made this work possible, and for answering our questions about the stellar isochrones. A.J.L. acknowledges Meredith Durbin for useful discussions and for making their code on Hess diagrams publicly available,  which we used to generate the CMDs in this paper.
 A.J.L. also thanks Wendy Freedman, Alex Ji, Barry Madore, and Kyle Rocha for stimulating discussions that helped inspire this work.   D.R.W. thanks the Minnesota Institute for Astrophysics for hosting his sabbatical visit during the writing of this paper.  A.J.L. and D.R.W. thank Quixotic Coffee in St. Paul, MN for providing a stimulating and convenient working environment.

A.J.L. thanks the LSSTC Data Science Fellowship Program, which is funded by LSSTC, NSF Cybertraining Grant \#1829740, the Brinson Foundation, and the Moore Foundation; her participation in the program has benefited this work.  Y.R. was supported by the National Natural Science Foundation of China (NSFC) through grant numbers 12133002 and 12203025.

This research has made use of NASA's Astrophysics Data System Bibliographic Services. This research has made use of the NASA/IPAC infrared Science Archive (IRSA), which is operated by the Jet Propulsion Laboratory, California Institute of Technology, under contract with the National Aeronautics and Space Administration. This paper uses data from UKIRT, which is owned by the University of Hawaii (UH) and operated by the UH Institute for Astronomy.

\end{acknowledgments}

\vspace{5mm}
\facilities{UKIRT (WFCam)}

\software{Astropy \citep{astropy13, astropy18, astropy22}, \texttt{imf} (\url{https://github.com/keflavich/imf}), \texttt{MATCH} \citep{dolphin02,dolphin12,dolphin13}, Matplotlib \citep{hunter07}, NumPy \citep{harris20}, Pandas \citep{pandas}, Scikit-learn \citep{pedregosa11}}

\appendix
\restartappendixnumbering

\section{Photometric Quality-metric cuts}\label{sec:error}

\begin{figure}\figurenum{A1}
\centering
\includegraphics[width=\columnwidth]{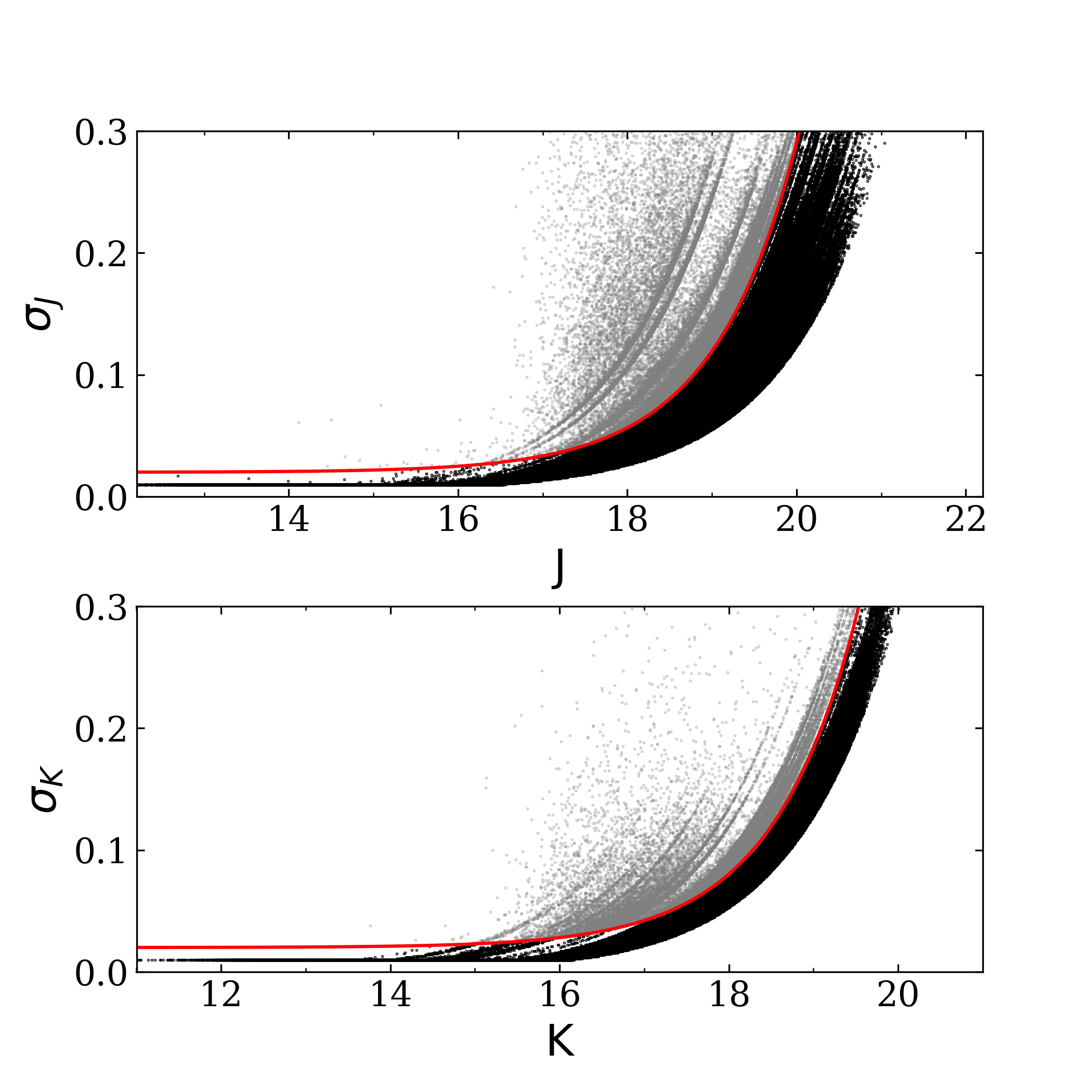}
\caption{Photometric uncertainty as a function of J- and K-band magnitudes. Quality cuts are shown in red, with the sources passing the given restriction in black and the sources not passing the restriction in grey.  }
\label{fig:error}
\end{figure}

We clean the UKIRT photometry used in the main text via the photometric uncertainty as a function of their J-band and K-band magnitudes. The functional forms of these functions are shown below and also plotted in Figure \ref{fig:error}:

\begin{equation}
    \sigma_J < 0.02 + 0.003 \times e^{m_J - 15.5} 
\end{equation}

\begin{equation}
    \sigma_K < 0.02 + 0.003 \times e^{m_K - 15}. 
\end{equation}

\section{Recovery of Oldest AGB Stars}
\label{sec:oldest_ages}

\begin{figure}\figurenum{B1}
\centering
\includegraphics[width=\columnwidth]{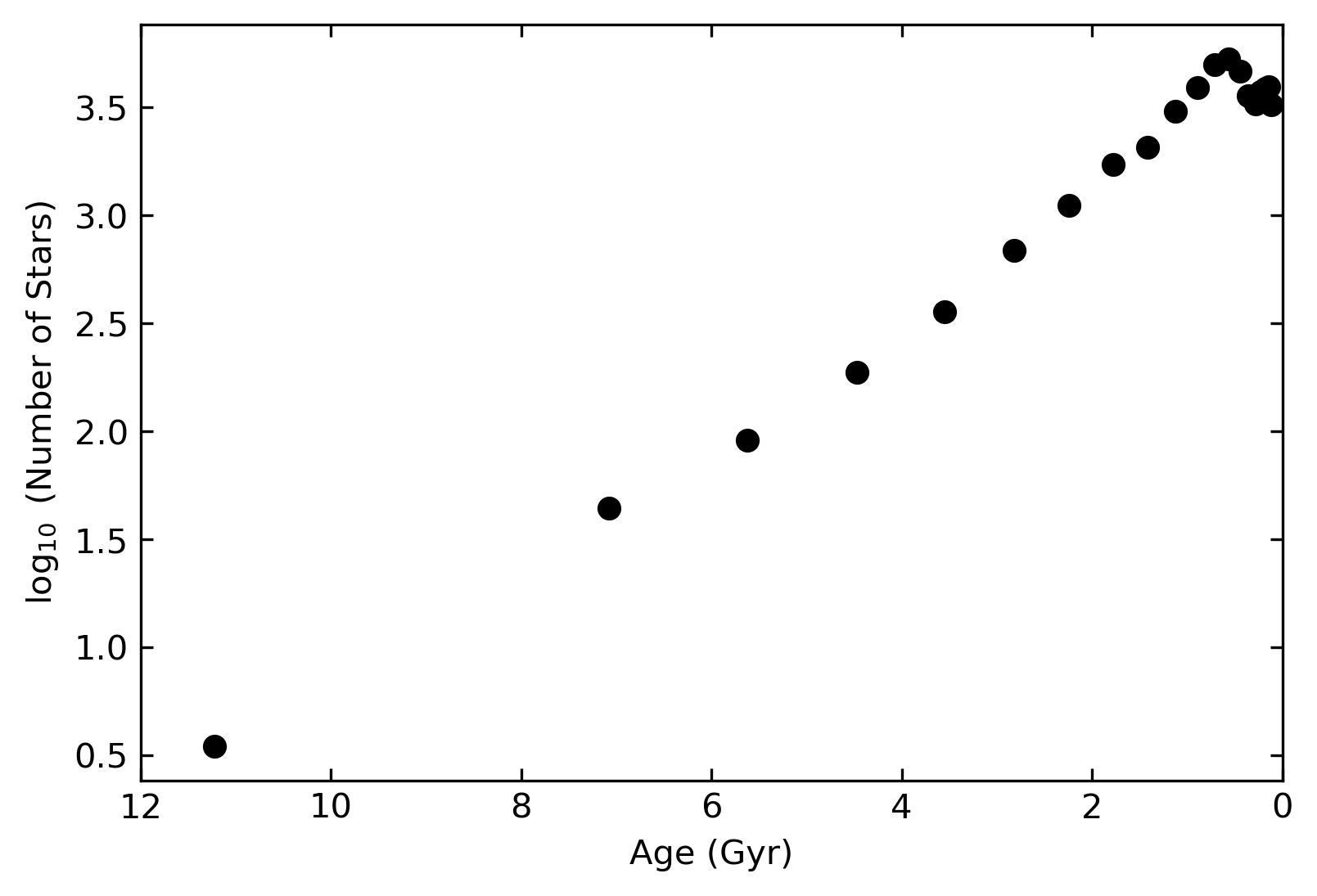}
\caption{Number of stars recovered per age bin vs. age for our full $t<14$~Gyr fits. We recover a negligible number of stars in the t = 11.2~Gyr age bin, as expected. }
\label{fig:n_stars}
\end{figure}

Throughout the paper, we limit our simulated and real SFHs to a lookback time of $\sim8$~Gyr ago.  This age was motivated by the depth of our ground-based data.  To verify that our AGB photometric selection criteria excluded stars with ages $t>8$~Gyr for our analysis in Section \ref{subsec:m31}, we show a plot of the number of stars recovered per age bin in Figure \ref{fig:n_stars}. This plot demonstrates that we are recovering a negligible number of stars with ages $t>8$~Gyr, as expected.    When we find a suitable NIR dataset, i.e., that has $>1000$ AGB stars and an optical CMD that reaches fainter than the oldest MSTO, we will extend the photometric selection criteria to fainter magnitudes to undertake an SFH comparison to older ages.

\begin{figure*}\figurenum{B2}
\gridline{\fig{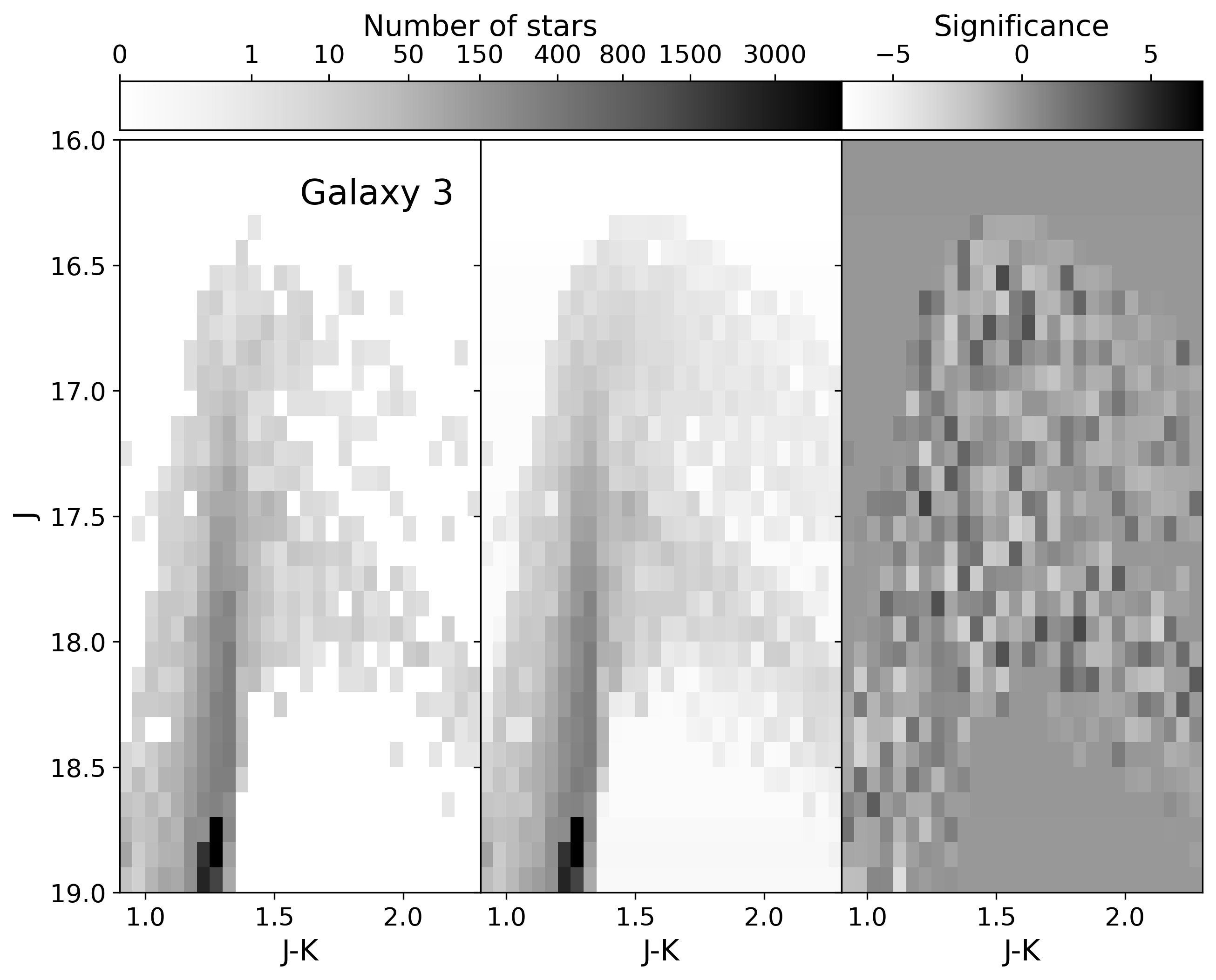}{0.333\textwidth}{} 
        \fig{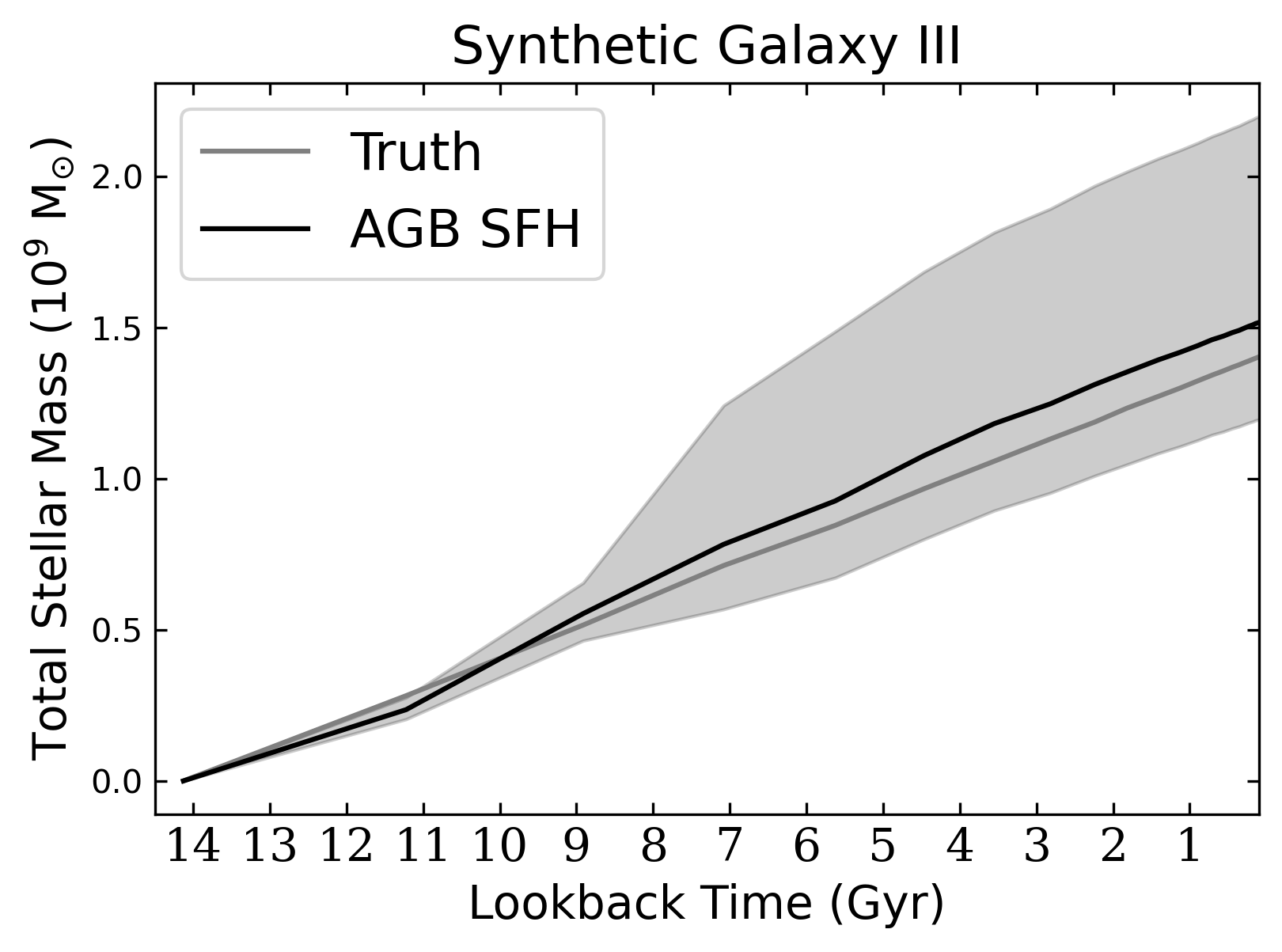}{0.333\textwidth}{}
        \fig{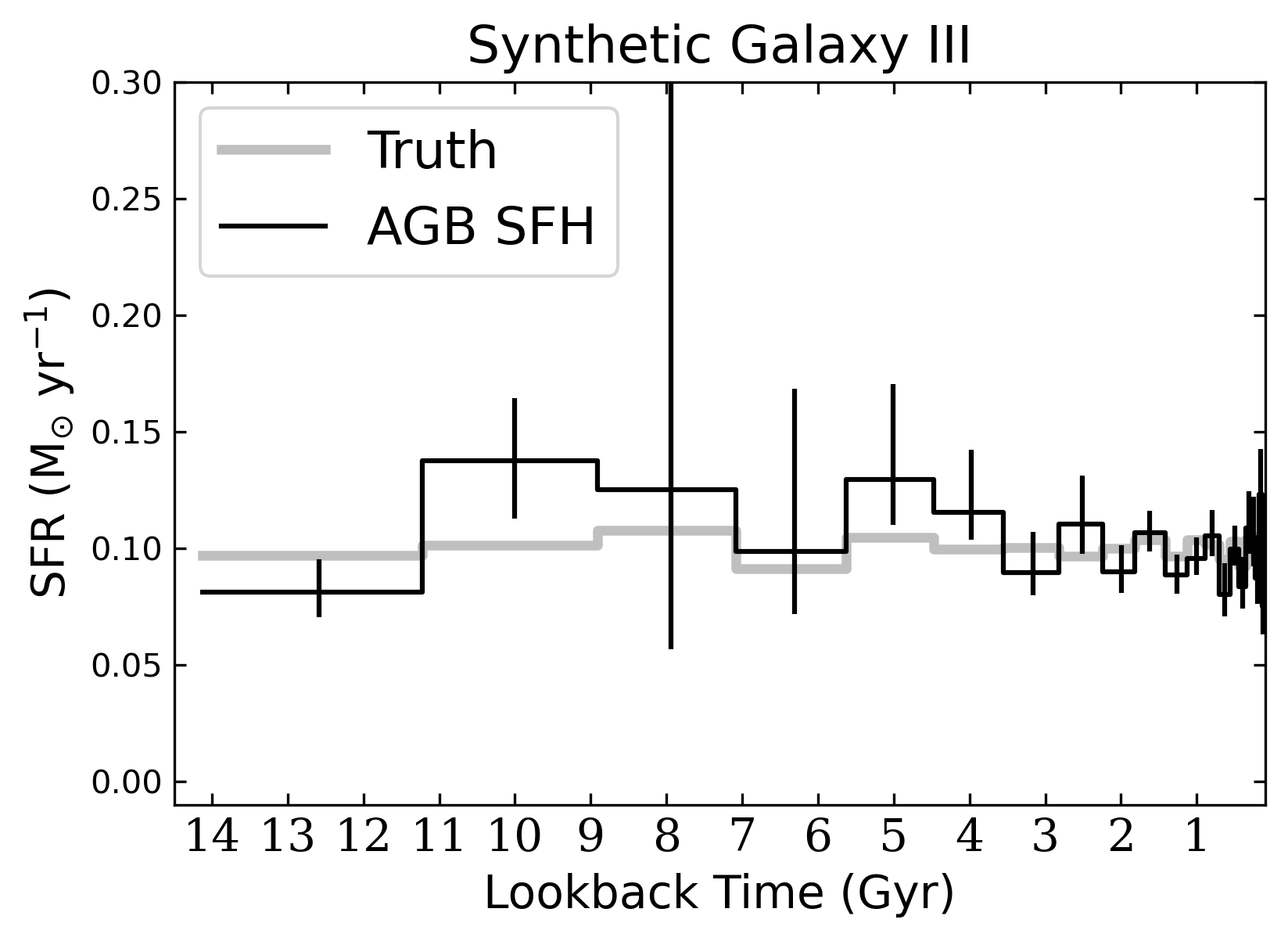}{0.333\textwidth}{}}
\caption{An example of recovering a SFH back to the oldest epochs of star formation, $\gtrsim13$~Gyr ago, using only AGB stars in the NIR. (Left) Density maps for the observed, best-fit, and residual CMDs, expressed in units of Poisson standard deviations. (Middle) Cumulative stellar mass formed as a function of time. (Right) Total star formation rate as a function of time.  The excellent agreement indicates that resolved AGB stars in the NIR, in principle, can be used to measure SFHs across all cosmic time.}\label{fig:mock_old} 
\end{figure*}

In the meantime, we use simulated data to show that, in principle, AGB star SFHs can be recovered back to the oldest epochs of star formation.  We adopt an age range of $\log(\rm{age}) = 8.0$ to $10.1$ with a spacing of 0.1 dex, and a metallicity range of [M/H] = -1.4 to 0.0 dex with a spacing of 0.2~dex.  The results for a a roughly constant star formation rate of $0.1~\rm{M_{\odot}/yr}$ from 100 Myr to 14.1 Gyr in the $J-K$ filter combination are shown in Figure \ref{fig:mock_old}.  The mock tests show excellent recovery of the input SFH.

\begin{figure*}\figurenum{B3}
\gridline{\fig{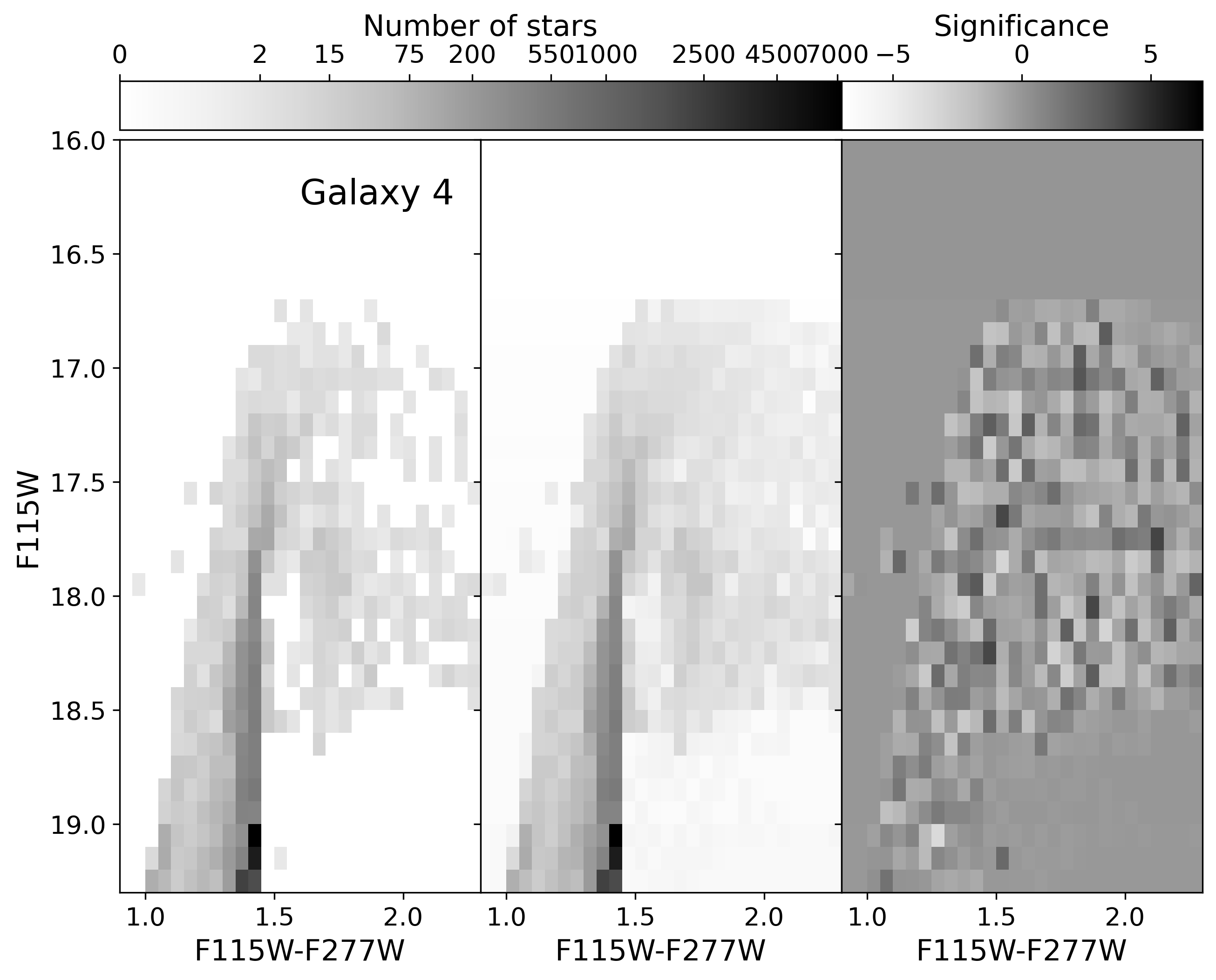}{0.333\textwidth}{} 
        \fig{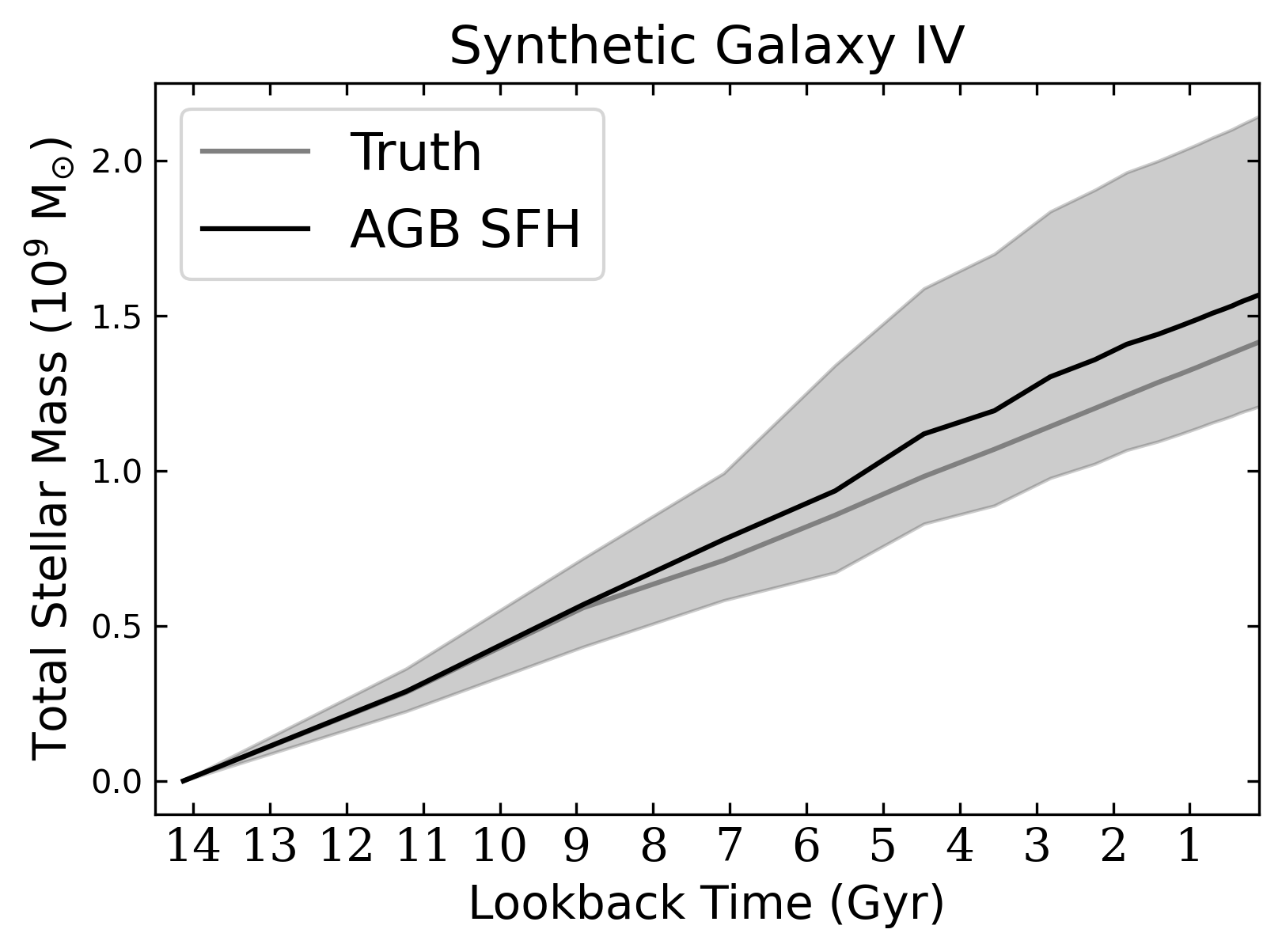}{0.333\textwidth}{}
        \fig{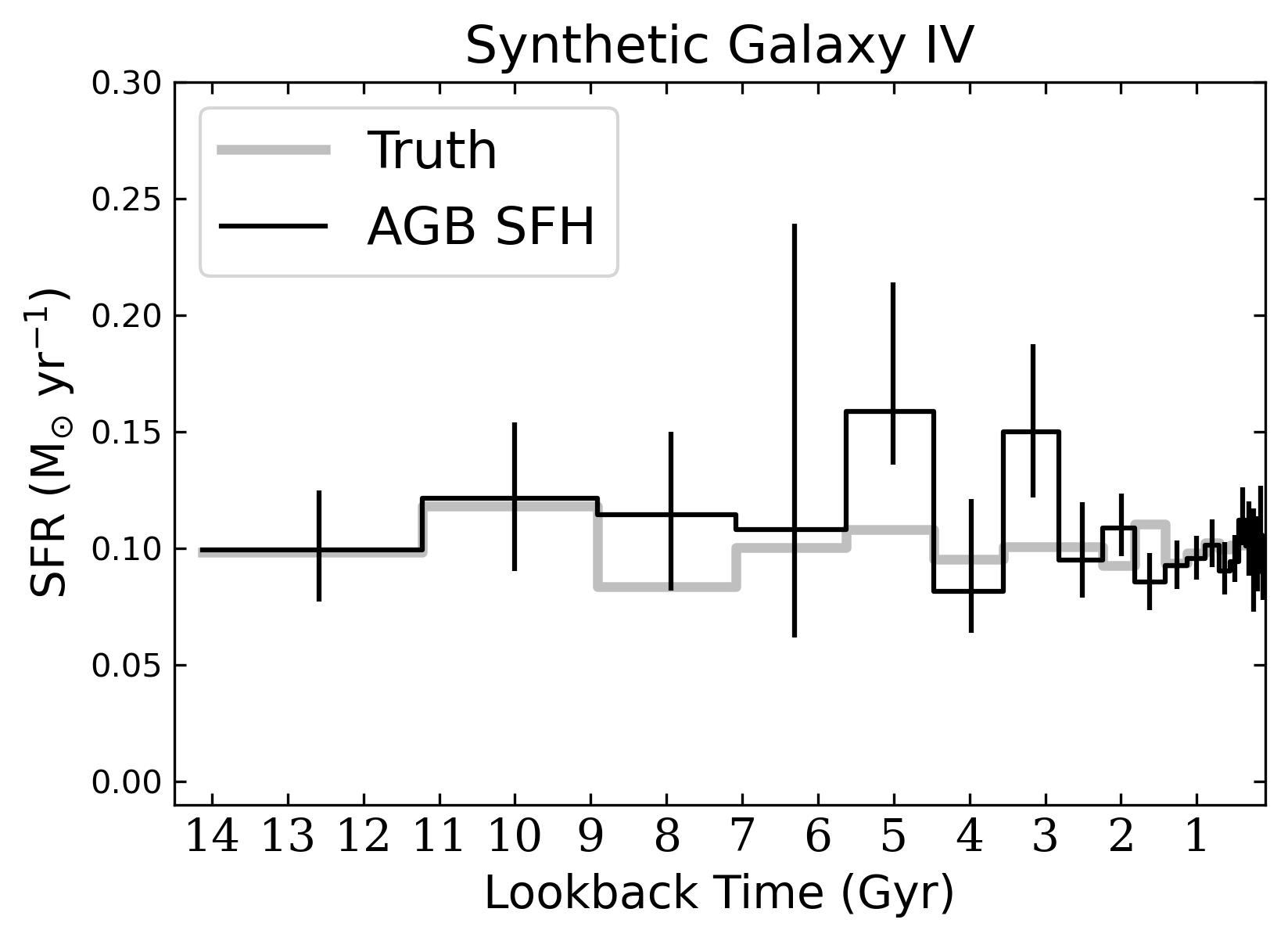}{0.333\textwidth}{}}
\caption{Same as the previous figure and test, only for the JWST filter combination $F115W-F277W$, which is recommended fore precision TRGB distances \citep[e.g.,][]{Newman2024}.  }\label{fig:mock_jwst} 
\end{figure*}

In Figure \ref{fig:mock_jwst}, we undertake the same test only using an example JWST filter set $F115W-F277W$.  The motivation for this filter combination is that it is particularly efficient for measuring precise TRGB distances with JWST \citep[e.g.,][]{Newman2024}. Such data would also well-suited for measuring AGB star SFHs.  As with the other simulated data experiments, the recovered SFH is in good agreement with the input SFH.  Deviations in the SFR at early time are not statistically significant and are similar to fluctuations in CMD-based SFH recovery discussed in \citet{dolphin02}, and further emphasize the importance of robust statistical uncertainties \citep[e.g.,][]{dolphin13}.  Our purpose was largely to demonstrate the utility of excellent JWST TRGB data for this additional use case.  We plan a more exhaustive exploration of JWST (as well as Roman and Euclid) filter combinations in a future paper.

\section{AGB Age-metallicity relation}\label{sec:amr}
\begin{figure}\figurenum{C1}
\centering
\includegraphics[width=\columnwidth]{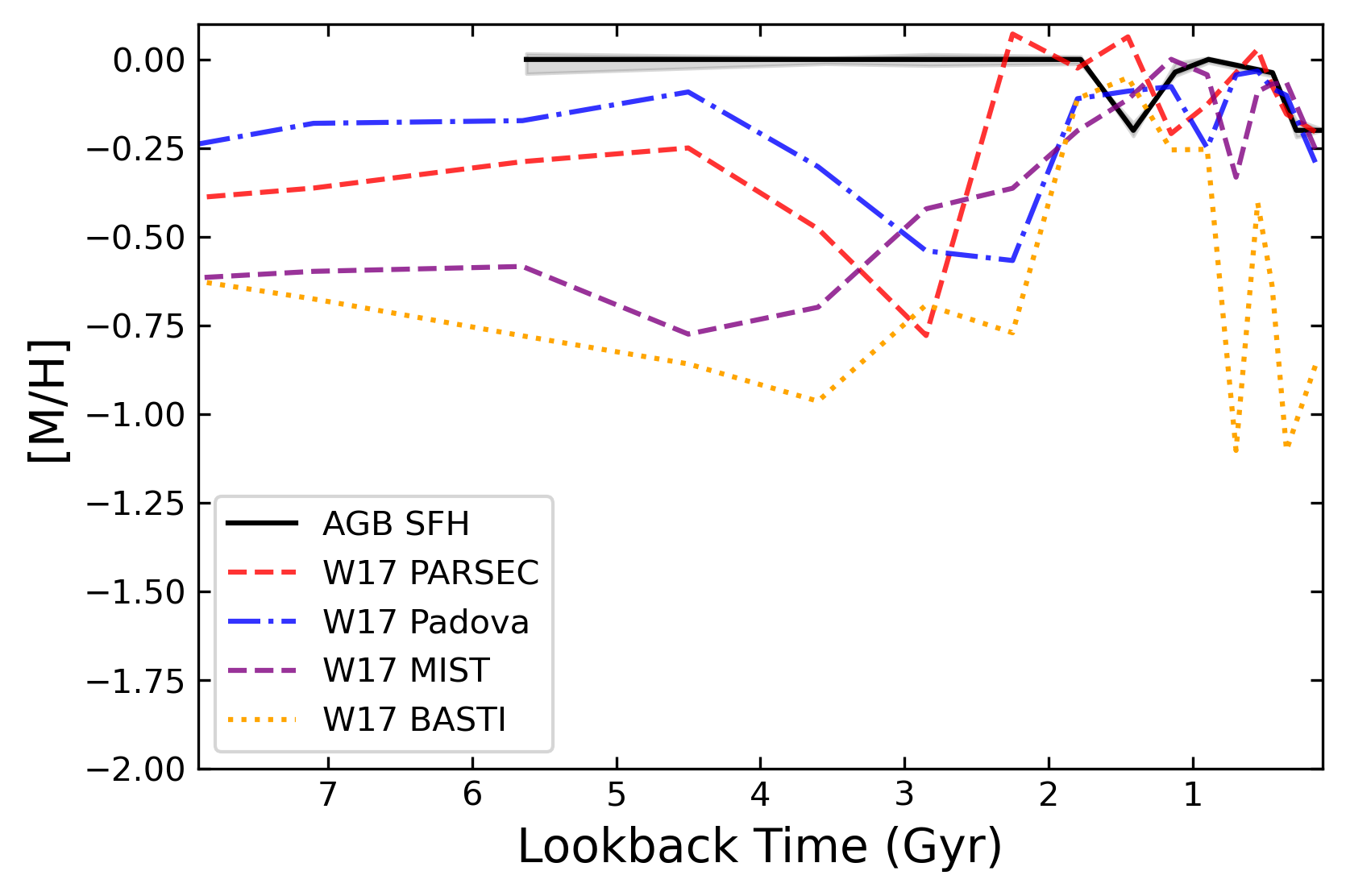}
\caption{Metallicity as a function of age for our fits, compared to all the model fits from \citetalias{williams17}.}
\label{fig:amr}
\end{figure}

Figure \ref{fig:amr} presents the best-fit AMR from our AGB star SFH measurements, compared with all the model fits from \citetalias{williams17} for data $d>11$~kpc. We find good overall agreement between their overall shapes, mainly the noticeable drop to [M/H]$\approx-0.5$~dex at $\sim2-3$~Gyr, albeit we recovered it at slightly earlier times, and only to [M/H]$\approx-0.25$~dex. We plan to further investigate how well AGB stars can recover the AMRs compared with fits derived from the oMSTO in future papers.

\section{How many stars are needed for an accurate AGB star SFH measurement?}\label{sec:how_many}

\begin{figure*}\figurenum{D1}
\gridline{\fig{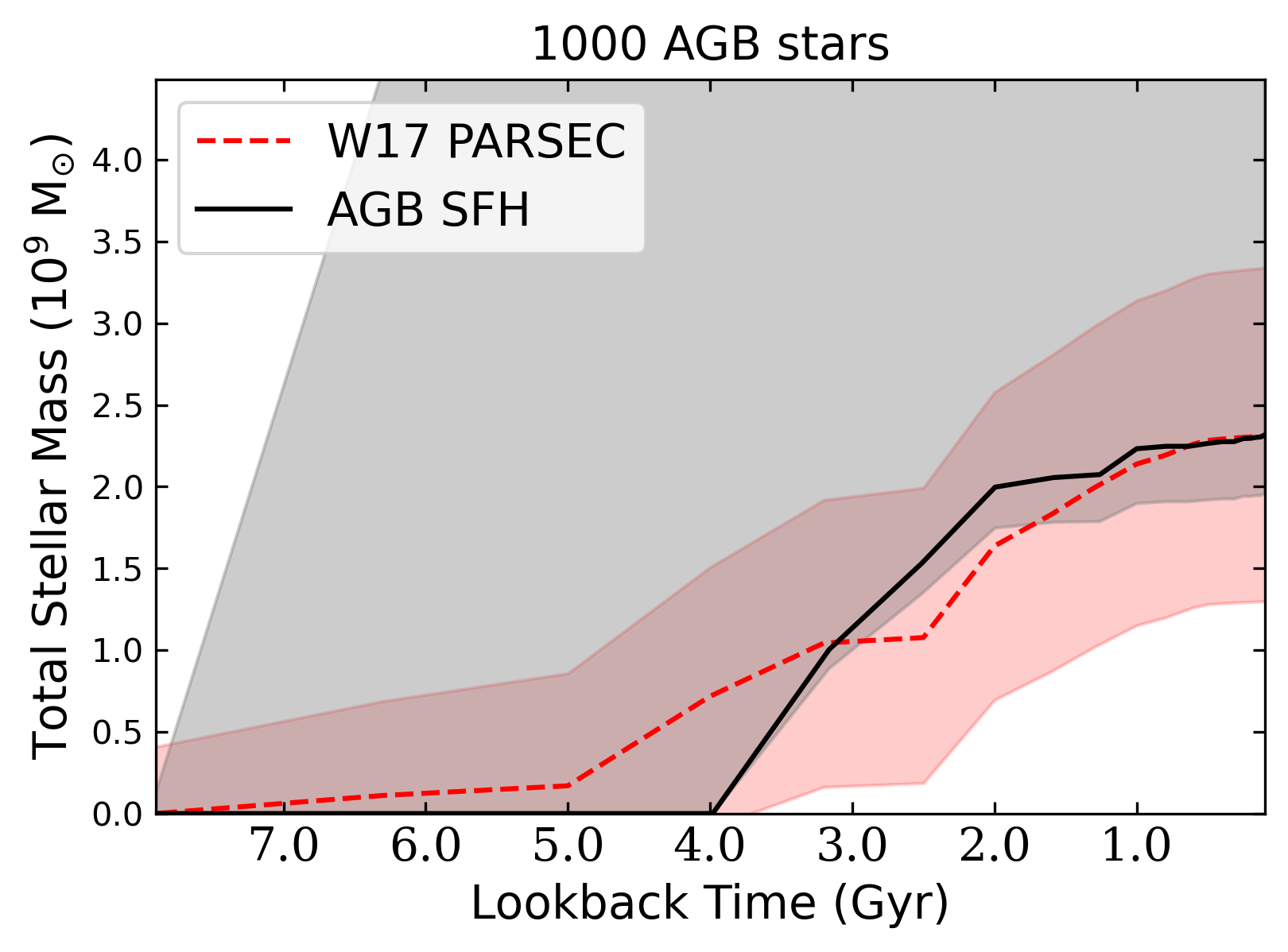}{0.5\textwidth}{} 
        \fig{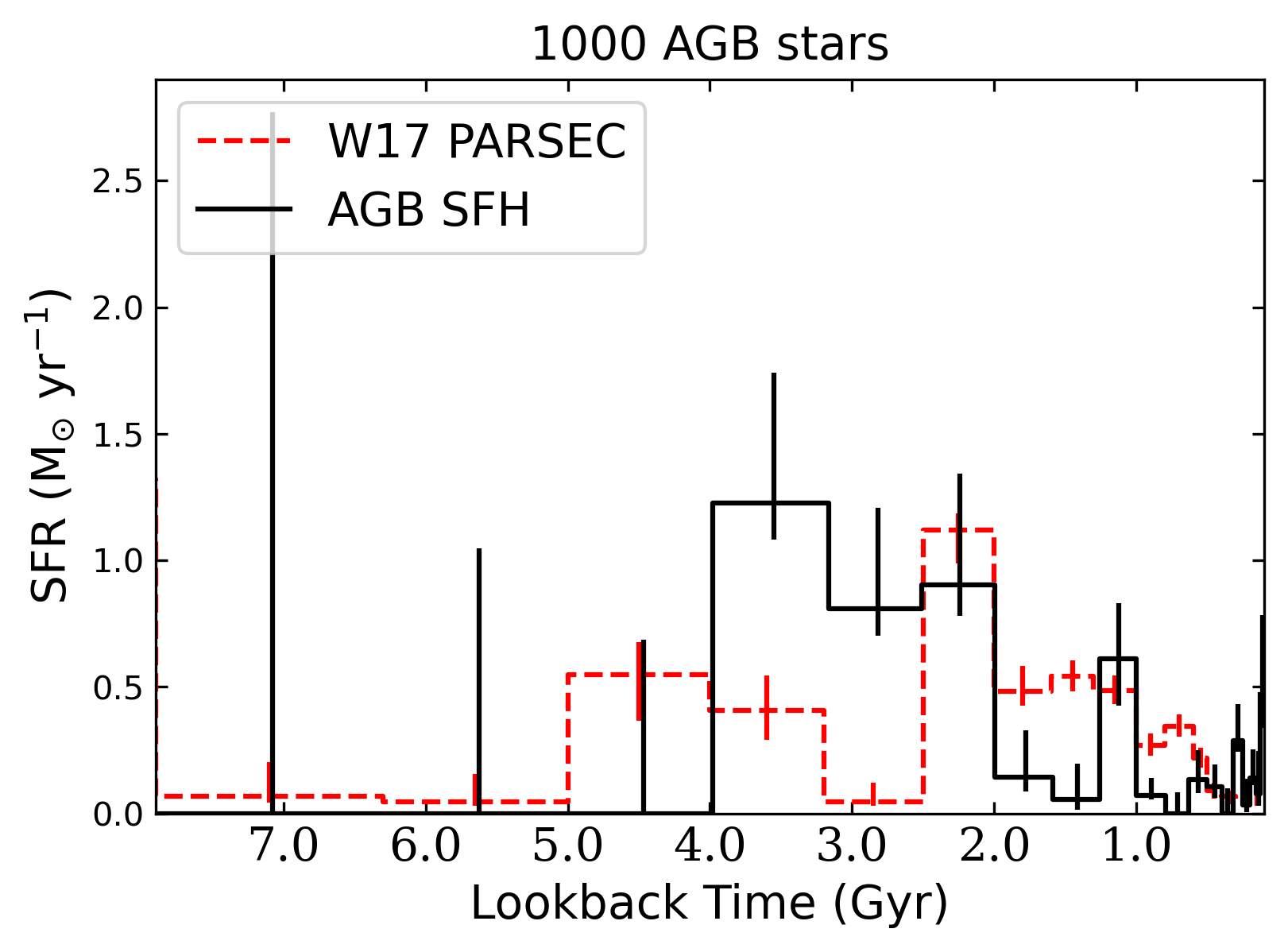}{0.5\textwidth}{}}
\gridline{\fig{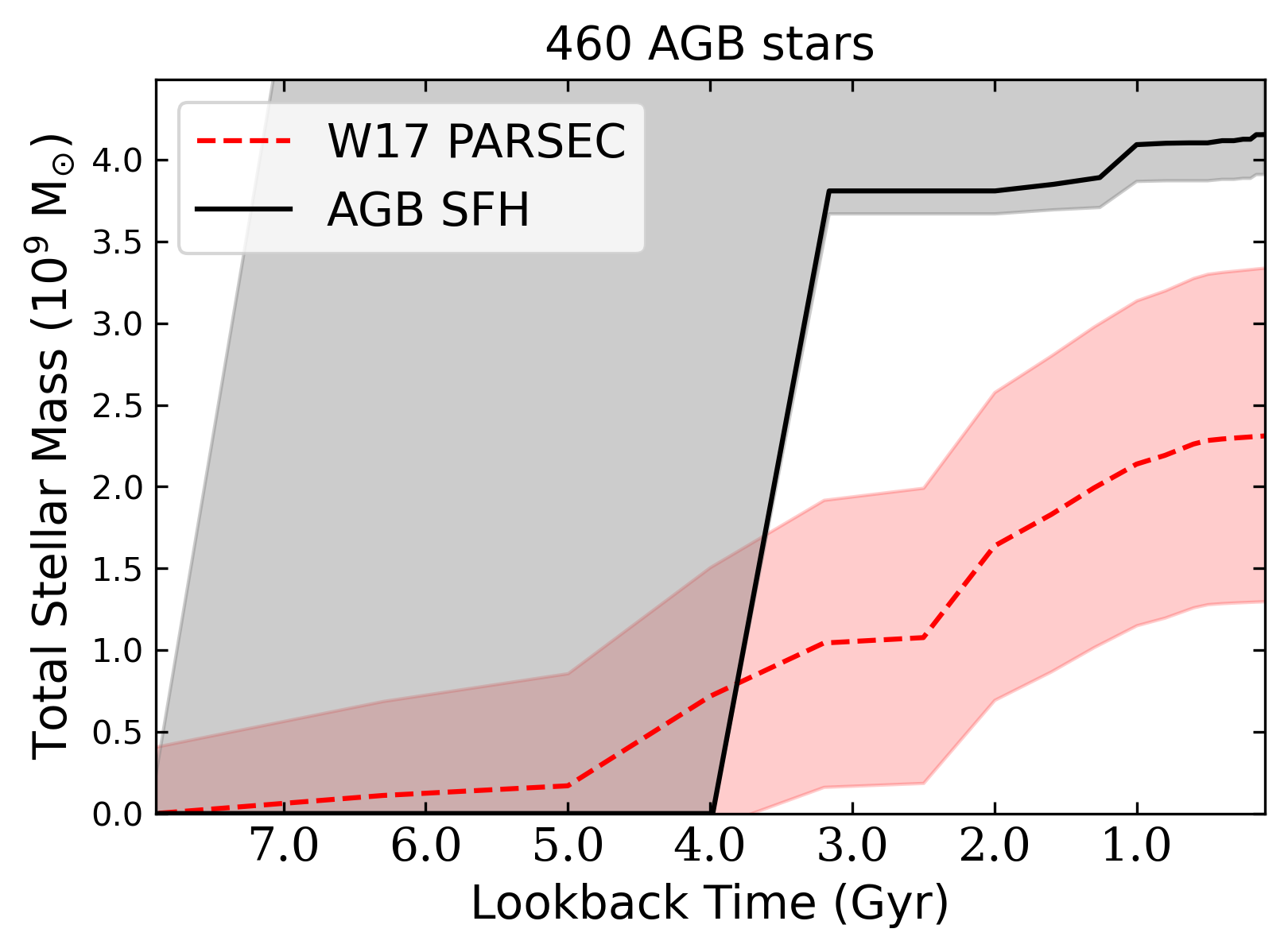}{0.5\textwidth}{} 
        \fig{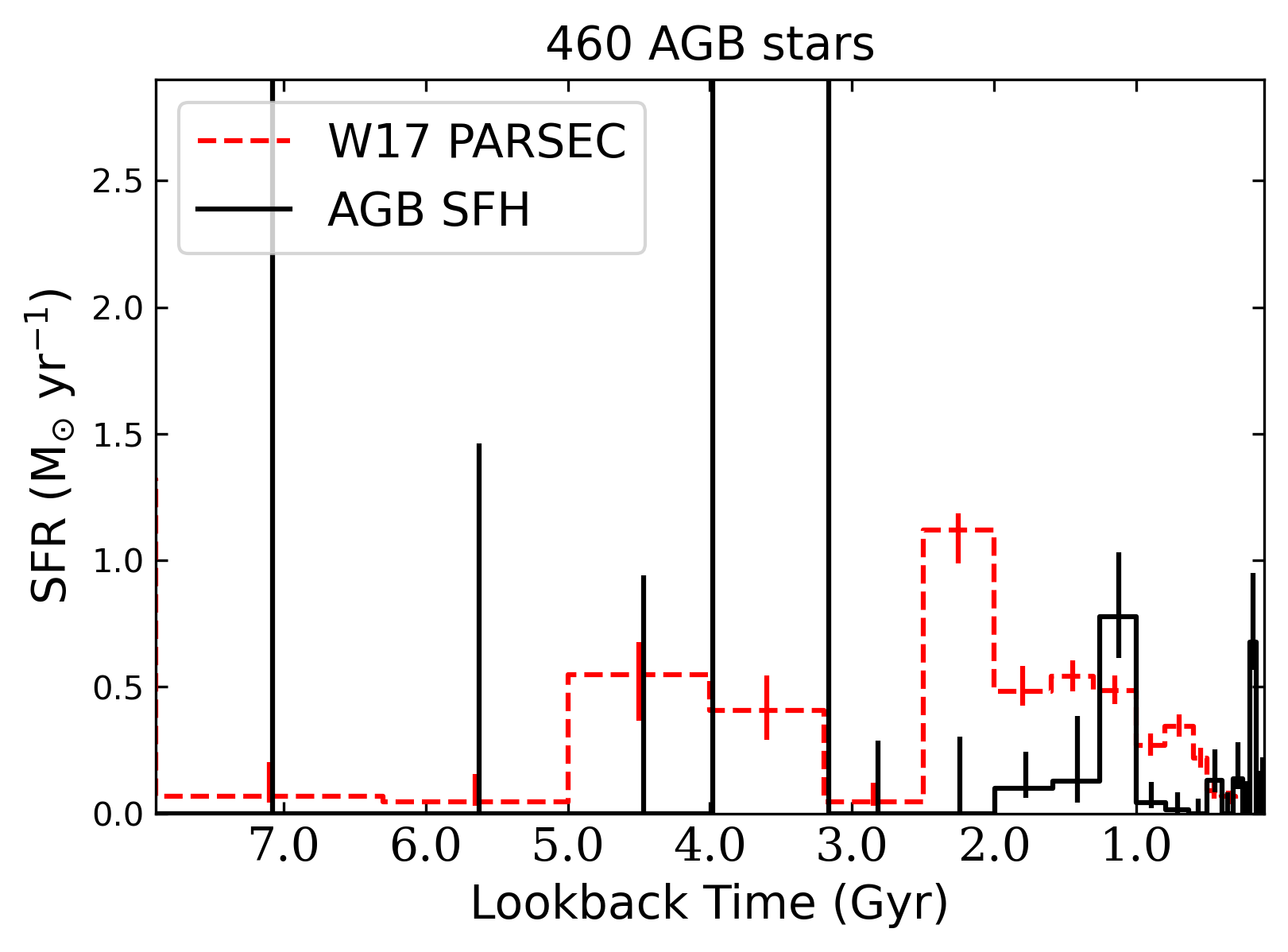}{0.5\textwidth}{}}
\caption{Accuracy and precision of AGB star SFH and SFR fits as a function of the number of randomly sampled AGB stars from the full sample of $\sim7700$~AGB stars used in Figure \ref{fig:disk_compare}.
} \label{fig:disk_uncertain} 
\end{figure*}

In this section, we investigate the effect of decreased sample size on the precision and accuracy of our AGB star SFH measurements. As described in \S \ref{subsubsec:phat_AGB}, a sample of 7700 AGB stars delivered small statistical uncertainties that agree well systematically with the measurement from \citetalias{williams17}. To test how the uncertainties and bias change as a function of sample size, first we randomly sample 3600 AGB stars, or about 1/2 of our main sample, and re-measure the AGB star SFH. We repeat this test until our sample size decreases to $\sim460$~AGB stars. In Figure \ref{fig:disk_uncertain}, we show two representative fits measured from random samples of 1000 AGB stars and 460 AGB stars.

As expected, the statistical uncertainties increase with decreased sample size. We note a sample of 1000 AGB stars deliver an AGB star SFH that agrees with (albeit with larger statistical errors) the AGB star SFH measured from the full sample, to within $1\sigma$. For a sample of 460 AGB stars, the measured AGB star SFH  begins to systematically deviate slightly from the AGB measured from 7700 AGB stars. Therefore, we conclude that accurate AGB star SFH measurements should be based on samples of at least $\sim1000$~AGB stars.

\clearpage\newpage





\begin{thebibliography}

\bibitem[Albers et al.(2019)]{Albers19} Albers, S.~M., Weisz, D.~R., Cole, A.~A., et al.\ 2019, \mnras, 490, 5538

\bibitem[Anand et al.(2024a)]{Anand24a} Anand, G.~S., Tully, R.~B., Cohen, Y., et al.\ 2024a, \apj, 973, 83

\bibitem[Anand et al.(2024b)]{Anand24b} Anand, G.~S., Tully, R.~B., Cohen, Y., et al.\ 2024b, arXiv:2408.16810


\bibitem[Astropy Collaboration et al.(2022)]{astropy22} Astropy Collaboration, Price-Whelan, A.~M., Lim, P.~L., et al.\ 2022, \apj, 935, 167

\bibitem[Astropy Collaboration et al.(2018)]{astropy18} Astropy Collaboration, Price-Whelan, A.~M., Sip{\H{o}}cz, B.~M., et al.\ 2018, \aj, 156, 123


\bibitem[Astropy Collaboration et al.(2013)]{astropy13} Astropy Collaboration, Robitaille, T.~P., Tollerud, E.~J., et al.\ 2013, \aap, 558, A33

\bibitem[Barmby et al.(2006)]{barmby06} Barmby, P., Ashby, M.~L.~N., Bianchi, L., et al.\ 2006, \apjl, 650, L45

\bibitem[Bekki et al.(2001)]{Bekki01} Bekki, K., Couch, W.~J., Drinkwater, M.~J., et al.\ 2001, \apjl, 557, L39


\bibitem[Bernard et al.(2015a)]{bernard15a} Bernard, E.~J., Ferguson, A.~M.~N., Chapman, S.~C., et al.\ 2015a, \mnras, 453, L113

\bibitem[Bernard et al.(2015b)]{bernard15b} Bernard, E.~J., Ferguson, A.~M.~N., Richardson, J.~C., et al.\ 2015b, \mnras, 446, 2789

\bibitem[Bhardwaj et al.(2016)]{Bhardwaj16} Bhardwaj, A., Kanbur, S.~M., Macri, L.~M., et al.\ 2016, \aj, 151, 88

\bibitem[Block et al.(2006)]{Block06} Block, D.~L., Bournaud, F., Combes, F., et al.\ 2006, \nat, 443, 832


\bibitem[Boyer et al.(2011)]{Boyer2011} Boyer, M.~L., Srinivasan, S., van Loon, J.~T., et al.\ 2011, \aj, 142, 103

\bibitem[Boyer et al.(2015)]{Boyer2015} Boyer, M.~L., McQuinn, K.~B.~W., Barmby, P., et al.\ 2015, \apjs, 216, 10

\bibitem[Boyer et al.(2024)]{Boyer24} Boyer, M.~L., Pastorelli, G., Girardi, L., et al.\ 2024, \apj, 973, 120




% \bibitem[Boyer et al.(2011)]{boyer11} Boyer, M.~L., Srinivasan, S., van Loon, J.~T., et al.\ 2011, \aj, 142, 103

% \bibitem[Boyer et al.(2019)]{boyer19} Boyer, M.~L., Williams, B.~F., Aringer, B., et al.\ 2019, \apj, 879, 109

\bibitem[Bressan et al.(2012)]{bressan12} Bressan, A., Marigo, P., Girardi, L., et al.\ 2012, \mnras, 427, 127

\bibitem[Brown et al.(2006)]{Brown2006} Brown, T.~M., Smith, E., Ferguson, H.~C., et al.\ 2006, \apj, 652, 323

\bibitem[Cardelli et al.(1989)]{cardelli89} Cardelli, J.~A., Clayton, G.~C., \& Mathis, J.~S.\ 1989, \apj, 345, 245

\bibitem[Cepa \& Beckman(1988)]{Cepa88} Cepa, J. \& Beckman, J.~E.\ 1988, \aap, 200, 21


\bibitem[Chen et al.(2015)]{chen15} Chen, Y., Bressan, A., Girardi, L., et al.\ 2015, \mnras, 452, 1068

\bibitem[Chen et al.(2014)]{chen14} Chen, Y., Girardi, L., Bressan, A., et al.\ 2014, \mnras, 444, 2525

\bibitem[Choi et al.(2002)]{Choi02} Choi, P.~I., Guhathakurta, P., \& Johnston, K.~V.\ 2002, \aj, 124, 310


\bibitem[Cioni \& Habing(2005)]{Cioni2005} Cioni, M.-R.~L. \& Habing, H.~J.\ 2005, \aap, 429, 837


\bibitem[Cioni et al.(2006)]{Cioni06} Cioni, M.-R.~L., Girardi, L., Marigo, P., et al.\ 2006, \aap, 448, 77

\bibitem[Cioni et al.(2008)]{cioni08} Cioni, M.-R.~L., Irwin, M., Ferguson, A.~M.~N., et al.\ 2008, \aap, 487, 131

\bibitem[Cole et al.(2014)]{Cole14} Cole, A.~A., Weisz, D.~R., Dolphin, A.~E., et al.\ 2014, \apj, 795, 54

\bibitem[Costa \& Frogel(1996)]{Costa96} Costa, E. \& Frogel, J.~A.\ 1996, \aj, 112, 2607


\bibitem[Crnojevi{\'c} et al.(2013)]{crnojevic13} Crnojevi{\'c}, D., Ferguson, A.~M.~N., Irwin, M.~J., et al.\ 2013, \mnras, 432, 832

\bibitem[Dalcanton et al.(2012a)]{dalcanton12a} Dalcanton, J.~J., Williams, B.~F., Melbourne, J.~L., et al.\ 2012a, \apjs, 198, 6


\bibitem[Dalcanton et al.(2012b)]{Dalcanton12b} Dalcanton, J.~J., Williams, B.~F., Lang, D., et al.\ 2012b, \apjs, 200, 18

\bibitem[Dalcanton et al.(2015)]{dalcanton15} Dalcanton, J.~J., Fouesneau, M., Hogg, D.~W., et al.\ 2015, \apj, 814, 3

\bibitem[Davidge(2003)]{Davidge03} Davidge, T.~J.\ 2003, \apj, 597, 289.

\bibitem[Davidge(2005)]{Davidge05} Davidge, T.~J.\ 2005, \aj, 130, 2087

\bibitem[Davidge(2014)]{davidge14} Davidge, T.~J.\ 2014, \apj, 791, 66

\bibitem[de Boer et al.(2012a)]{deBoer12a} de Boer, T.~J.~L., Tolstoy, E., Hill, V., et al.\ 2012a, \aap, 539, A103

\bibitem[de Boer et al.(2012a)]{deBoer12b} de Boer, T.~J.~L., Tolstoy, E., Hill, V., et al.\ 2012b, \aap, 544, A73

\bibitem[de Boer et al.(2014)]{deBoer14} de Boer, T.~J.~L., Tolstoy, E., Lemasle, B., et al.\ 2014, \aap, 572, A10


\bibitem[de Boer et al.(2011)]{deBoer11} de Boer, T.~J.~L., Tolstoy, E., Saha, A., et al.\ 2011, \aap, 528, A119



\bibitem[de Grijs \& Bono(2014)]{deGrjis14} de Grijs, R. \& Bono, G.\ 2014, \aj, 148, 17

\bibitem[de Vaucouleurs et al.(1991)]{deVaucouleurs91} de Vaucouleurs, G., de Vaucouleurs, A., Corwin, H.~G., et al.\ 1991, Third Reference Catalogue of Bright Galaxies. Springer, New York, NY (USA), 1991


\bibitem[Dey et al.(2023)]{Dey2023} Dey, A., Najita, J.~R., Koposov, S.~E., et al.\ 2023, \apj, 944, 1

\bibitem[Dierickx et al.(2014)]{Dierickx14} Dierickx, M., Blecha, L., \& Loeb, A.\ 2014, \apjl, 788, L38


\bibitem[Dolphin(2002)]{dolphin02} Dolphin, A.~E.\ 2002, \mnras, 332, 91

\bibitem[Dolphin(2012)]{dolphin12} Dolphin, A.~E.\ 2012, \apj, 751, 60

\bibitem[Dolphin(2013)]{dolphin13} Dolphin, A.~E.\ 2013, \apj, 775, 76

\bibitem[Dolphin et al.(2003)]{Dolphin2003} Dolphin, A.~E., Saha, A., Skillman, E.~D., et al.\ 2003, \aj, 126, 187

\bibitem[Dorman et al.(2015)]{Dorman2015} Dorman, C.~E., Guhathakurta, P., Seth, A.~C., et al.\ 2015, \apj, 803, 24


\bibitem[Draine et al.(2014)]{draine14} Draine, B.~T., Aniano, G., Krause, O., et al.\ 2014, \apj, 780, 172

\bibitem[D'Souza \& Bell(2018)]{dsouza18} D'Souza, R. \& Bell, E.~F.\ 2018, Nature Astronomy, 2, 737

% \bibitem[Durbin et al.(2023)]{durbin23} Durbin, M.~J., Beaton, R.~L., Monson, A.~J., et al.\ 2023, \aj, 166, 236

% \bibitem[Escala et al.(2020)]{escala20} Escala, I., Gilbert, K.~M., Kirby, E.~N., et al.\ 2020, \apj, 889, 17

\bibitem[Fardal et al.(2008)]{fardal08} Fardal, M.~A., Babul, A., Guhathakurta, P., et al.\ 2008, \apjl, 682, L33




\bibitem[Fardal et al.(2006)]{fardal06} Fardal, M.~A., Babul, A., Geehan, J.~J., et al.\ 2006, \mnras, 366, 1012

\bibitem[Fardal et al.(2012)]{fardal12} Fardal, M.~A., Guhathakurta, P., Gilbert, K.~M., et al.\ 2012, \mnras, 423, 3134

% \bibitem[Feast \& Whitelock(1992)]{feast92} Feast, M.~W. \& Whitelock, P.~A.\ 1992, \mnras, 259, 6

\bibitem[Ferguson et al.(2002)]{ferguson02} Ferguson, A.~M.~N., Irwin, M.~J., Ibata, R.~A., et al.\ 2002, \aj, 124, 1452

\bibitem[Ferguson et al.(2005)]{ferguson05} Ferguson, A.~M.~N., Johnson, R.~A., Faria, D.~C., et al.\ 2005, \apjl, 622, L109

\bibitem[Frogel \& Blanco(1983)]{Frogel83} Frogel, J.~A. \& Blanco, V.~M.\ 1983, \apjl, 274, L57

\bibitem[Gallart et al.(1996)]{Gallart96} Gallart, C., Aparicio, A., Bertelli, G., et al.\ 1996, \aj, 112, 1950

\bibitem[Gallart et al.(2005)]{gallart05} Gallart, C., Zoccali, M., \& Aparicio, A.\ 2005, \araa, 43, 387

\bibitem[Geha et al.(2015)]{Geha2015} Geha, M., Weisz, D., Grocholski, A., et al.\ 2015, \apj, 811, 114

% \bibitem[Girardi et al.(2020)]{girardi20} Girardi, L., Boyer, M.~L., Johnson, L.~C., et al.\ 2020, \apj, 901, 19

% \bibitem[Girardi et al.(2000)]{girardi00} Girardi, L., Bressan, A., Bertelli, G., et al.\ 2000, \aaps, 141, 371

% \bibitem[Girardi et al.(2008)]{girardi08} Girardi, L., Dalcanton, J., Williams, B., et al.\ 2008, \pasp, 120, 583.


% \bibitem[Goldman et al.(2022)]{goldman22} Goldman, S.~R., Boyer, M.~L., Dalcanton, J., et al.\ 2022, \apjs, 259, 41

\bibitem[Gordon et al.(2006)]{Gordon06} Gordon, K.~D., Bailin, J., Engelbracht, C.~W., et al.\ 2006, \apjl, 638, L87


\bibitem[Gregersen et al.(2015)]{gregersen15} Gregersen, D., Seth, A.~C., Williams, B.~F., et al.\ 2015, \aj, 150, 189

\bibitem[Habing \& Olofsson(2003)]{habing03} Habing, H.~J. \& Olofsson, H.\ 2003, Asymptotic Giant Branch stars (New
York: Springer)

\bibitem[Hamedani Golshan et al.(2017)]{hamedani17} Hamedani Golshan, R., Javadi, A., van Loon, J.~T., et al.\ 2017, \mnras, 466, 1764

\bibitem[Hamren et al.(2015)]{hamren15} Hamren, K.~M., Rockosi, C.~M., Guhathakurta, P., et al.\ 2015, \apj, 810, 60

\bibitem[Harmsen et al.(2023)]{harmsen23} Harmsen, B., Bell, E.~F., D'Souza, R., et al.\ 2023, \mnras, 525, 4497

\bibitem[Harris et al.(2020)]{harris20} Harris, C.~R., Millman, K.~J., van der Walt, S.~J., et al.\ 2020, \nat, 585, 357

\bibitem[Hidalgo et al.(2011)]{Hidalgo2011} Hidalgo, S.~L., Aparicio, A., Skillman, E., et al.\ 2011, \apj, 730, 14

\bibitem[Hidalgo et al.(2018)]{Hidalgo2018} Hidalgo, S.~L., Pietrinferni, A., Cassisi, S., et al.\ 2018, \apj, 856, 125


\bibitem[Howley et al.(2008)]{Howley2008} Howley, K.~M., Geha, M., Guhathakurta, P., et al.\ 2008, \apj, 683, 722


\bibitem[Hodgkin et al.(2009)]{hodgkin09} Hodgkin, S.~T., Irwin, M.~J., Hewett, P.~C., et al.\ 2009, \mnras, 394, 675

\bibitem[Hoyt et al.(2024)]{Hoyt2024} Hoyt, T.~J., Jang, I.~S., Freedman, W.~L., et al.\ 2024, arXiv:2407.07309


\bibitem[Huang et al.(2020)]{huang20} Huang, C.~D., Riess, A.~G., Yuan, W., et al.\ 2020, \apj, 889, 5

\bibitem[Huang et al.(2024)]{huang24} Huang, C.~D., Yuan, W., Riess, A.~G., et al.\ 2024, \apj, 963, 83

\bibitem[Hunt et al.(2024)]{Hunt24} Hunt, L.~K., Annibali, F., Cuillandre, J.-C., et al.\ 2024, arXiv:2405.13499

\bibitem[Hunter(2007)]{hunter07} Hunter, J.~D.\ 2007, Computing in Science and Engineering, 9, 90

\bibitem[Ibata et al.(2004)]{Ibata04} Ibata, R., Chapman, S., Ferguson, A.~M.~N., et al.\ 2004, \mnras, 351, 117

\bibitem[Iben \& Renzini(1983)]{Iben83} Iben, I. \& Renzini, A.\ 1983, \araa, 21, 271

\bibitem[Irwin et al.(2004)]{irwin04} Irwin, M.~J., Lewis, J., Hodgkin, S., et al.\ 2004, \procspie, 5493, 41

\bibitem[Jarrett et al.(2003)]{Jarrett03} Jarrett, T.~H., Chester, T., Cutri, R., et al.\ 2003, \aj, 125, 525

\bibitem[Javadi et al.(2011)]{javadi11} Javadi, A., van Loon, J.~T., \& Mirtorabi, M.~T.\ 2011, \mnras, 414, 3394

\bibitem[Jeffery et al.(2011)]{Jeffery11} Jeffery, E.~J., Smith, E., Brown, T.~M., et al.\ 2011, \aj, 141, 171

\bibitem[Jung et al.(2012)]{Jung12} Jung, M.~Y., Ko, J., Kim, J.-W., et al.\ 2012, \aap, 543, A35

\bibitem[Lazzarini et al.(2022)]{lazzarini22} Lazzarini, M., Williams, B.~F., Durbin, M.~J., et al.\ 2022, \apj, 934, 76

\bibitem[Lee(1996)]{Lee96} Lee, M.~G.\ 1996, \aj, 112, 1438

\bibitem[Lee et al.(2021)]{Lee21} Lee, A.~J., Freedman, W.~L., Madore, B.~F., et al.\ 2021, \apj, 907, 112

\bibitem[Lee(2023)]{lee23} Lee, A.~J.\ 2023, \apj, 956, 15

\bibitem[Lee et al.(2024)]{lee24} Lee, A.~J., Freedman, W.~L., Madore, B.~F., et al.\ 2024, arXiv:2408.03474

\bibitem[Li et al.(2024)]{Li24} Li, S., Riess, A.~G., Casertano, S., et al.\ 2024, \apj, 966, 20

\bibitem[Liu et al.(2024)]{Liu2024} Liu, Z., Wang, J., Jing, Y., et al.\ 2024, Research in Astronomy and Astrophysics, 24, 085005

\bibitem[Lewis et al.(2015)]{Lewis15} Lewis, A.~R., Dolphin, A.~E., Dalcanton, J.~J., et al.\ 2015, \apj, 805, 183

\bibitem[Kroupa(2001)]{kroupa01} Kroupa, P.\ 2001, \mnras, 322, 231

\bibitem[Madore \& Freedman(2020)]{Madore20} Madore, B.~F. \& Freedman, W.~L.\ 2020, \apj, 899, 66

\bibitem[Maraston et al.(2006)]{Maraston06} Maraston, C., Daddi, E., Renzini, A., et al.\ 2006, \apj, 652, 85

\bibitem[Marigo et al.(2013)]{marigo13} Marigo, P., Bressan, A., Nanni, A., et al.\ 2013, \mnras, 434, 488

\bibitem[Marigo \& Girardi(2007)]{Marigo07} Marigo, P. \& Girardi, L.\ 2007, \aap, 469, 239

\bibitem[Marigo et al.(2008)]{marigo08} Marigo, P., Girardi, L., Bressan, A., et al.\ 2008, \aap, 482, 883

\bibitem[Marigo et al.(2017)]{marigo17} Marigo, P., Girardi, L., Bressan, A., et al.\ 2017, \apj, 835, 77

\bibitem[McConnachie(2012)]{McConnachie2012} McConnachie, A.~W.\ 2012, \aj, 144, 4

\bibitem[McConnachie et al.(2004)]{McConnachie2004} McConnachie, A.~W., Irwin, M.~J., Lewis, G.~F., et al.\ 2004, \mnras, 351, L94

\bibitem[McConnachie et al.(2005)]{mcconnachie05} McConnachie, A.~W., Irwin, M.~J., Ferguson, A.~M.~N., et al.\ 2005, \mnras, 356, 979

\bibitem[McConnachie et al.(2018)]{mcconnachie18} McConnachie, A.~W., Ibata, R., Martin, N., et al.\ 2018, \apj, 868, 55



\bibitem[McKinney(2010)]{pandas} McKinney, W.\ 2010, Proceedings of the 9th Python in Science Conference, 51-56

\bibitem[McQuinn et al.(2015)]{McQuinn2015} McQuinn, K.~B.~W., Cannon, J.~M., Dolphin, A.~E., et al.\ 2015, \apj, 802, 66

\bibitem[McQuinn et al.(2017)]{McQuinn2017} McQuinn, K.~B.~W., Boyer, M.~L., Mitchell, M.~B., et al.\ 2017, \apj, 834, 78



\bibitem[McQuinn et al.(2024a)]{McQuinn2024a} McQuinn, K.~B.~W., B. Newman, M.~J., Savino, A., et al.\ 2024a, \apj, 961, 16

\bibitem[McQuinn et al.(2024b)]{McQuinn2024b} McQuinn, K.~B.~W., Mao, Y.-Y., Tollerud, E.~J., et al.\ 2024b, \apj, 967, 161

\bibitem[McQuinn et al.(2024c)]{McQuinn2024c} McQuinn, K.~B.~W., Newman, M.~J.~B., Skillman, E.~D., et al.\ 2024, arXiv:2409.19050



\bibitem[McQuinn et al.(2010)]{McQuinn10} McQuinn, K.~B.~W., Skillman, E.~D., Cannon, J.~M., et al.\ 2010, \apj, 721, 297



\bibitem[Melbourne \& Boyer(2013)]{Melbourne2013} Melbourne, J. \& Boyer, M.~L.\ 2013, \apj, 764, 30

\bibitem[Melbourne et al.(2010)]{melbourne10} Melbourne, J., Williams, B., Dalcanton, J., et al.\ 2010, \apj, 712, 469

\bibitem[Melbourne et al.(2012)]{Melborne2012} Melbourne, J., Williams, B.~F., Dalcanton, J.~J., et al.\ 2012, \apj, 748, 47

\bibitem[Monachesi et al.(2012)]{Monachesi2012} Monachesi, A., Trager, S.~C., Lauer, T.~R., et al.\ 2012, \apj, 745, 97

\bibitem[Monaco et al.(2009)]{Monaco09} Monaco, L., Saviane, I., Perina, S., et al.\ 2009, \aap, 502, L9

\bibitem[Monelli et al.(2010)]{Monelli10} Monelli, M., Hidalgo, S.~L., Stetson, P.~B., et al.\ 2010, \apj, 720, 1225

\bibitem[Mu{\~n}oz et al.(2018)]{Munoz18} Mu{\~n}oz, R.~R., C{\^o}t{\'e}, P., Santana, F.~A., et al.\ 2018, \apj, 860, 66

\bibitem[Nally et al.(2024)]{Nally24} Nally, C., Jones, O.~C., Lenki{\'c}, L., et al.\ 2024, \mnras, 531, 183


\bibitem[Newman et al.(2024)]{Newman2024} Newman, M.~J.~B., McQuinn, K.~B.~W., Skillman, E.~D., et al.\ 2024, arXiv:2406.03532

\bibitem[Nieten et al.(2006)]{nieten06} Nieten, C., Neininger, N., Gu{\'e}lin, M., et al.\ 2006, \aap, 453, 459

\bibitem[Norris et al.(1981)]{Norris1981} Norris, J., Cottrell, P.~L., Freeman, K.~C., et al.\ 1981, \apj, 244, 205


% \bibitem[Olszewski et al.(1996)]{olszewski96} Olszewski, E.~W., Suntzeff, N.~B., \& Mateo, M.\ 1996, \araa, 34, 511

\bibitem[Parada et al.(2021)]{Parada21} Parada, J., Heyl, J., Richer, H., et al.\ 2021, \mnras, 501, 933

\bibitem[Pastorelli et al.(2019)]{pastorelli19} Pastorelli, G., Marigo, P., Girardi, L., et al.\ 2019, \mnras, 485, 5666

\bibitem[Pastorelli et al.(2020)]{pastorelli20} Pastorelli, G., Marigo, P., Girardi, L., et al.\ 2020, \mnras, 498, 3283

\bibitem[Pedregosa et al.(2011)]{pedregosa11} Pedregosa, F., Varoquaux, G., Gramfort, A., et al.\ 2011, Journal of Machine Learning Research, 12, 2825

\bibitem[Reid \& Mould(1984)]{Reid84} Reid, N. \& Mould, J.\ 1984, \apj, 284, 98

\bibitem[Rejkuba et al.(2022)]{rejkuba22} Rejkuba, M., Harris, W.~E., Greggio, L., et al.\ 2022, \aap, 657, A41

\bibitem[Ren et al.(2021)]{ren21} Ren, Y., Jiang, B., Yang, M., et al.\ 2021, \apj, 907, 18

\bibitem[Ren et al.(2024)]{ren24} Ren, Y., Jiang, B., Wang, Y., et al.\ 2024, \apj, 966, 25

\bibitem[Rezaeikh et al.(2014)]{rezaeikh14} Rezaeikh, S., Javadi, A., Khosroshahi, H., et al.\ 2014, \mnras, 445, 2214

\bibitem[Richardson et al.(2008)]{richardson08} Richardson, J.~C., Ferguson, A.~M.~N., Johnson, R.~A., et al.\ 2008, \aj, 135, 1998

\bibitem[Richardson et al.(2011)]{Richardson11} Richardson, J.~C., Irwin, M.~J., McConnachie, A.~W., et al.\ 2011, \apj, 732, 76


\bibitem[Richer et al.(1984)]{Richer84} Richer, H.~B., Crabtree, D.~R., \& Pritchet, C.~J.\ 1984, \apj, 287, 138


\bibitem[Ripoche et al.(2020)]{Ripoche20} Ripoche, P., Heyl, J., Parada, J., et al.\ 2020, \mnras, 495, 285

\bibitem[Rose et al.(2005)]{Rose05} Rose, J.~A., Arimoto, N., Caldwell, N., et al.\ 2005, \aj, 129, 712


\bibitem[Rosenfield et al.(2014)]{Rosenfield2014} Rosenfield, P., Marigo, P., Girardi, L., et al.\ 2014, \apj, 790, 22

\bibitem[Rosenfield et al.(2016)]{rosenfield16} Rosenfield, P., Marigo, P., Girardi, L., et al.\ 2016, \apj, 822, 73

\bibitem[Savino et al.(2025)]{Savino25} Savino, A., Weisz, D.~R., Dolphin, A.~E., et al.\ 2025, \apj, 979, 205


\bibitem[Savino et al.(2022)]{Savino2022} Savino, A., Weisz, D.~R., Skillman, E.~D., et al.\ 2022, \apj, 938, 101



\bibitem[Savino et al.(2023)]{Savino2023} Savino, A., Weisz, D.~R., Skillman, E.~D., et al.\ 2023, \apj, 956, 86



\bibitem[Schlafly \& Finkbeiner(2011)]{schlafly11} Schlafly, E.~F. \& Finkbeiner, D.~P.\ 2011, \apj, 737, 103

\bibitem[Schlegel et al.(1998)]{schlegel98} Schlegel, D.~J., Finkbeiner, D.~P., \& Davis, M.\ 1998, \apj, 500, 

\bibitem[Sharina et al.(2006)]{Sharina06} Sharina, M.~E., Afanasiev, V.~L., \& Puzia, T.~H.\ 2006, \mnras, 372, 1259

\bibitem[Skillman et al.(2017)]{Skillman17} Skillman, E.~D., Monelli, M., Weisz, D.~R., et al.\ 2017, \apj, 837, 102525

\bibitem[Skillman et al.(2003)]{Skillman2003} Skillman, E.~D., Tolstoy, E., Cole, A.~A., et al.\ 2003, \apj, 596, 253

\bibitem[Skillman et al.(2017)]{Skillman2017} Skillman, E.~D., Monelli, M., Weisz, D.~R., et al.\ 2017, \apj, 837, 102



\bibitem[Skrutskie et al.(2006)]{skrutskie06} Skrutskie, M.~F., Cutri, R.~M., Stiening, R., et al.\ 2006, \aj, 131, 1163

\bibitem[Tang et al.(2014)]{tang14} Tang, J., Bressan, A., Rosenfield, P., et al.\ 2014, \mnras, 445, 4287

% \bibitem[Tolstoy et al.(2009)]{tolstoy09} Tolstoy, E., Hill, V., \& Tosi, M.\ 2009, \araa, 47, 371

\bibitem[Velguth et al.(2024)]{Velguth24} Velguth, B.~N., Bell, E.~F., Smercina, A., et al.\ 2024, \apj, 974, 189

\bibitem[Weisz et al.(2011)]{Weisz2011} Weisz, D.~R., Dalcanton, J.~J., Williams, B.~F., et al.\ 2011, \apj, 739, 5

\bibitem[Weisz et al.(2014)]{Weisz2014} Weisz, D.~R., Dolphin, A.~E., Skillman, E.~D., et al.\ 2014, \apj, 789, 147

\bibitem[Weisz et al.(2023a)]{Weisz23a} Weisz, D.~R., Savino, A., \& Dolphin, A.~E.\ 2023a, \apj, 948, 50

\bibitem[Weisz et al.(2023b)]{Weisz23b} Weisz, D.~R., McQuinn, K.~B.~W., Savino, A., et al.\ 2023b, \apjs, 268, 15

\bibitem[Williams et al.(2009)]{Williams09} Williams, B.~F., Dalcanton, J.~J., Seth, A.~C., et al.\ 2009, \aj, 137, 41

\bibitem[Williams et al.(2017)]{williams17} Williams, B.~F., Dolphin, A.~E., Dalcanton, J.~J., et al.\ 2017, \apj, 846, 145



\bibitem[Williams et al.(2023)]{williams23} Williams, B.~F., Durbin, M., Lang, D., et al.\ 2023, \apjs, 268, 48

\bibitem[Williams et al.(2014)]{williams14} Williams, B.~F., Lang, D., Dalcanton, J.~J., et al.\ 2014, \apjs, 215, 9

\bibitem[Wood et al.(1985)]{Wood85} Wood, P.~R., Bessell, M.~S., \& Paltoglou, G.\ 1985, \apj, 290, 477

\bibitem[Zgirski et al.(2021)]{Zgirski21} Zgirski, B., Pietrzy{\'n}ski, G., Gieren, W., et al.\ 2021, \apj, 916, 19

\end{thebibliography}
\end{document}